\documentclass[apj]{emulateapj}

\shorttitle{Globular Cluster Systems}
\shortauthors{Harris}

\begin{document}

\title{Globular Cluster Systems in Giant Ellipticals:  the Mass/Metallicity Relation\footnote{Based on 
observations with the NASA/ESA {\sl Hubble Space Telescope}, obtained at the 
Space Telescope Science Institute, which is operated by the Association of 
Universities for Research in Astronomy, Inc., under NASA contract NAS 5-26555.}
}

\author{William E. Harris}
\affil{Department of Physics \& Astronomy, McMaster University,
  Hamilton ON L8S 4M1}
\email{harris@physics.mcmaster.ca}

\begin{abstract}
Data from the Hubble Space Telescope taken with the ACS/WFC
camera have been used to investigate the globular cluster (GC) populations
around six giant elliptical galaxies that are $\sim 40$ Mpc
distant.  From these six fields, imaged in $B$ and $I$, a total of more than 15000
candidate GCs have been measured, of which $8000$ or more
are high-probability globular clusters.  The data  reach a 
limiting magnitude near $M_I \simeq -8$, about 0.4 mag fainter
than the GC luminosity function turnover point, and thus
thoroughly cover the bright half of the GC population. 
Most of the individual GCs on these images are marginally resolved nonstellar objects,
so King-model profiles convolved with the stellar point-spread functions
are used to measure their individual total magnitudes, colors, and
linear effective radii.  
The classic bimodal form of the GC color-magnitude distribution shows
up unambiguously in all the galaxies, allowing an accurate definition of
the mean colors along each of the two sequences
as a function of magnitude (the 
\emph{mass/metallicity relation} or MMR).  
The blue, metal-poor cluster sequence shows a clearly defined but
nonlinear MMR:  in this particular photometric dataset the mean GC color changes
smoothly from a near-vertical sequence at low luminosity
$(M_I \gtrsim -9.5$) to an increasingly redward slope
at higher luminosity.  By contrast, the red, metal-rich sequence
shows little trace of a MMR and is nearly vertical at all
luminosities.  The form and slope of the MMR along either sequence
do not depend strongly on either cluster size $r_h$ or
galactocentric distance $R_{gc}$.

All the observed features of the present data
agree with the interpretation that the MMR is created
primarily by GC \emph{self-enrichment}, along the lines of
the quantitative model of \citet{bai08}.  During the protocluster
formation stage, the more massive GCs are better able to hold
back the enriched products of the earliest supernovae and to
seed the lower-mass stars still in formation. The ``threshold''
mass at which this effect should become noticeable is near 1 million Solar masses,
which is closely consistent with the transition region that is seen in
the data.  More generally, the data favor models in which the 
star formation efficiency in a protocluster is roughly independent
of mass, and in which the gas retention efficiency is a strong
function of mass.

Correlation of the median scale sizes $r_h$ of the GCs with
other parameters shows that the metal-poor clusters are
consistently 17\% larger than those of the metal-rich clusters,
and that this difference holds at all galactocentric distances
and luminosities.  At the same time, cluster size scales
with halo location as $r_h \sim R_{gc}^{0.11}$, indicating
that \emph{both} metallicity and the external tidal environment 
play roles in determining the scale size of a given cluster.
Lastly, both the red and blue GC components show 
\emph{metallicity gradients} with galactocentric distance that are shallow but real:
heavy-element abundance scales as $Z \sim R_{gc}^{-0.1}$ for
both types.
\end{abstract}


\section{Introduction}

Much of the recent observational work on the populations of
globular clusters (GCs) in other galaxies has concentrated
on their metallicity distribution function (MDF).
A near-universal result, confirmed from both photometric
and spectroscopic samples of GCs in many galaxies of all types,
is that the MDF is bimodal, strongly suggestive of two major
epochs of cluster formation \citep[see, e.g.][for an extensive range
of results and discussions]{zepf93,larsen01,har06,pen06,kun07,str07,weh08,
pen08,wat09}.
This bimodality paradigm has steadily been reinforced 
as the quality and size of the databases have increased.

A more recent discovery is the existence of intriguing second-order
structure in the MDF.  In some large galaxies a correlation between
color and luminosity has been found along the bluer, more metal-poor
GC sequence \citep{har06,str06,mie06}.  Since the integrated color of
old star clusters depends predominantly on their metallicity, this correlation
represents a \emph{mass/metallicity relation} or MMR
(the trend is also referred to colloquially as a blue tilt, but ``MMR'' is
more accurate and more general). 
The essence of the trend is that 
the more massive blue GCs are progressively more heavy-element enriched.  
Just as intriguingly, the redder and more metal-rich GC sequence has
not shown any definite evidence for a MMR even in galaxies where the
blue-sequence MMR is strongly present.
Early hints of this effect were, in fact, clearly noticed by
\citet{ostrov98} and \citet{dirsch03} from wide-field photometric
surveys of the GC system in the Fornax cD NGC 1399.  There, they 
pointed out that the highest-luminosity GCs showed a broad unimodal
color distribution, as if the blue and red components had merged together 
at the top end.  

To make the
effect more challenging to understand, some large galaxies such as NGC 4472
may not show the blue-sequence MMR \citep{str06,mie06} even though the measured
sample of GCs seems large enough to reveal it, if present.  M87, at first claimed
to have a MMR \citep{str06}, does not show it according to a larger set of data
analyzed by \citet{wat09}.  As will be discussed more extensively below,
a critical factor in deciding the presence of an MMR is not
just the total measured population of GCs, but also the highest luminosity (mass) to which
the sample reaches.  These factors are also crucial in deciding whether
or not the MMR might be \emph{linear} in color versus magnitude (as is assumed
in most previous papers), or \emph{nonlinear}, which in turn is a
central issue for the model interpretations that are now starting
to appear.

Early descriptive interpretations for the physical cause of the MMR have been suggested in the
discovery papers \citep{har06,str06,mie06}, but the most promising
direction at present for producing the effect
is organized around the idea of GC self-enrichment \citep{str08,bai08}.
The basic approach is that clusters 
forming within massive protocluster gas clouds 
can hold on to some portion of their 
SNe ejecta during the first round of star formation and thus
enrich the still-forming lower-mass stars in the cluster. The
higher the proto-GC mass, the higher the gas retention, thus leading to a MMR.
The quantitative model of \citet{bai08} shows that the amount
of self-enrichment should be negligible for protoclusters with 
$M \lesssim 10^6 M_{\odot}$, which are small enough that
SN ejecta easily escape their potential well.
Thus for clusters smaller than this threshold, 
the cluster metallicity should be uncorrelated with mass; that is, 
\emph{the GC sequences should be vertical below this transition-mass point.} 

In addition, the Bailin/Harris model
predicts that a \emph{red-sequence} MMR should exist, though 
this should become noticeable only at still higher
GC mass and with a smaller amplitude because the effects of internal
self-enrichment will be a smaller fraction of the initial (pre-enriched)
cluster metallicity at any mass.  There is even a hint that this 
red-sequence MMR has already been observed in the extremely populous
cluster systems in NGC 3311 and NGC 4874 \citep{weh08,har09,bai08}.

At the same time as the interpretive models have advanced, the state of
the observations is still somewhat confused.  The initial claim of
\citet{har06} was that, at masses less than $\sim 10^6 M_{\odot}$,
the blue sequence is vertical (i.e., color is
uncorrelated with luminosity) 
and that only the higher-mass range showed the correlation.  
Other papers discuss the MMR in terms of a 
linear relation between color and metallicity extending
continuously downward  
\citep{str06,spi06,can07,deg07,weh08,spi08,coc09}.

Another, and extremely important, source of confusion is understanding why
various individual galaxies do not show an obvious MMR; for example, comments have been raised
that the Milky Way does not have the effect.  However, the Milky Way 
does not have enough clusters to reveal a MMR with any statistical
confidence \emph{whether or not} it is present. The effect of
small-number statistics is especially important if the MMR affects
only the higher-mass GCs, which are the rarest.

With the benefit of hindsight,
it is easy to understand why these correlations were not noticed 
earlier: the MMR slope is relatively modest, so 
big statistical samples of clusters \emph{and} precise, metallicity-sensitive
photometry are both needed to see it.  Few such samples existed till
quite recent years.  
One way to put this into perspective is to note
that the basic phenomenon of bimodality is a \emph{first-order}
feature of the GC metallicity distribution that becomes obvious even with samples
as small as $\sim 10^2$ clusters (the prime example is the Milky Way itself, which was the first
galaxy in which the bimodality paradigm became clear 
\citep{zin85}).  By contrast, the MMR is a \emph{second-order} feature and thus in rough terms
can be expected to require samples of $\sim 10^3$ clusters to become clearly
detectable.

In giant ellipticals it is possible to
measure hundreds of GCs well above the million-Solar-mass level 
where the MMR is agreed to be most visible; for example, the GC sample to be described
in this paper (see below) has 750 clusters brighter than $M_I = -10.5$,
whereas in the Milky Way there is only one such cluster, $\omega$ Centauri.
The same comparison is true of dwarf galaxies and most disk galaxies, where
any attempts to search for the MMR are severely hampered by small-number
statistics even with high-quality photometry or spectroscopy.
In other words, \emph{for
most galaxies on an individual basis, the existence of a MMR is
not decidable}.  \citet{mie06}, in their study of GCs in 76 Virgo
galaxies, worked around this issue by combining their galaxies into four
luminosity groups (dwarfs through supergiants) to gain increased statistical
weight.  Even so, the dwarf sample contains too few GCs in the critical
high-luminosity regime to clearly reveal a color/luminosity slope there.

A challenge to the existence of the effect itself has been presented by \citet{kun08}.
His claim is that the MMR is spurious, an artifact of the combination
of aperture photometry, low signal-to-noise, and a mass/radius relation for GCs.  These concerns
will be dealt with in this paper in the sections below.  It is, 
however, already worth emphasizing
that the three discovery papers \citep{har06,str06,mie06} employed three
different photometric measurement procedures over a wide variety of galaxies,
yielding similar MMR slopes.  In addition, \emph{ground-based} photometric
studies of
the GCSs in distant galaxies, in which the individual GCs are completely
unresolved and any differential-size issues become moot, also show
the MMR \citep{dirsch03,weh08}.  Nevertheless, given particularly
the growing state of the theoretical modelling, 
the data need to be put onto a better foundation.

The largest GCS populations of all, and in principle the best for
a clear measurement of the MMR, are found in giant and supergiant ellipticals,
many of which are cD's or Brightest Cluster
Galaxies (BCGs) at the centers of rich clusters
\citep{h01,hpm95,blake99,har06}. 
In this paper, I present new and homogeneous photometry for
six gE's with globular cluster systems that are among the largest known.

Throughout this paper, a distance scale $H_0 = 70$ km s$^{-1}$
Mpc$^{-1}$ is used to convert redshifts and angular diameters to true 
distances and linear radii.

The plan for this rather long paper is as follows:  Section 2 introduces
the set of galaxies being studied and the goals for the measurement,
while Section 3 steps through the complete process of measurement of the
cluster magnitudes, colors, and linear sizes, and Section 4 presents
an overview of the final color-magnitude data.
Section 5 describes the complete analysis of the GC effective radii
(scale sizes) and traces their correlations with external parameters
such as metallicity, galactocentric distance, and metallicity.
Section 6 presents the full results for the mass/metallicity relation
along both the blue and red GC sequences, and its implications for
the theoretical models.  Lastly, Section 7 presents new results for
the spatial metallicity gradients in the red and blue GC subsystems.
The discussion concludes with a summary of the findings in Section 8,
and some concerns in Section 9 that may affect both the basic nature
of the observations and its links with theory.

\section{The Data Sample}

The raw data are for six galaxies drawn from the giant ellipticals
observed in HST program GO 9427 (Harris, PI).  These six,
listed in Table 1, are all imaged with the ACS/WFC in the two broadband
filters $F435W$ ($B$) and $F814W$ ($I$), and are the ones in this particular
program that turned out to have the largest GC populations.  
As noted in Table 1, most are at distances $\sim 40$ Mpc.
The data from this program also contained three other giant E targets,
NGC 5322, 5557, and 7049:  these three had distinctly smaller
GC populations, and as will be described below, they were used instead
to help define the level of field contamination.\footnote{NGC 7626 
was not included in the 2006 paper but is added here
since the raw images are completely homogeneous with the other five.  The full
data analysis for NGC 7626 will be described in Harris et al. (2009, in preparation)
along with four other giant ellipticals that have $B-$band images from ACS/WFC
and $I-$band images from archival WFPC2 exposures.}  

In the original data paper for this program \citep{har06}, 
both PSF-fitting and fixed-aperture photometry were used to measure
the GCs.
However, neither approach is strictly correct, because
many of the GCs are not quite small enough in linear size
to appear starlike on the images, and furthermore
their intrinsic sizes are not
identical from cluster to cluster.  Various tests in the 
literature \citep[e.g.][]{lar99,jor05,mie06,har09}
demonstrate that when the FWHM of the star cluster is smaller than about 10\% of
the FWHM of the stellar point-spread function (PSF), its size can no longer be 
determined from these conventional
imaging techniques and it can be treated as starlike. But if it is bigger,
its size needs to be individually accounted for to determine
a correct total magnitude.  For purposes of comparative discussion,
we can usefully distinguish four different regimes of resolution:

\begin{itemize}
\item{} \emph{Well resolved}:  The FWHM or effective diameter of the GC is
much larger than the PSF,
$FWHM(GC) >> FWHM(PSF)$.  This is the case, for example, for HST imaging
of clusters in M31 and other Local Group galaxies \citep[e.g.][]{bar07}.
\item{} \emph{Partially resolved}: The GC and stellar profiles have
comparable widths, $FWHM(GC) \sim FWHM(PSF)$.  This is the case for
HST imaging of 
many of the GCs in the Virgo galaxies \citep{jor05,mie06} at $d = 16$ Mpc
and for NGC 5128 \citep{gom07,mcl08} at $d = 4$ Mpc, for example.
\item{} \emph{Marginally resolved}:  The GC is barely large
enough to be measured relative to the PSF, $FWHM(GC) \sim 0.1 - 0.3~FWHM(PSF)$.
The GCs analyzed in the present paper fall into this category.
\item{} \emph{Unresolved}:  $FWHM(GC) < 0.1~FWHM(PSF)$, for all practical
purposes making it starlike and allowing normal PSF-fitting photometry.
GCs in much more distant galaxies such as Coma \citep{har09}, even with
HST resolution, fall into this category.
\end{itemize}

\begin{figure}
\plotone{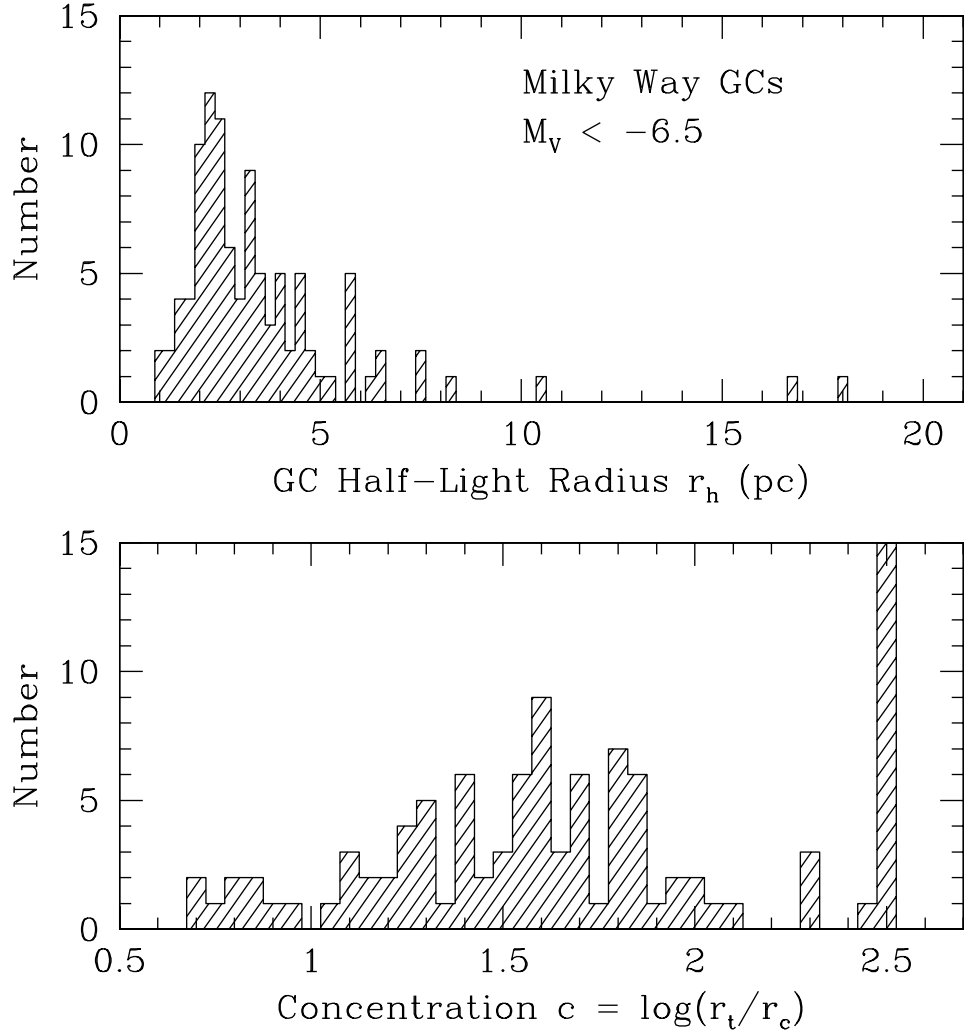}
\caption{\emph{Upper panel:} Distribution of half-light radii for Milky Way globular clusters.
All 102 such clusters with luminosities brighter than $M_V=-6.5$ are included here,
to roughly match the luminosity range of the GCs observed in the six target
galaxies of this study.  Ninety percent of them have $r_h < 6$ parsecs.
Data are from \citet{har96}.
\emph{Lower panel:} Distribution of King-model central concentration $c = {\rm log} (r_t/r_c)$
for the same sample of Milky Way clusters above.  The bin at $c=2.5$ is populated by
the core-collapsed GCs, for which $c=2.5$ is arbitrarily assigned.
}
\label{rh_mwgc}
\end{figure}

The mean half-light radius of classic GCs is typically about 3 pc,
and 90\% of them have $r_h \lesssim 6$ parsecs (Figure \ref{rh_mwgc}).
Thus at $d \simeq 40$ Mpc, they will subtend a typical angular diameter
$\simeq 0.03''$, which equals about 30\% of the $0.1''$ FWHM of stars on the
HST ACS camera.  This comparison means that we should expect the
GCs in these target galaxies to fall in the \emph{marginally resolved}
or (for the smallest ones) \emph{unresolved} regime.  
Although the differential aperture-size corrections to their photometry can be
expected to be small, the purpose of the present paper is to determine
these corrections and to construct a new database of photometry within
which any MMR can be more securely determined.

\section{Photometric Techniques}

\subsection{Defining the PSF}

Each target galaxy discussed in this paper is surrounded by some thousands of barely-nonstellar
globular clusters plus a variety of `field' contaminants (a relatively few foreground
stars plus numerous faint, small background galaxies).  As will be seen 
below, the GC populations clearly outnumber all other types of objects.
The first task for the photometry is to derive individual profile-size measurements
of each detected object.
PSF-convolved model GC profiles must be matched to
each individual object and the optimum fit selected.  Several codes have
been developed and used in the literature to do this, including the ISHAPE code
of \citet{lar99} and \citet{larsen01} which has 
been used for young and old star clusters in a wide range of distant
galaxies; the KINGPHOT code of \citet{jor05} used principally for the GCs in
Virgo galaxies; and the code of \citet{mcl08} which can fit any one of
five standard GC models and has been used for clusters in M31 and
NGC 5128.  

The fitting code adopted for this work is ISHAPE.   This code
gives a fast, well tested, effective procedure for fitting \emph{marginally} resolved GCs like 
these, where we are primarily interested in determining the cluster size $r_{eff}$
and cannot expect to obtain precise solutions for more subtle structural 
parameters such as ellipticity, orientation, or core radius.  These latter
quantities become dramatically easier to measure for star clusters in nearby galaxies, such as
for Virgo, NGC 5128, or the Local Group members where the degree of resolution
is much higher.  ISHAPE and its parent package BAOLAB also contain highly 
effective tools for testing the results, including profile-subtracted images,
and the ability to generate simulated populations of PSF-convolved objects
of arbitrary size and ellipticity.

It is possible to construct PSFs for HST images through
TinyTim modelling, but difficulties arise
for appropriate choices of a diffusion kernel
\citep[see, e.g.][for recent extensive discussions in similar studies]{spi06,geo08}.
To avoid any such uncertainties in the modelling
chain I have employed purely empirical PSFs constructed for each individual
frame.  Fortunately, the true FWHM of the stellar PSF on the ACS/WFC frames
is known beforehand to fall in the
range $0.09''-0.10''$ on series of images that have been properly coadded.
This \emph{a priori} constraint helps to select out candidate starlike objects.

\begin{figure}
\plotone{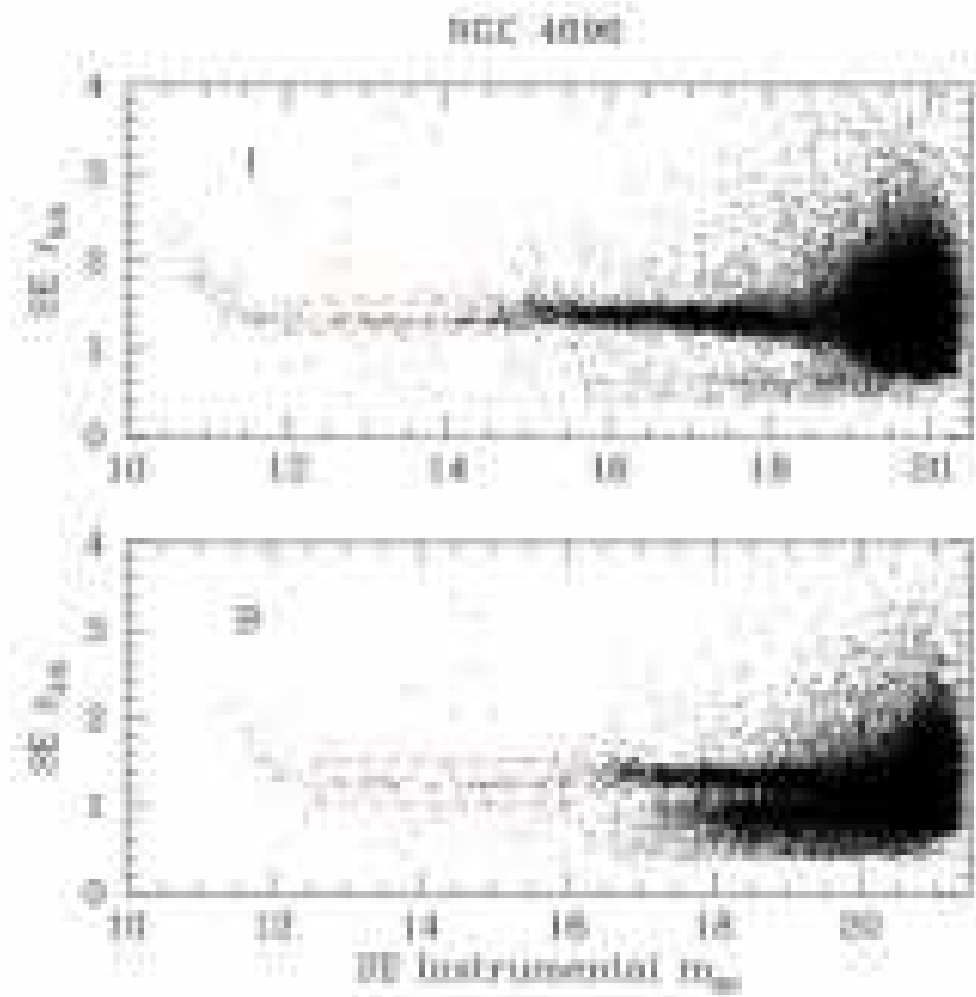}
\caption{Selection of candidate stars for defining the stellar point spread function.
The data shown here are SExtractor measurements of the half-light radius $r_{0.5}$
versus an aperture magnitude, for all detected objects in the NGC 4696 galaxy field.
No attempt has been made to cull out nonstellar objects or detections due bad pixels,
clumps of noise, or other artifacts.  The magnitude scales have arbitrary zeropoints
and do not represent the true $B,I$ magnitudes.  Note in each panel the narrow sequence of
``starlike'' sources; at the very bright end ($m_I < 12, m_B<12.5$) these become saturated 
and the sequences curve upward.  The dashed boxes mark out the objects that were
chosen as candidate PSF stars for further investigation (see text).
}
\label{reff_se}
\end{figure}

All exposures for each target field and in each filter were registered
and combined through \emph{multidrizzle},
as described in more detail 
in \citet{har06}.  To build a preliminary catalog of detected objects, the SExtractor
code \citep{ber96} was used with a generously low threshold, and from this, a diagnostic
graph of object half-light radius $r_{1/2}$ versus total magnitude was plotted.
An example for the NGC 4696 field is shown in Figure \ref{reff_se}.
The next stage was an iterative process of identifying genuine starlike, isolated 
objects across the image from which the PSF could be defined.  Objects selected
as candidate PSF stars were those within the region marked out in the Figure,
falling along the lower envelope of the $r_{1/2}$ distribution.  Moderately
bright, unsaturated stars fall within this region, whereas saturated stars
fall along the upturn region at the bright end of the distribution.  The
candidates were then inspected through IRAF/imexamine, their radial profiles
and FWHMs measured, and ones with FWHM falling in the range $1.8 - 2.0$ px 
(that is, $0.09'' - 0.10''$) were selected.  This subsample was put into
the IRAF/daophot/psf routine, inspected at finer detail, and any with
remaining excess nonstellar wings or faint neighbors rejected.  The final
numbers of objects used in each field, along with the final FWHM of the adopted
point-spread function, are listed in Table 2.  In all cases, the stars used to
define the final PSF were widely distributed in location, and thus the
PSF represents an average across the field (though no differences in the PSF
shape or width at the level of $\pm 0.1$ px as a function of position were found;
see Harris et al. 2006 for additional discussion).

Other characteristics of the raw image data including the
internal photometric precision and completeness are described in much more
detail in \citet{har06}.  Briefly, the exposures for each galaxy were
originally designed to reach an intrinsic limiting magnitude  $M_I \simeq -8$ at 50\% detection
completeness, so that the turnover point (peak frequency) of the globular cluster
luminosity function (GCLF) at $M_I \simeq -8.4$ would be reached at almost 100\% completeness.
The data then thoroughly sample about 4 magnitudes of the GCLF, essentially 
its entire bright half.  The internal photometric measurement uncertainty
is smaller than $\sigma(B-I) = \pm 0.10$ mag 
for $M_I < -8.5$ as well.  This color spread is
approximately the intrinsic Gaussian width
of the blue sequence and is smaller than the width of the red sequence,
allowing the bimodal sequences to be well resolved over the full
magnitude range of interest here.

\subsection{Measuring Effective Radius and Aperture Corrections}

As mentioned above, the single most important structural parameter we are after
is the effective or half-light radius of the GCs around these giant ellipticals.
This quantity is what determines the amount of ``aperture correction'' to apply
to the measured magnitudes. In addition, the cluster scale sizes are
interesting in their own right as possible
functions of galactocentric position and GC metallicity and will be discussed
separately below.  

Within ISHAPE, $r_{eff}$ is found by fitting the PSF-convolved cluster
profile to each object in the image, then varying the assumed $r_{eff}$ till a
best fit is achieved.  
The code also returns a $(S/N)$ ratio, which depends mainly on object brightness
and can be used to assess the quality of the result.
For any objects judged to be starlike, a value $r_{eff} = 0$
is returned.  There are other structural parameters that can be solved for in
principle, including the ellipticity $e$, orientation angle $\theta$, and central
concentration ratio, but these are easy to determine only for much more well
resolved objects than we work with here.
 
ISHAPE uses a $10 \times$-subsampled version of the IRAF/daophot PSF as a basis for
its image convolution.  For the intrinsic GC profile, it can use a variety of analytic
functions; I adopt the \citet{kin62} model since that model was defined originally to match
real GCs and has been thoroughly tested in other star-cluster programs
of this type \citep[e.g.][]{lar99,larsen01,spi06,geo08}.  The GCs in the target galaxies here
are not nearly well enough resolved to allow ISHAPE to solve independently for
the cluster concentration ratio $c \equiv \rm{log} (r_t/r_c)$ 
where $r_t, r_c$ are the tidal and core radii, so a fiducial $c = 1.5$
(``King30'' in the internal notation of ISHAPE) was
adopted for all the solutions, following \citet{lar99} and \citet{larsen01}.  
This $c-$value matches the mean
King-model concentration for the Milky Way, M31, and NGC 5128 globular clusters,
among others \citep{har96,bar07,mcl08}.  In Fig.~\ref{rh_mwgc}, the $c-$distribution
for the Milky Way clusters is shown; 83\% of the non-core-collapsed clusters
are in the range $1 < c < 2$.  In addition, ISHAPE was run with the assumption $e=0$,
because the vast majority of known globular clusters have quite modest
projected ellipticities, mostly $e < 0.1$ \citep[e.g.][]{har02,bar07}.
Through a set of simulations to be described below,
the sensitivity and accuracy of the $r_{eff}$ solution to the object's brightness
and assumed $c$, $e$ were tested.   
In practice, ISHAPE returns the best-fit
FWHM for the GC model profile, which must then be converted to an actual effective
or half-light radius through a multiplicative factor that depends
on $c$; for the specific case of the 
King30 model, $r_{eff} \equiv r_h = 1.48 FWHM$.

\begin{figure}
\plotone{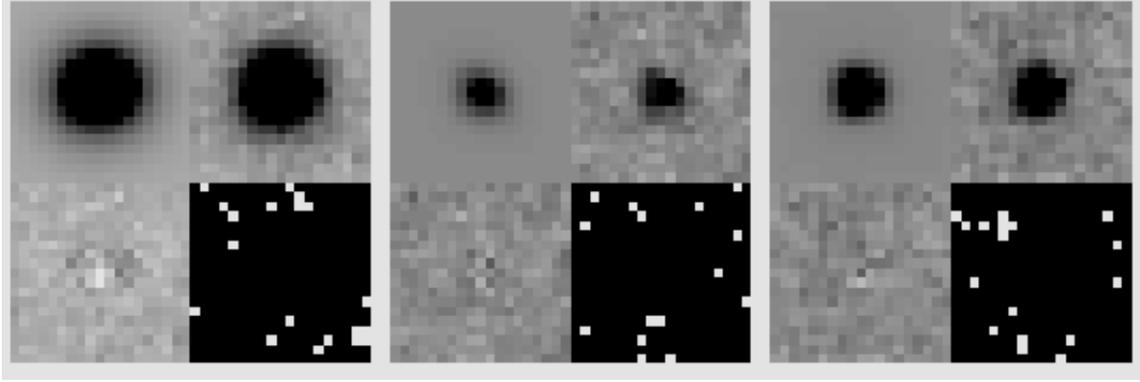}
\caption{Three sample illustrations of the \emph{ISHAPE} profile fits to
objects in the NGC 4696 field.  In each panel the upper right quadrant is
the real object, the upper left quadrant shows the best-fitting model
(that is, the King30 profile convolved with the PSF),
and the lower left quadrant shows the residuals (real $-$ model).
The lower right quadrant shows the pixel weight map.
(a) \emph{Left panel:} A bright nonstellar object:  the best-fit
solution is for FWHM $a = 1.3$ px, ellipticity $b/a = 0.91$,
and $S/N = 440$.  In this case the intrinsic size of the object is
nearly as large as the PSF width of $\simeq 1.8$ px.
(b) \emph{Middle panel:} A faint nonstellar object, for which
$a = 0.82$ px, $b/a = 0.50$, and $S/N = 24$.  Note that the object looks
much rounder than its intrinsic shape $(b/a)=0.5$ because it has been convolved with the
circularly symmetric PSF, whose FWHM is more than twice as wide.
(c) A starlike object, for which $a = 0$, $b/a = 1$, and $S/N = 108$.  
}
\label{ishapefits}
\end{figure}

In all cases except the few objects that happened to be affected by extreme crowding
(neighbors within about 3 px), the residual maps plainly showed good fits to the PSF-convolved
model.  Some sample ISHAPE fits to three different objects are shown in Figure \ref{ishapefits}.

Knowing the value of $r_h$, a self-consistent procedure for defining the total 
magnitude of the object must be constructed.  The procedure adopted here was to
start with a \emph{fixed-aperture} magnitude for the object using $r_{ap} = 2.5$ px,
with the standard tools in daophot \citep{ste87}.  This fiducial 2.5-px radius
is slightly larger than the stellar FWHM of $\simeq 2$ pc but yet small enough to 
avoid crowding issues.  This magnitude was then corrected to ``large'' aperture 
with a curve of growth (COG) of enclosed light versus aperture size;
for the purposes of this paper, I adopt $r_{max} = 20$ px $= 1''$, at which
the curves-of-growth for starlike and marginally resolved objects have
accurately converged.  
The key point is that each value of effective radius has its own curve of growth, with more extended
objects having relatively more of their light at larger radius.  

\begin{figure}
\plotone{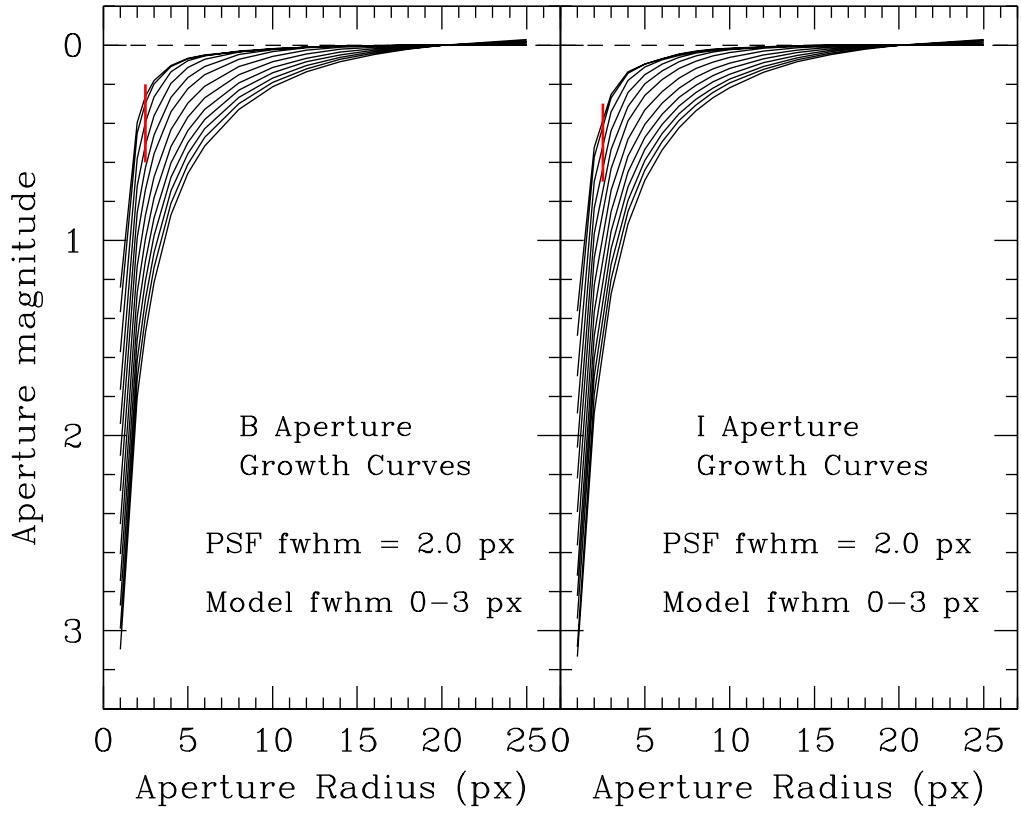}
\caption{Curves of growth for simulated globular clusters.  For each 
of the two graphs, the uppermost curve shows the magnitude enclosed within
an aperture of radius $r$ for a \emph{starlike} object.  All the other curves
in each set show the curve of growth for simulated GCs of various sizes, convolved
with the same stellar PSF profile.  All the curves are plotted to converge at
$r=20$ px (1 arcsecond).  The more extended the GC profile is, the more gradually
the curve of magnitude versus $r$ increases:  from top to bottom, the assumed GC profile FWHM 
values are $(0.0, 0.2, 0.4, 0.6, 0.8, 1.0, 1.25, 1.50, 1.75, 2.0, 2.25, 2.50, 2.75, 3.0)$.
The \emph{vertical bar} plotted at $r=2.5$ px is the adopted radius for our
fixed-aperture photometry, and it covers the typical range of GC profile widths for
the GCs in our data.
}
\label{growthcurves}
\end{figure}

A complete grid of COGs generated through
the ISHAPE simulation tools, for values of FWHM ranging from 0 to 3 pixels,
is shown in Figure \ref{growthcurves}.  This range correponds to $r_{eff} \simeq
0 - 30$ pc linear radius, a generous coverage of the sizes of known 
GCs and even UCDs.  More realistically, we would expect the GCs in these galaxies
to have radii $\lesssim 6$ pc = 0.6 px if they at all resemble 
known GCs.  As can be seen from the Figure, this primary range
of interest is marked by the solid vertical bar in each panel.
The figure shows that the COGs converge extremely well to the 
same total magnitude at the large (20-px)
radius and the aperture corrections are small.  It can readily be seen that
this procedure is equivalent to making 
the 2.5-px aperture magnitude brighter \emph{by the amount needed to shift
it onto the starlike curve}($r_{eff}=0$); then, the correction from 2.5 px to
20 px aperture radius is made along the starlike curve, which is the upper
envelope of the grid of tracks in the Figure.

\begin{figure}
\plotone{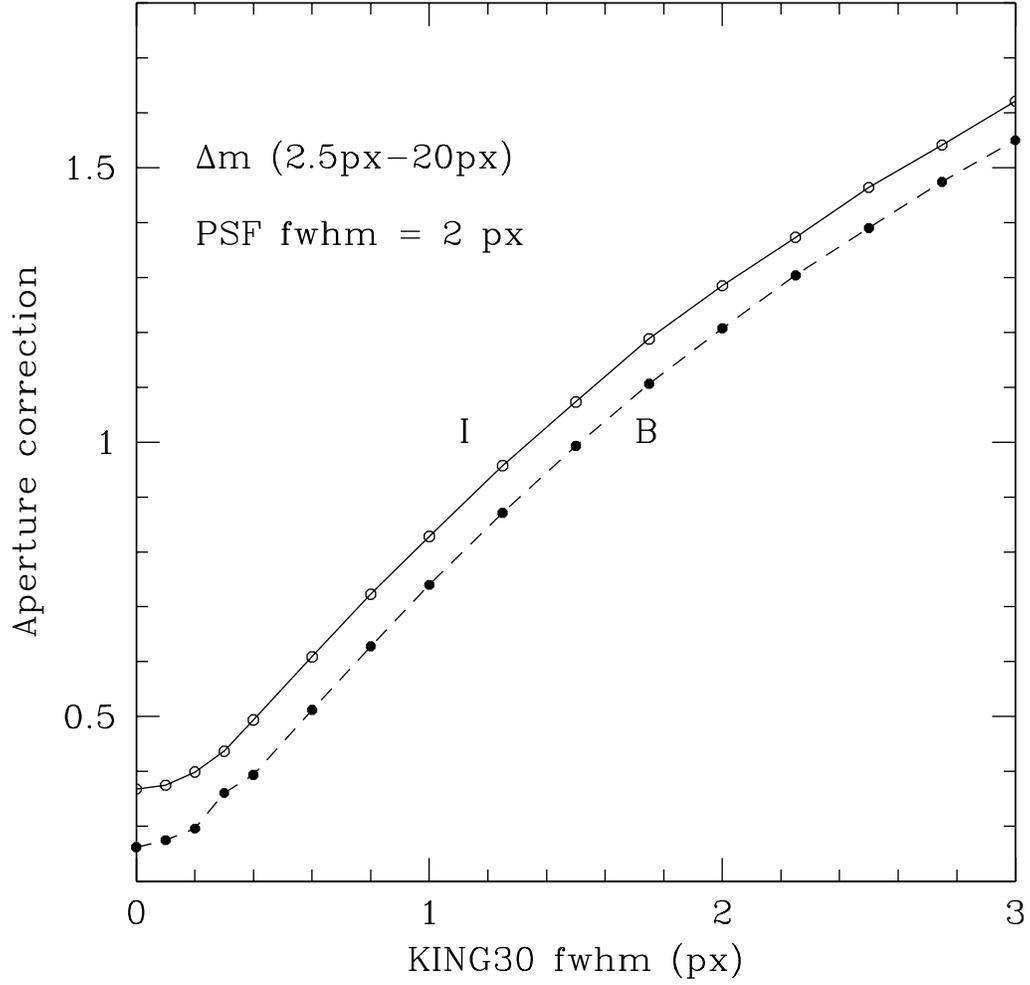}
\caption{Aperture corrections plotted as a function of intrinsic size of
the individual globular cluster.  These are generated from the family
of curves plotted in the previous figure, where  the magnitude enclosed within
$r=20$ px is defined as the ``total'' magnitude of the object.  To
obtain this total magnitude, the magnitude within a 2.5-px aperture is measured
and then the aperture correction from this graph is subtracted.  The data for
our actual GCs fall in the regime FWHM $ \lesssim 0.6$ px, i.e. the 
lower left part of the curves.
}
\label{apcor}
\end{figure}

Figure \ref{apcor} shows the magnitude correction $\Delta m (2.5px-20px)$ as a 
function of object FWHM for the
two filters.  Again, the regime of interest here is
almost entirely at the lower end ($r_{eff} \lesssim 0.6$px) where the 
aperture corrections are quite small.  In practice, for about a third of the objects
in the measured sample, the sizes turn out to be indistinguishable from stars,
though most of these starlike objects are also very faint and of no interest
in the following discussion.  In addition, the correction curves in $B$ and $I$
are nearly parallel, so that the net aperture correction to the \emph{color} $(B-I)$
is very small.

\begin{figure}
\plotone{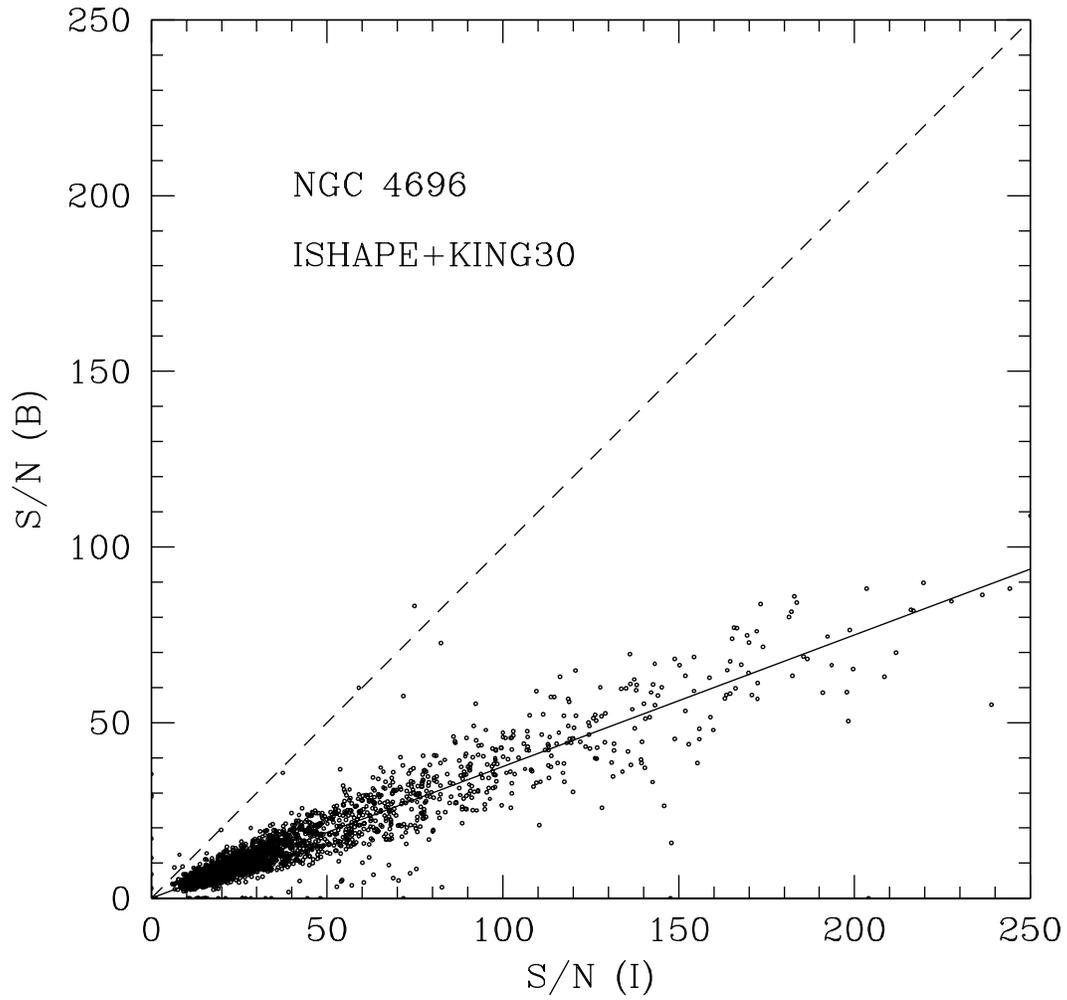}
\caption{Scatter plot of the measured \emph{ISHAPE} signal-to-noise in
$B$ versus the S/N for the same objects in $I$.  Ideal 1:1 agreement
would fall along the dashed line, while the solid line shows the actual
correlation where S/N($B$) $\simeq 0.4$ S/N($I$).  Although the data
for only the NGC 4696 field are shown, these results are typical of 
all six of the galaxy fields.
}
\label{sn_compare}
\end{figure}

The images in the $I-$band go slightly deeper than in the $B-$band for objects
with GC-like intermediate colors, so that the $S/N$ of the
ISHAPE solutions is higher in $I$. This comparison is illustrated in
Figure \ref{sn_compare} for the NGC 4696 field, in which $(S/N)_B \simeq 0.4 (S/N)_I$.
The results for all the other fields are quite similar.  
The COG
appropriate for each filter was used to
define the total $B$ and $I$ magnitudes for each object.
Finally, the $S/N-$weighted mean of $r_{eff}(I)$ and
$r_{eff}(B)$ was adopted as the fiducial effective radius. 
The color index of each GC follows directly from $(B-I) = B_{tot} - I_{tot}$.

\subsection{Tests of the GC Size Measurement}

\subsubsection{Sensitivity to Central Concentration}

\begin{figure}
\plotone{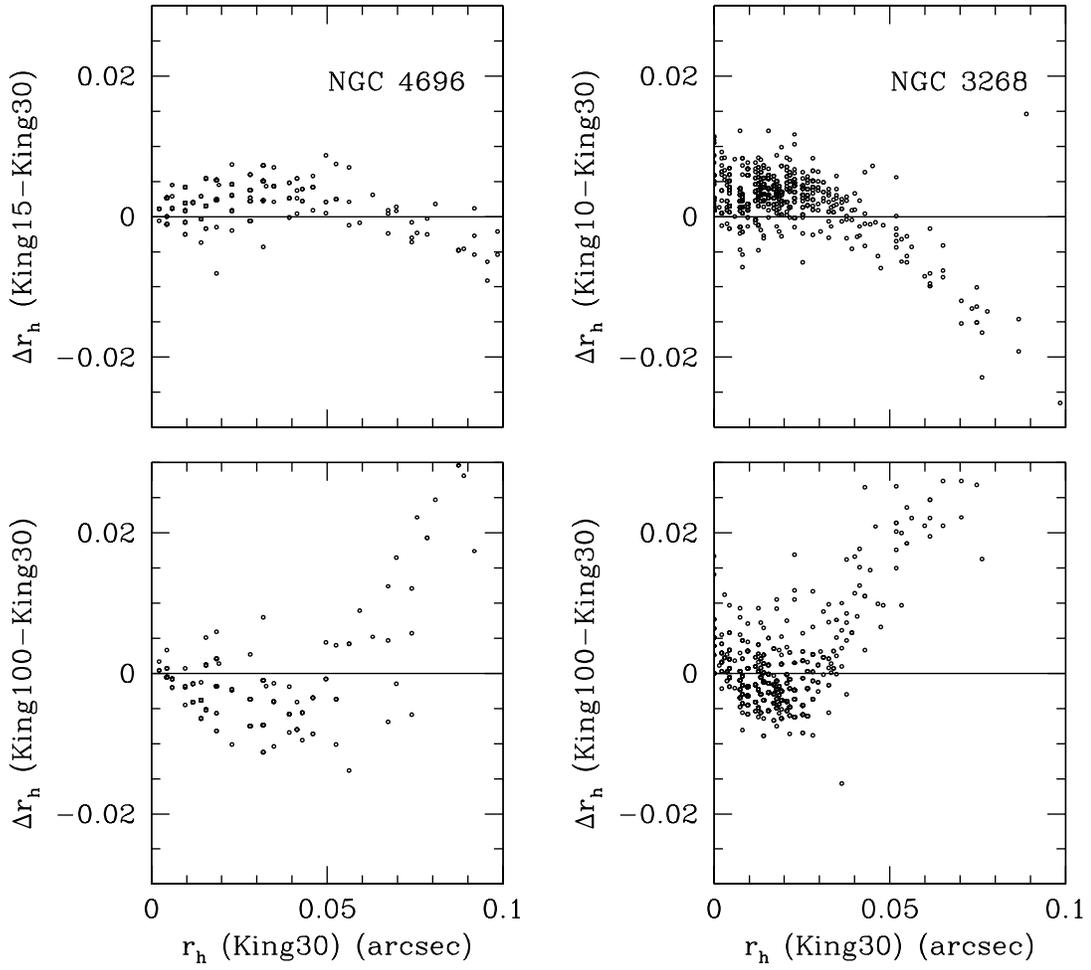}
\caption{Tests of the sensitivity of our GC size measurement to the assumed
central concentration $c$ of the King model.  The panels on the left show
tests in the NGC 4696 field of
the differences in measured $r_h$ relative to the baseline King30 model,
for (King15-King30) and (King100-King30).  On the right, similar results
are shown from the NGC 3268 field for the comparison (King10-King30)
and (King100-King30).  The units of both axes are in arcseconds.
}
\label{ctest}
\end{figure}

Although the King30 model accurately represents the mean for real GCs,
the individual $r_t/r_c$ ratios can differ between GCs 
by factors of 3 or more in either direction, or $\pm 0.5$ in $c$ 
(see Fig.~\ref{rh_mwgc}).  To test the sensitivity of the solutions to the
adopted $c$, I repeated the solutions on several of the galaxy fields
adopting much higher and lower $c-$values:  in the ISHAPE notation, these were King10 ($c=1.0$), 
King15 ($c=1.2$), and King100 ($c=2.0$).  Sample differences
for two of the fields are shown in Figure \ref{ctest}.   These tests show that
for clusters with intrinsically large radii $r_h \gtrsim 0.05'' = 1$ px,
the choice of $c$ is important and systematic differences in the deduced cluster
size occur that grow with $r_h$.
However, for these rare big objects $c$ can be allowed to be a free parameter within
ISHAPE and explicitly solved for.  At smaller radii $(r_h \lesssim 0.05'')$ the
systematic differences are less than $0.01''$.  Said differently, for these tiny objects $c$
cannot be solved for as a free parameter, but making the wrong assumption
for $c$ leads to only very small systematic errors in the deduced size relative
to the correct value:  if the assumed $c$ is too small, $r_h$ is 
very slightly overestimated,
and if $c$ is too large, $r_h$ is slightly underestimated 
\citep[see also][for similar tests and conclusions]{geo08}.
The numerical reason why $r_h$ is rather insensitive to $c$ in this small-object regime
is simply that the conversion factor
$f(c)=(FWHM/r_h)$ depends on $c$ itself, and so an initial overestimate of the FWHM is mostly
compensated for by an opposite change in $f(c)$ to reach an equally good fit.
In this ``marginally resolved'' regime, the object-to-object rms
scatter and systematic shift together add up to an uncertainty $\sigma(r_h) = \pm0.003''$
due solely to the intrinsic uncertainty in $c$.

\subsubsection{Sensitivity to Cluster Ellipticity}

\begin{figure}
\plotone{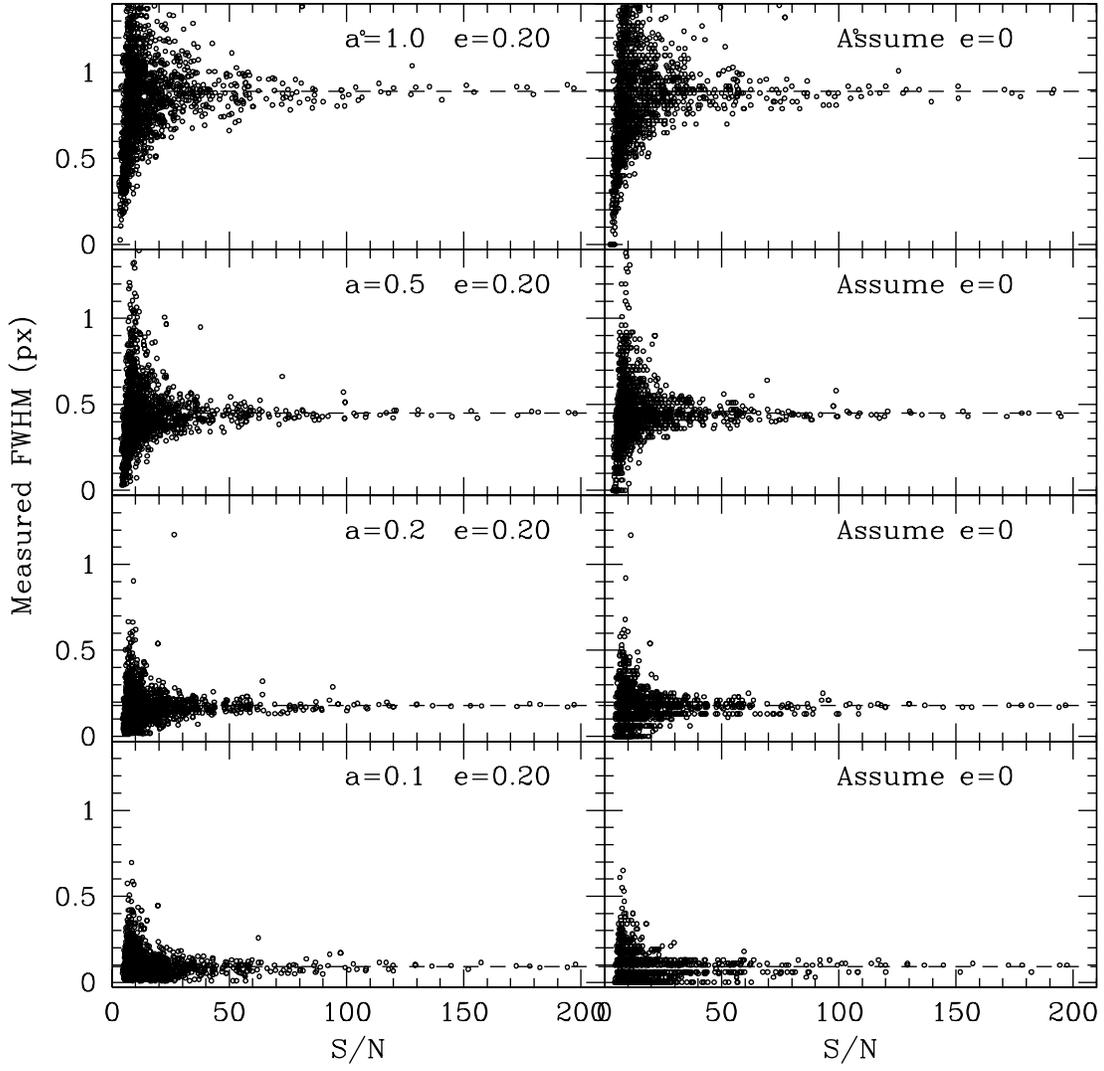}
\caption{Tests of the sensitivity of our solutions for GC size to
the assumed ellipticity $e$.  The simulated clusters have an assumed
King30 profile with FWHM$=a$, convolved with a model PSF with FWHM=2.0 pixels.
In each graph the correct (input) value of the intrinsice FWHM of the
clusters is shown as the
horizontal dashed line.  In the left-hand set of graphs, ISHAPE was allowed
to solve for both the FWHM and the ellipticity, while in the right-hand
set, it was forced to assume $e=0$ and solve only for the FWHM.
}
\label{ellip}
\end{figure}

The assumption $e \equiv 0$ is
clearly an idealized case.  If a cluster is actually elliptical, how much of an
effect will that assumption have on the deduced $r_h$?  
Figure \ref{ellip} shows the results of an
extensive set of simulations.  Here, artificial clusters of intrinsic shape
$e = (1-b/a) = 0.2$ over a wide range of brightness \emph{and} a wide range
of sizes were convolved with
the PSF, put onto an image with realistic background noise, then remeasured with ISHAPE.
It should be noted that the adopted value $e=0.2$, though modest, is already much larger 
than the ellipticities of most real GCs, so that any bias that shows up in these
tests will be an upper limit to the average for real clusters.

In Figure \ref{ellip}, the \emph{measured} FWHM from ISHAPE is plotted as a
function of $S/N$ (essentially the brightness of the object). 
In the left panel, ISHAPE is allowed to solve for $e = 1 - (b/a)$ as a free parameter,
and in the right panel, the code is required to assume 
$e=0$.  The four different rows show
input objects from big to small, with semimajor axis from 1.0 down to 0.1 px (that is,
from 50\% down to 5\% of the PSF width).  In both
cases, the true value of the FWHM is accurately returned for $S/N \gtrsim 50$ and
for intrinsic sizes bigger than about 10\% of the PSF FWHM.  For fainter and smaller
objects than these limits, the solution for the object size becomes indeterminate.

\begin{figure}
\plotone{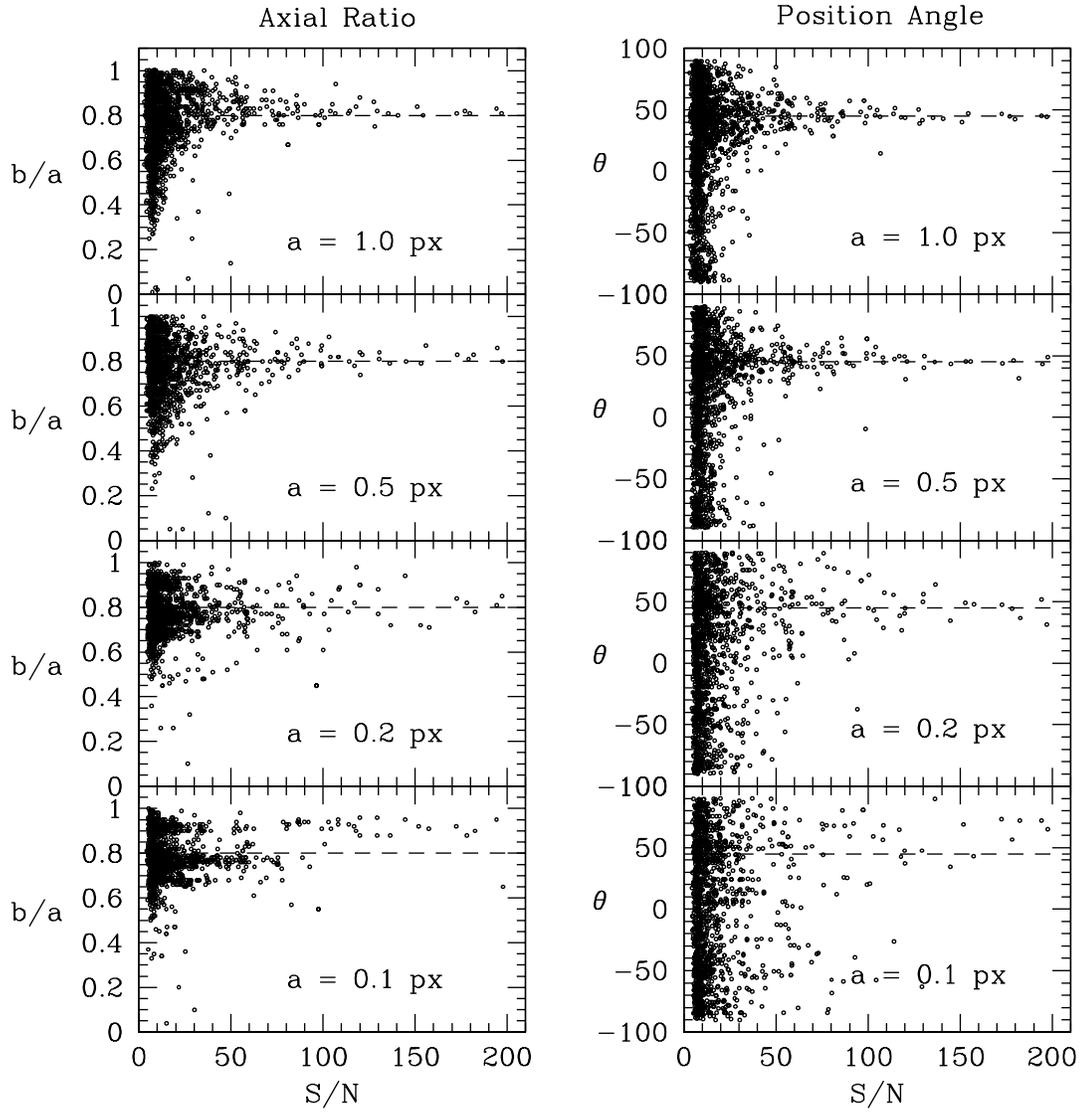}
\caption{Tests of ISHAPE solutions on simulated clusters for the object shape 
$(b/a) = (1-e)$ and orientation angle $\theta$ as functions of
object brightness and size.  The correct (input) values are
$(b/a) = 0.8$ and $\theta=45^o$, as shown by the horizontal
dashed lines.
}
\label{test_theta}
\end{figure}

Additional tests of the accuracy of the solutions are shown in Figure \ref{test_theta}.
Here, the results from the same simulations are shown for the measured
axial ratio $b/a$ and for the orientation angle $\theta$.  For these simulations
the ``right'' answers (the input values) are $(b/a = 0.8, \theta=45^o)$.  
The results of these tests show once again that for the brighter objects with 
$S/N > 50$ \emph{and} intrinsic sizes FWHM $\gtrsim 0.2$ px (about 10\% of the PSF size), 
accurate answers are returned.  For the faintest objects, however,
the spread of solutions for $FWHM, (b/a)$, and especially $\theta$ rapidly blows up.
For the smallest objects as well (FWHM less than 10\% of the PSF width), the solutions
for object shape and orientation are unreliable.

The random measurement uncertainty for $r_h = 1.48~FWHM$ in the regime $S/N > 50$ and $r_h < 0.5$ px
is, from the simulations, $\sigma(r_h) = 0.0025''$.  No systematic errors in the solutions
for the object size, shape, and orientation occur until
$FWHM < 0.1$ px, at which stage the object can be treated as virtually starlike.  

\subsubsection{Intrinsic Scatter and Total Uncertainty in Size Measurements} 

\begin{figure}
\plotone{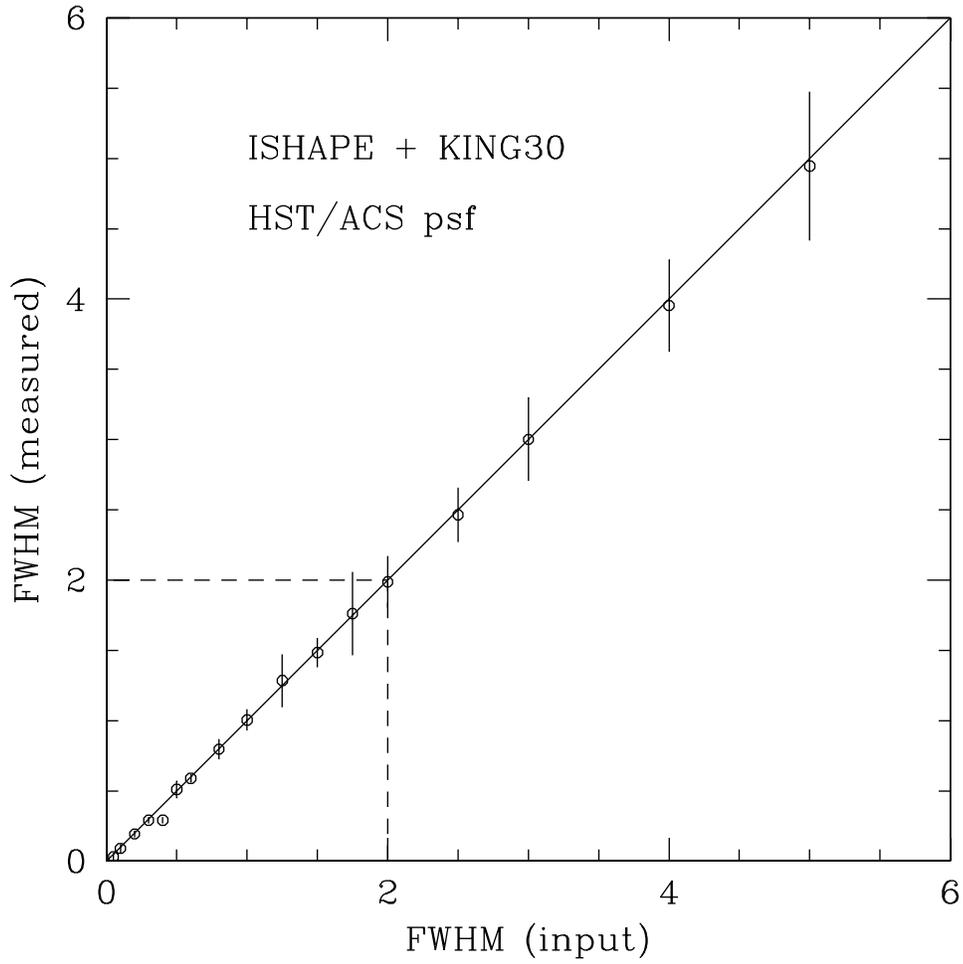}
\caption{Tests of the \emph{ishape} measurements for simulated globular
clusters of various intrinsic sizes.  KING30 model GCs of an assumed
input FWHM (measured in pixel units, where 1 px $= 0.05''$) and a range
of luminosities are convolved
with the stellar point-spread function and inserted into the image on
top of realistic background noise.  These simulated objects are then
measured with \emph{ISHAPE} and the measured FWHM is plotted against
the input value.  The dashed lines show the FWHM of the assumed stellar
PSF at 2.0 px.  These tests show that accurate values of the GC sizes
can be recovered down to FWHM $\simeq 0.2$ px, or 10\% of the stellar FWHM.
}
\label{fwhmtest}
\end{figure}

A convenient summary of the range of object sizes that ISHAPE can be expected 
to solve for is shown in Figure \ref{fwhmtest}.  This is the result of a separate
set of simulations where the input objects all have $(S/N) > 50$ and a range of
sizes extending from ``starlike'' (FWHM=0) up to FWHM=5 px.  Just as in the
previous tests, the model cluster has a King30 profile and it is convolved with
a PSF with width 2.0 px.  Each point shows the average
of many dozens of simulated objects, with the error bar representing the standard
deviation of the individual scatters around the means.  The graph shows that for
these relatively bright objects, the measured sizes are \emph{systematically} accurate
down to 10\% of the PSF size, in agreement with the findings of \citet{lar99}.

The consistent message from all these tests is that we can accurately recover
the true size and shape of the object if it is reasonably bright
\emph{and} large enough to satisfy FWHM $\gtrsim 0.1$ FWHM(PSF).
For objects in this ``well conditioned'' range,
the total uncertainty in our size measurements can be summarized as follows. \\
\noindent (a) The random uncertainty in the object size due only to the ISHAPE fitting procedure,
as noted above and shown from the scatter in Figs.~\ref{ellip} and \ref{fwhmtest}, translates
to a mean of $\pm 0.0025''$. \\
\noindent (b) The additional uncertainty due to the possible range of $c-$values 
(Fig.~\ref{ctest}) averages $\pm 0.003''$. \\
\noindent (c) Finally, there is an external uncertainty due to the PSF size itself.
The numbers listed in Table \ref{psftab} show that for each galaxy field and filter,
the actual PSF width is internally uncertain by typically $\pm0.0033''$ in $B$ and
$\pm 0.0018''$ in $I$.  These translate to an ISHAPE fitting uncertainty in
$r_h$ of about $\pm 0.005''$. 

Summing these three independent effects in quadrature, we obtain a net random
uncertainty per object of $\sigma(r_h) = \pm 0.006''$ or $\pm 0.12$ pixel.
For a galaxy at $d \sim 40$ Mpc, this converts to a true uncertainty in
scale size of typically $\sigma(r_h) = \pm 1.1$ parsecs due solely to the GC profile-fitting
procedure.  

\begin{figure}
\plotone{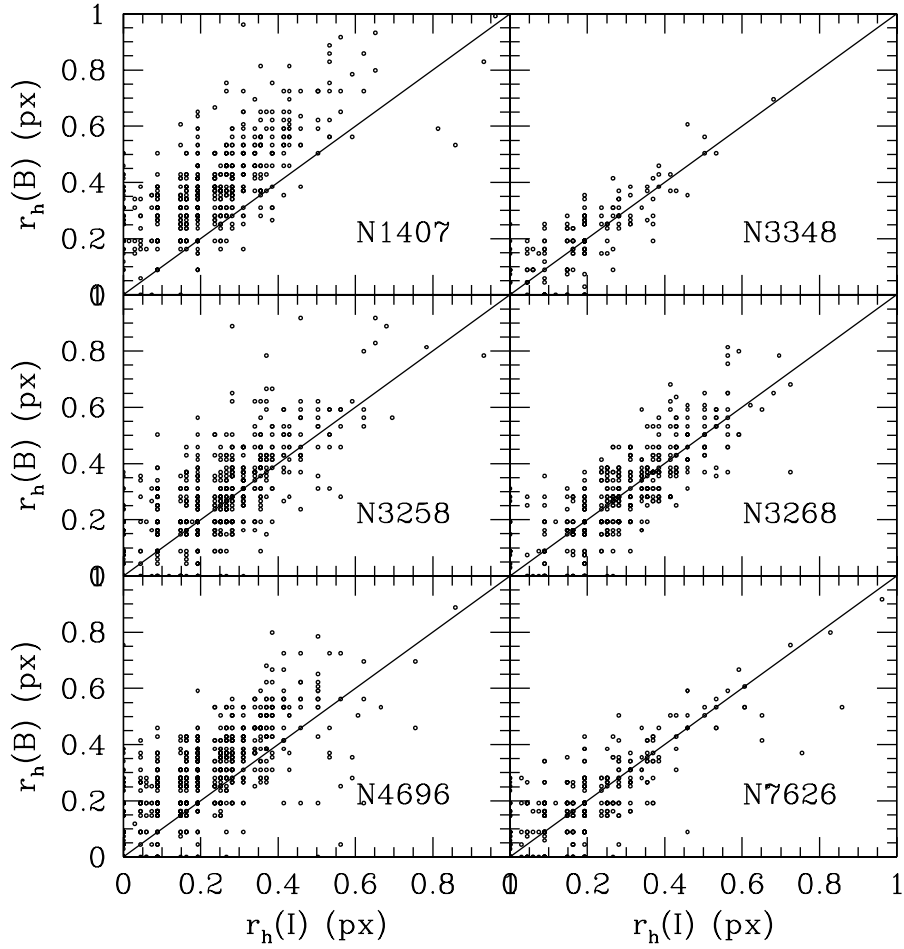}
\caption{Comparisons of the \emph{ishape} measurements of object size
for the real objects in each of the six galaxy fields.   Here the
best-fit half-light radius $r_h = 1.48~fwhm$ from ISHAPE is shown
in $B$ versus $I$ for all objects with ISHAPE total $S/N > 50$.
Ideal one-to-one agreement is shown by the solid lines.
}
\label{fwhmcomp}
\end{figure}

A comparison of the measured object sizes, done independently on the $B$ and $I$
images, is shown in Figure \ref{fwhmcomp} for each of the six galaxy fields.
In most cases, the mutual agreement between the two filters is good and
well within the expected differences due to PSF-size uncertainty.  The one exception
is for NGC 1407, where $r_h(B)$ is systematically bigger than $r_h(I)$ by
$\sim 0.2$ px, an
offset almost twice as large as the expected uncertainty due to the PSF sizes
mentioned above.
The first source of this anomaly to suspect would be that the PSF size 
in this particular field is too big in $I$,
or too small in $B$, or some combination of these; however, the PSFs in both
filters have very much the same intrinsic widths as in the other five
fields (Table 2) and were derived with exactly the same careful iterative procedure.
Further tests have not revealed any clear solution and I have chosen to
leave the results as they are, at very least to ensure that the final procedures
for all the fields are homogeneous.
The net effect on the mean cluster sizes for NGC 1407 is not as large
as this comparison might indicate, because the final $r_h$ value is the
weighted average from the two filters and the $I$ filter carries most
of the weight (see Fig.~\ref{sn_compare}).

Fig.~\ref{fwhmcomp} also allows us to make one
last test of the internal measurement uncertainties.  The FWHM measurements are
done separately for the same objects on the $B$ and $I$ frames, with 
independently determined point-spread functions, so the scatter in the $B,I$
size measurements provides a reasonable estimate of the measurement uncertainty.
The rms scatter is closely similar over all six fields, and averages
$\sigma(r_h) = 0.16$ px, equivalent to $\pm 0.008''$.  Accounting for
the fact that the final $r_h$ values are the average of $B$ and $I$ then
reduces the uncertainty to $\pm 0.006''$, in close agreement with the
estimate given above.

\section{The Color-Magnitude Distribution}

\begin{figure}
\plotone{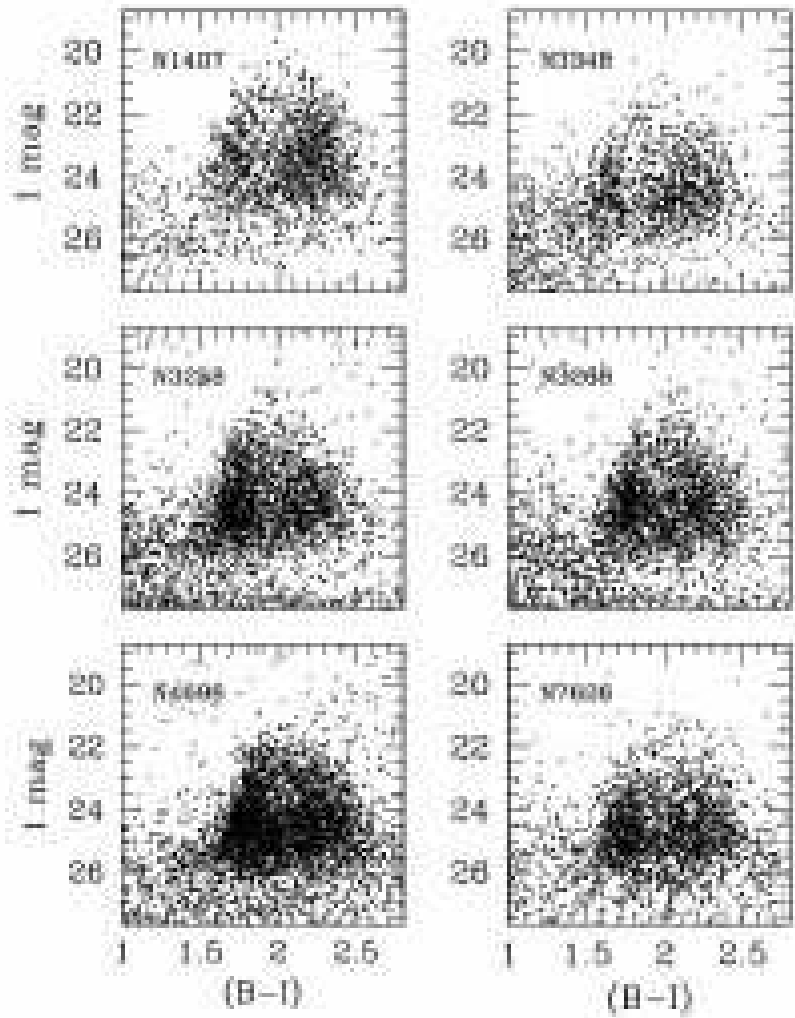}
\caption{The color-magnitude data for the measured objects in each of
six target galaxy fields.  Each individual object is plotted with its fully size-corrected
$(B,I)$ magnitudes as described in the text.  The majority of the objects in
each field are the globular cluster populations.  The main source of field contamination
is predominantly from small, faint background galaxies.
}
\label{cmd6}
\end{figure}

	With the object size measurement and aperture photometry complete, we now have
a database of $(B,I)$ photometry where the magnitudes of each individual object 
were determined starting from small-aperture photometry and then fully size-corrected to
total magnitudes.  The resulting distributions in the raw color-magnitude
plane $(I, B-I)$ are shown in Figure \ref{cmd6}.

\begin{figure}
\plotone{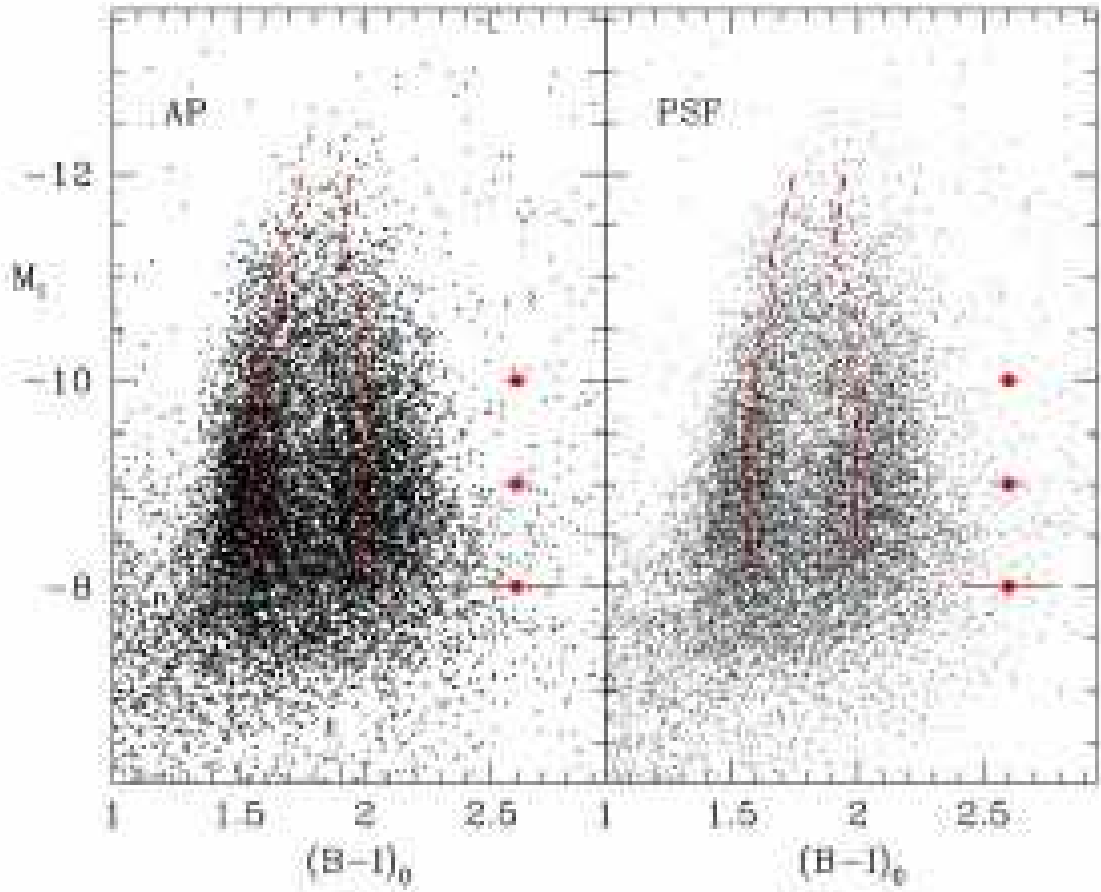}
\caption{The color-magnitude data for all six galaxies, now converted to
luminosity and intrinsic color $(M_I, (B-I)_0)$ and with all fields combined.
The \emph{left panel} shows the size-corrected aperture photometry from the
current discussion, while the \emph{right panel} shows the data for the
same fields based solely on PSF fitting as published earlier in
\citet{har06}.  The mean lines through the
blue and red globular cluster sequences for the newer data at left are
shown by the heavy dashed lines.  The \emph{same lines} are repeated in
the right panel to indicate the small offsets between the two datasets.
Sample errorbars showing the measurement uncertainties of the photometry
are shown for $M_I = -8, -9, -10$.
}
\label{cmd_oldnew}
\end{figure}

In each galaxy the bimodal GC sequences are evident, but the first question
to deal with is to ask how these new, recalibrated data match up with 
the previous photometry based primarily on PSF-fitting \citep{har06}.
A graphical comparison is shown in Figure \ref{cmd_oldnew}.
The huge statistical weight of the combined data, compared with any
other GCS study, can be realized by noting that this diagram contains
15000 objects that were measured by \emph{both} psf and corrected-aperture
methods, most of them globular clusters.

Partially anticipating the results of the later discussion, we use the
mean lines for the blue and red GC sequences in Fig.~\ref{cmd_oldnew}
(calculated as described later in Section \ref{bimodal}) to note that
both the old and new datasets define closely similar bimodal sequences.  
In other words,
completely reworking the data according to a technically improved     
procedure has yielded no first-order change.  The 
major reason for this similarity is that the cluster sizes
are small enough that
the individual aperture-size corrections relative to the fiducial
starlike curve (Fig.~\ref{growthcurves}) are modest. When they
are differenced between $B$ and $I$, the aperture-size corrections to the 
\emph{color indices} are even less.

To a finer level of detail, comparison of the two datasets shows that
the final colors in the present size-corrected data are a few hundredths
of a magnitude bluer than
in the older data, particularly in the bright range $M_I \lesssim -10$.
As will be seen in the next sections, this offset
is \emph{not} due to some
particularly strong size/luminosity relation, because the characteristic
sizes of the GCs have large scatter at any luminosity, and they are all
individually accounted for.
The net result is that the deduced MMR on the blue sequence (the slope of the 
correlation between luminosity and color) is still present but at
slightly lower amplitude than in the earlier data.  These outcomes will be
discussed in more detail below.  First, however, we look more
thoroughly at the distribution of sizes of the clusters themselves,
which have several points of interest in their own right and are an
important preamble to the features of the MMR.

\section{Analysis of the Cluster Sizes}

The characteristic (half-light) radius of a GC is an astrophysically valuable
and frequently used
quantity because it is expected to stay nearly invariant with internal
dynamical evolution well after the initial cluster formation period
\citep[e.g.][]{spi72,aar98,baum02,tre07}, and so the measured $r_h$
should represent an intrinsic structural property built in at an
early time.  It has long been realized that
$r_h$ shows large cluster-to-cluster differences but on the average it
increases systematically with (three-dimensional) Galactocentric distance
$r_{gc}$ in the Milky
Way \citep{vdb91} following a rough scaling rule $r_h \sim r_{gc}^{0.5}$. 
This trend is generally interpreted as meaning that proto-GCs in
the process of formation
will take up systematically larger scale sizes if forming
within shallower surrounding potentials.  Similar, though weak, correlations
in the observed $r_h$ over a large range in
\emph{projected} (two-dimensional) galactocentric distance $R_{gc}$ 
have been found in other galaxies such as the edge-on disk galaxies
NGC 5866 \citep{can07} and
NGC 4594 \citep{spi06}, and the nearby gE NGC 5128 \citep{gom07}, typically
scaling as $r_h \sim R_{gc}^{0.2 \pm 0.1}$.
A similar trend can also be found within the composite data for 
the Virgo galaxies \citep{jor05},
though the data in this case cover a much smaller range in $R_{gc}$.

In addition, differences in the mean $r_h$ between red (metal-rich)
clusters and blue (metal-poor) clusters \emph{at the same galactocentric
distance} have been noted \citep[see][for a sampling 
of the key results]{kun98,kun99,larsen01,jor05,spi06,gom07}.  
The effect claimed in these papers is for the metal-richer clusters to be about 20\%
smaller than the blue ones.  Although considerable scatter in $r_h$ 
is found in each subgroup, this mean difference has persisted as the size and
quality of the available measurements have increased.  This size difference
has been suggested to be the result of a geometric projection effect
\citep{lar03} whereby the metal-richer clusters lie in a more centrally
concentrated distribution within the galaxy, so that a higher proportion of
them are at smaller $R_{gc}$ and thus physically smaller because of the
$r_h/R_{gc}$ correlation.  Alternately, \citet{jor04} suggests that the
difference is an intrinsic function of metallicity and the result of
metallicity-dependent stellar evolution times.  

Data that cover a large
range in $R_{GC}$ and that are based on large statistical samples of clusters
are needed to sort out the interpretations.  If the difference is due to a projection
effect along the lines suggested by
\citet{lar03}, then it should become much smaller at larger distances outward into
the halo.  

\subsection{The GC Size Distribution}

\begin{figure}
\plotone{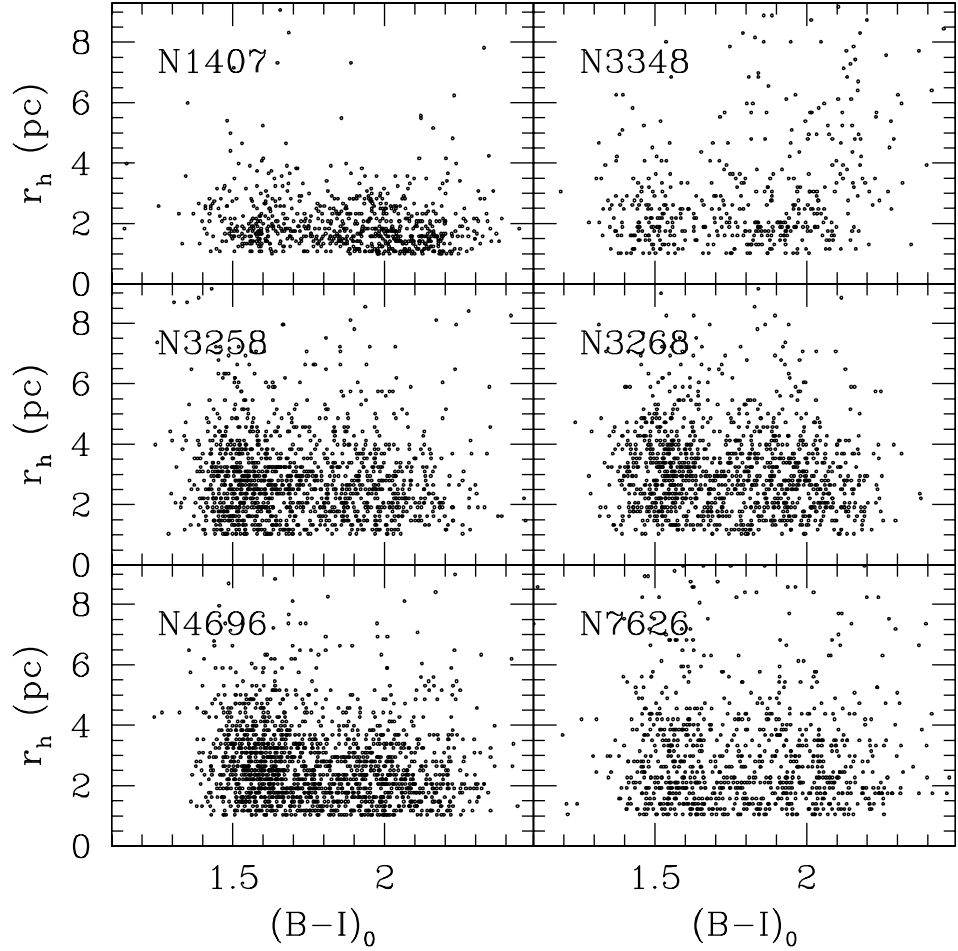}
\caption{Measured half-light radii for the individual globular cluster candidates
in the six galaxy fields.  Only the objects with $r_h > 1$ pc are plotted, though
they have not been selected by $(S/N)$.  The two metallicity groups (blue and red
clusters) can be seen in each sample.  Over all six fields, the total is 6476 objects.
}
\label{rh6}
\end{figure}

The database of $r_h$ values from the six giant ellipticals in this paper clearly does not
have highly reliable \emph{individual} measurements of cluster size.  But it has
the compensating advantages of \emph{large sample size} and
coverage of a \emph{large range in galactocentric distance} $(R_{gc}/R_{eff})$.
The measurements for our six fields are displayed individually in Figure \ref{rh6}.
Here, the half-light sizes are shown for all objects with $r_h > 1$ pc regardless of
their internal quality (that is, not selected by $(S/N)$;
the particular cutoff of 1 pc will be justified below).
In every graph, the bimodal nature of the color distribution is evident, as well as a
consistent pattern for the blue GCs to extend up to larger sizes than do
the red GCs.\footnote{The mean $r_h$ values for NGC 1407 are noticeably smaller, a likely
result of the discrepancy between the PSF sizes in $B$ and $I$ as discussed above.
However, this galaxy provides less than 10\% of the total GC population in the 
study, and whether or not it is excluded has no important effect on
the subsequent discussion.}

\begin{figure}
\plotone{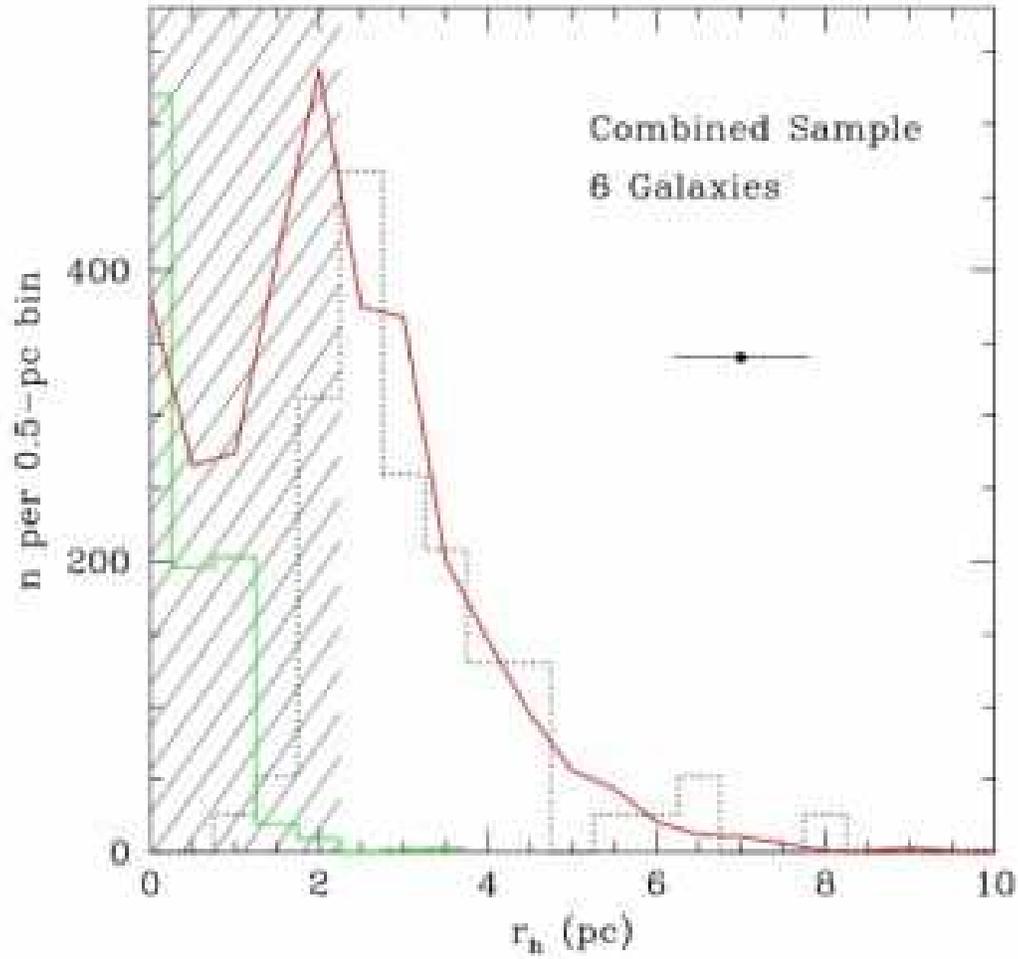}
\caption{Histogram (solid broken curve) of the best-quality 
size measurements ($S/N > 50$) for 3330 objects as determined
by ISHAPE, over all six galaxy fields combined.  All measurements of the object size are
converted to half-light radius in parsecs.  The shaded area covering $r_h \lesssim 2$
pc indicates the approximate size range within which ISHAPE cannot easily distinguish starlike
from nonstellar objects.  For comparison, the block histogram  on the left side 
of the graph peaking at $r_h = 0$
shows the distribution of measured radii that ISHAPE returned for 5300 simulated
\emph{starlike} objects at an assumed distance of 40 Mpc; 
all of these by definition have true radii $r_h = 0$.  Finally, the dotted
histogram shows the size distribution for the 66 globular clusters in the Milky Way
that are more luminous than the GC luminosity function turnover point and no further
than 20 kpc from the Galactic center (see text).
}
\label{histo_rh}
\end{figure}

If the GCs in these giant ellipticals are basically the same kind of star cluster as
the classic Milky Way clusters, then we would expect their size distribution as a
whole to resemble the Milky Way's.  In Figure \ref{histo_rh}, the combined results for
all six fields are shown in histogram form.  Here, the database is sampled in a different
way than in the preceding Figure:  objects of all colors are included regardless of their
$r_h$ value, but only the highest-quality measurements ($S/N > 50$) are included.
This selection leaves 3330 objects over all half-light radii.  
For $r_h \lesssim 2$ pc (the shaded region, equivalent to $0.01''$ at a distance of 40 Mpc),
we approach the limit of measurement
of the profile fitting code and any smaller values for the intrinsic size returned
by ISHAPE become very uncertain.  
The typical internal uncertainty \emph{per object}, from the arguments
in the previous section, is $\pm 1$ pc, 
shown as the error bar at upper right.  In other words, for 
the extreme case $r_h < 1$ pc,
GCs cannot be distinguished from stars
even under the most optimistic reading of the data.
However, it should be noted that GCs this compact physically appear
to be quite rare (there are only two this small in the Milky Way).

While keeping in mind the measurement limits mentioned above,
the total histogram shown in Fig.~\ref{histo_rh} clearly has two major components:
the dominant GC population peaking at $r_h \simeq 2$ pc;
then a dip near $r_h = 1$ pc
and a secondary peak at $r_h = 0$ which should consist of all the
starlike or near-starlike field contaminants plus the most
compact GCs.  As a further test,
ISHAPE was run on a simulated population
of 5300 \emph{starlike} objects ($r_h \equiv 0$) over a wide range of brightnesses
mimicking the range of the real data.  The size measurements of these
stars are shown in Fig.~\ref{histo_rh} as the green histogram.  The width 
of this histogram should give a reasonable estimate of the internal 
precision of the profile-fitting results.  The vast
majority of these simulated stars fall within $r_h < 1$ pc and confirm the 
empirical internal error estimate of $\pm 1$ pc deduced above. 

These various tests of the data show that
the total spread of the major GC component that continues
on upward past $r_h \sim 2$ pc
is too large to be explained simply by measurement scatter.
It is completely consistent with the interpretation
that we have genuinely resolved the intrinsic size distribution of the globular clusters
around these galaxies.  A final key test is to compare it with the Milky Way GC sizes.
In Fig.~\ref{histo_rh}
the Milky Way data from \citet{har96} are shown in the dotted histogram for
the 66 clusters with luminosities $M_V < -7.3$ (i.e., brighter than the turnover point
of the GC luminosity function), and no further from the Galactic center than 20 kpc.
Selecting the Milky Way sample this way makes it closely comparable to
the GC sample that we have in the giant elliptical data.  
The overall shapes of the two datasets resemble each other remarkably well,
particularly their extension to large radii.

A second-order difference is that the Milky Way $r_h$ distribution peaks at 
a point about 0.5 pc bigger than the peak for the giant elliptical data.
This offset corresponds to an angular size difference of $\Delta r_h \simeq 0.0025''$ at $d \sim 40$ Mpc
and is entirely within the internal uncertainties of the measurements
in this paper.  \emph{If} the offset
is physically real, however, it is likely to be because the gE's have a much higher proportion
of red (metal-rich) GCs than does the Milky Way, and as will be seen below, the red GCs
are systematically smaller than the blue GCs by $\sim 0.5$ pc (see Table \ref{rhmed} for
the median values in the Milky Way).

\subsection{Correlations of GC Size with Other Factors}

Having presented the overall size distribution, we now investigate correlations
with the three properties that have been suggested in the previous
literature to influence GC scale size:  luminosity, metallicity, and galactocentric
distance.  
Differences in GC scale size with \emph{metallicity} have drawn the most attention
in recent work, primarily from the HST-based photometry of GCSs in nearby
galaxies \citep[][among several others]{kun99,larsen01,jor05}.  
These same studies have not, however, been used to track size correlations
with either luminosity or spatial location very effectively, because
(a) as will be seen below, the size difference with cluster luminosity
becomes obvious only at the top end of the luminosity range where GCs are
rare; and (b) these previous studies were almost entirely restricted to
the inner-halo clusters within about 1.5 effective radii of the
galaxy spheroid light, so that only 
a very limited run of spatial location could be traced.  The database in
this study has the advantage that all three factors can be traced at
least to some degree.

\begin{figure}
\plotone{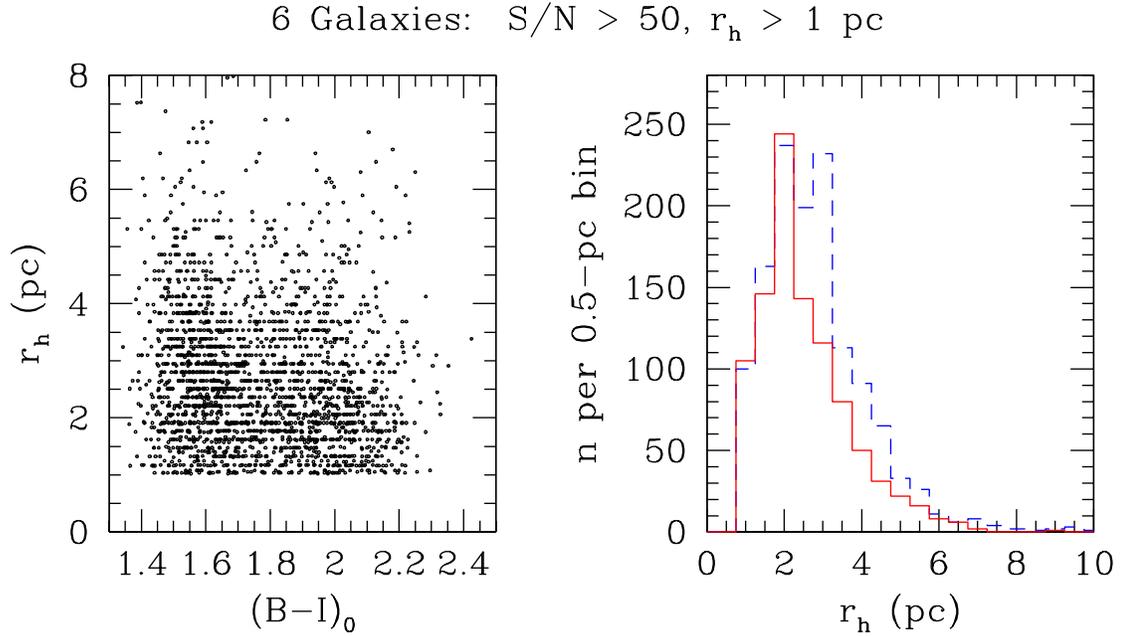}
\caption{\emph{Left panel:} Measured half-light radii for the combined globular cluster
populations in all six galaxies.  The best-quality 2495 objects with sizes 
larger than $r_h = 1$ pc \emph{and} with $(S/N) > 50$ are shown.   
\emph{Right panel:} Histograms of the $r_h$ values for the metal-poor GC
population ($1.3 < (B-I)_0 < 1.8$, solid line) and the metal-rich 
population ($1.8 < (B-I)_0 < 2.4$ dashed line).
}
\label{rh_color}
\end{figure}

In Figure \ref{rh_color}, the $r_h$ distributions are
shown now for the 2454 GCs bigger than 1 pc and with $(S/N) > 50$.  The right
panel shows the histograms separately for the metal-poor GCs (those with
$(B-I)_0 < 1.8$) and the metal-rich ones ($(B-I)_0 > 1.8$).  The expectations
from the previous literature are borne out here, that the lower-metallicity 
GCs are systematically bigger on average by $\sim 0.5 - 1$ pc, or roughly $15-20$\%.
It is highly unlikely that this difference is just an artifact of measurement,
because all the objects in the database, red and blue alike, were measured
through both $B$ and $I$ filters and appropriately $S/N-$weighted.  Both types
of clusters also appear at all galactocentric distances and over the full range
of GC luminosities allowed by the magnitude limits of the photometry.

One \emph{caveat} in these kinds of comparisons is that the sample has
been deliberately truncated for $r_h < 1$ pc.  This should, however, have only
a small effect on such diagnostic quantities as the median size of a subsample,
because at least in the Milky Way virtually no clusters have such
ultra-compact structures (Fig.~\ref{histo_rh}).  

The distribution of sizes versus cluster \emph{luminosity}, shown separately
for the blue and red GCs, is in Figure \ref{rh_allmag} and the median sizes
in half-magnitude bins are listed in Table \ref{rhmed}.  As in the previous figure, the
2454 objects with $(S/N) > 50$ and $r_h > 1$ pc are shown.  For luminosities higher
than $M_I \simeq -11$ (corresponding to about $10^6 L_{\odot}$) the clusters show 
a weak trend to become larger with increasing luminosity. 
For the lower-luminosity range $M_I \gtrsim -11$,
any trend of median cluster size with
luminosity is modest at best and perhaps not significant.
Both of these results agree with what has been found in the Local Group
galaxies \citep[e.g.][]{bar07}.  

On average, the median GC size of the blue
clusters is 15--20\% bigger than the red clusters, at any luminosity.
A standard Kolmogorov-Smirnov two-sample test also shows that their
size distributions are different at $> 99$\% significance.

In the previous literature, correlations of size with galactocentric distance
have also been suggested.  Even within the Milky Way, which is the only galaxy
for which we can explore that correlation in three dimensions, the trend
is relatively modest, and when it is projected onto two dimensions for any other
galaxy it can be expected to be even shallower.  Nevertheless,
the data in this paper can be used to gain some idea of
any such correlation because the range of $R_{gc}$ is quite large
for each of the six galaxies.  To normalize all the data
to a common system and allow treatment of all the galaxies together, 
we express the projected $R_{gc}$ value for each cluster in units of $R_{eff}$, the effective
radius of the galaxy's spheroid as determined from its surface brightness
profile (see Table \ref{basicdata} and the NED).

A schematic representation of the combined effects of spatial location and metallicity
together is shown in Figure \ref{rh_rad}, where size versus color is plotted in three
different spatial zones:  the inner halo ($R_{gc} < R_{eff}$), the mid-halo
($R_{eff} < R_{gc} < 2 R_{eff}$), and the outer halo ($R_{gc} > 2 R_{eff}$, 
extending out as far as $5 R_{eff}$).  
A more quantitative comparison is shown in Figure \ref{rh_rgc2} and
Table \ref{rhmed_rgc}.
Here, the median values in bins of 
$\Delta R = 0.5$ are plotted separately for the two metallicity subgroups.  
These show a weak trend for both types of 
clusters to have systematically larger scale sizes farther from their galactic
center.  Furthermore, the slopes of both relations are quite similar:  least-squares
solutions for the medians give
\begin{equation}
r_{h,med} \, (pc) \, = \, (2.236 \pm 0.079) \, + \, (0.203 \pm 0.030) (R_{gc}/R_{eff})
\end{equation}
for the blue clusters, and
\begin{equation}
r_{h,med} \, (pc) \, = \, (1.881 \pm 0.043) \, + \, (0.181 \pm 0.016) (R_{gc}/R_{eff})
\end{equation}
for the red clusters.  Consistent with what was found earlier, the blue GCs are
larger by 17\% than the red GCs, and this difference does not change significantly
with $R_{gc}$ bin.  

A power-law formulation of the same trend \citep{vdb91,spi06,gom07}
has perhaps a stronger physical motivation and
yields an equally good representation of the data.  The best-fit equations
in this form (for the blue and red clusters respectively) are also shown
in Fig.~\ref{rh_rgc2} and are given by
\begin{equation}
r_{h,med} \, (pc) \, = \, 2.53 (R_{gc}/R_{eff})^{0.11} \, ,
\end{equation}
\begin{equation}
r_{h,med} \, (pc) \, = \, 2.15 (R_{gc}/R_{eff})^{0.11}  \, .
\end{equation}

There are only two other galaxies beyond the Local Group for which GC-size
measurements cover a similarly large run in spatial location:  NGC 5128
\citep{gom07}, with data extending out to $\sim 8 R_{eff}$; and NGC 4594
\citep{spi06}, with data out to $\sim 6 R_{eff}$.  Both of these galaxies are
relatively nearby and thus their GCs are better resolved, but the sample sizes
are about one order of magnitude smaller than the one used here.  Nevertheless,
these two studies find power-law dependencies
for $r_h(R_{gc})$ with very much the same power-law slopes $\sim 0.1 - 0.2$
as found here.  In short,
all the current data suggest a consistent pattern for a shallow but
steady outward increase in cluster scale size regardless of metallicity.

\subsection{Interpretation}

\begin{figure}
\plotone{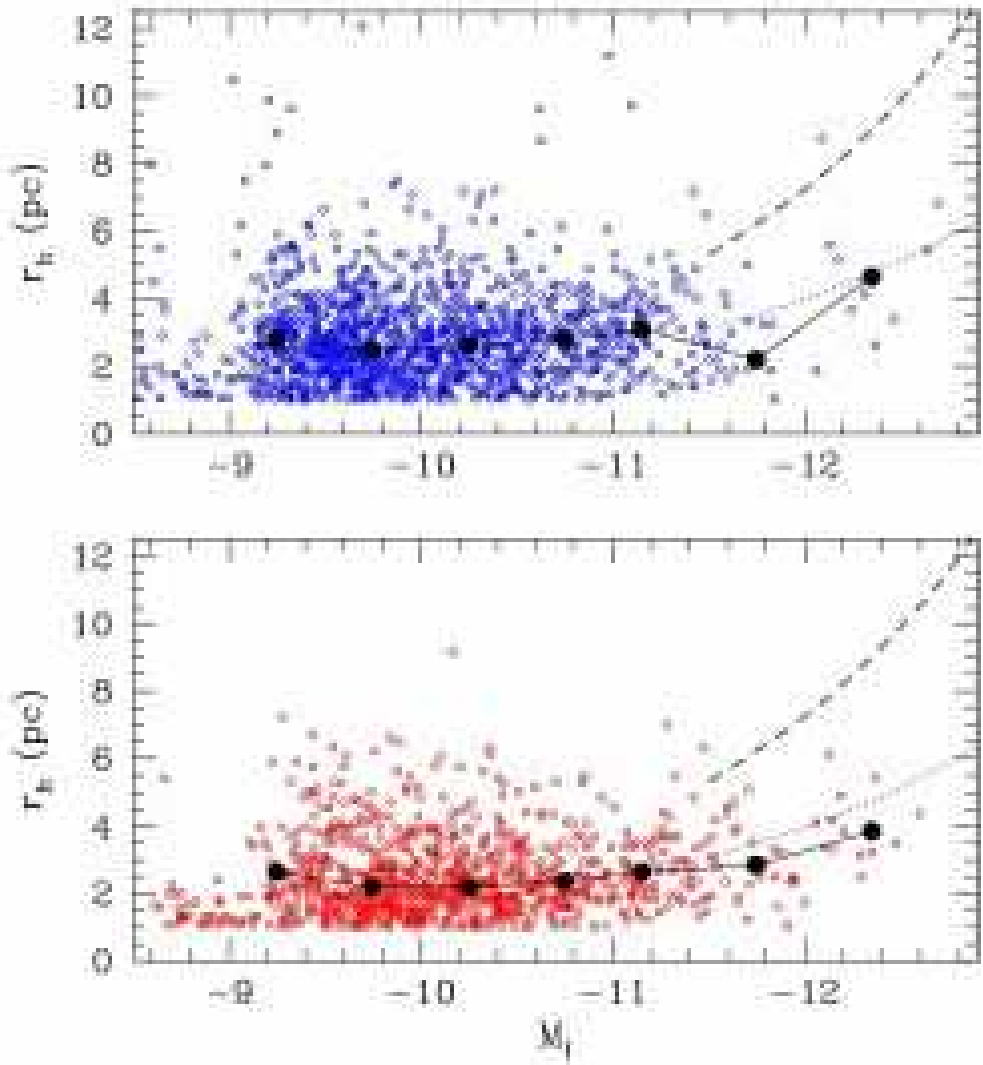}
\caption{\emph{Upper panel:} Half-light radii for the metal-poor globular
clusters, plotted versus cluster luminosity $M_I$.
Data are the same as in the previous figure.  Median values
of $r_h$ from Table 3 in half-magnitude intervals are shown as the large dots.
\emph{Lower panel:} The same data for the metal-rich globular clusters.
On average the metal-poor clusters have median sizes about 15\% larger
than do the metal-rich ones, at any luminosity.
The heavy dashed lines at the right in each panel show the mean size/luminosity
relation for UCDs and dE nuclei, from \citet{evs08}.
The \emph{dotted} lines show the \emph{lower envelope} of the same
UCD/dE,N relation.
}
\label{rh_allmag}
\end{figure}

\begin{figure}
\plotone{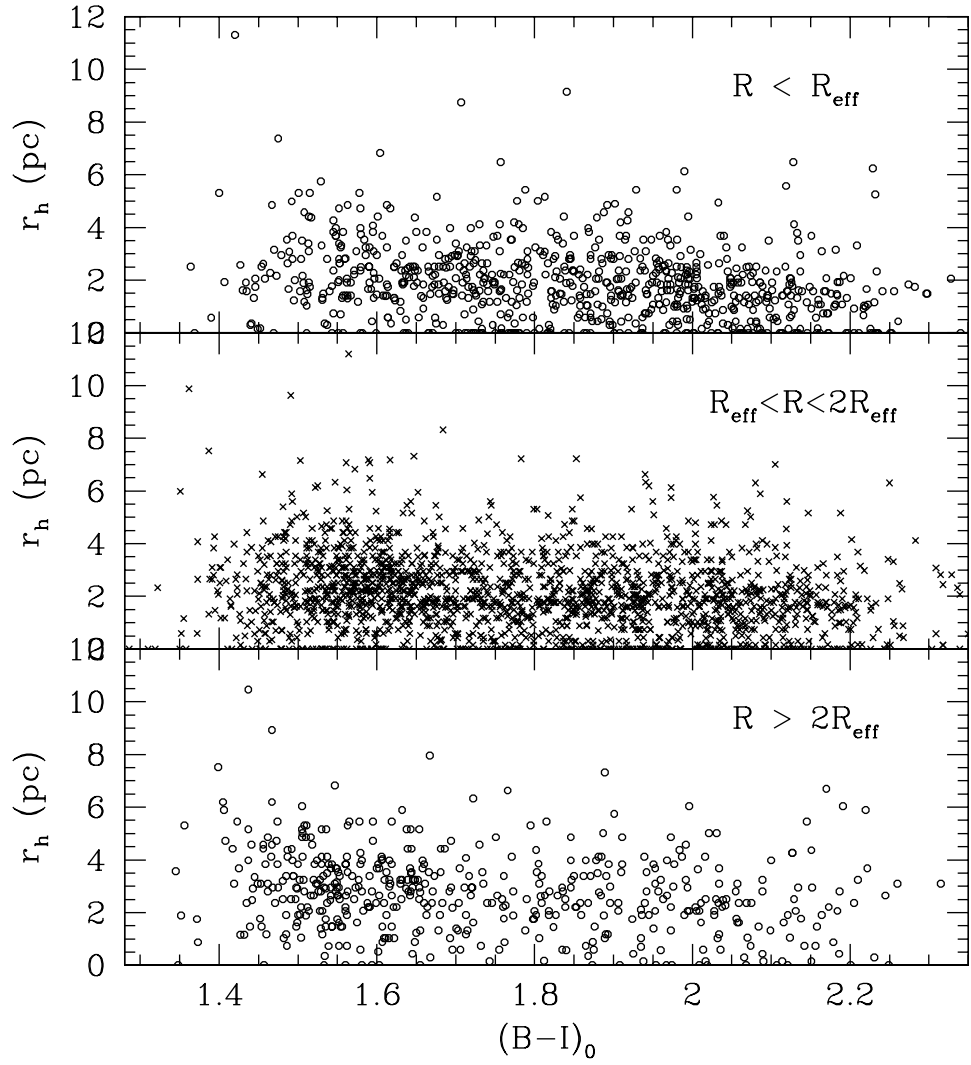}
\caption{The half-light radii for globular clusters plotted as a function
of color (metallicity), for three radial zones of galactocentric distance.
}
\label{rh_rad}
\end{figure}

\begin{figure}
\plotone{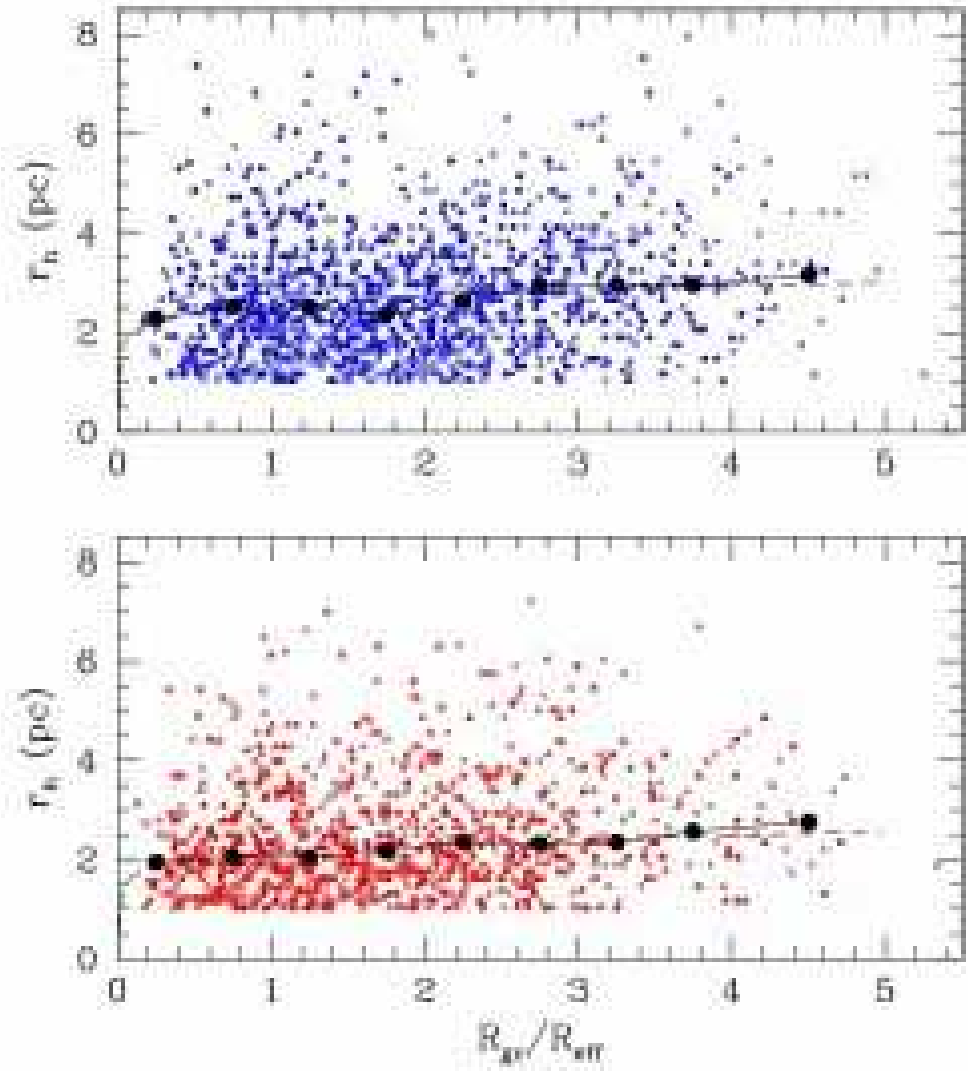}
\caption{The half-light radii for globular clusters plotted as a function
of galactocentric distance.  \emph{Upper panel:}  Blue, metal-poor GCs defined
as in the previous figures; \emph{Lower panel:}  Red, metal-rich GCs.
The large connected dots show the empirical trend of the median $r_h$ value. 
The \emph{dashed lines} give the power-law
fits to the median points, with equations as given in the text.
}
\label{rh_rgc2}
\end{figure}

A brief summary of the findings for the cluster scale sizes in
these six giant ellipticals can be made as follows:
\begin{itemize}
\item{} The characteristic half-light radius depends weakly and 
nonlinearly on total luminosity:  for the lower-luminosity range that
contains the vast majority of the clusters, i.e. 
the range $M_I \gtrsim -11$ or $L \lesssim 10^6 L_{\odot}$,
the median $r_h$ for clusters of all metallicities remains largely independent
of luminosity, whereas for the bright end $M_I \lesssim -11$ the median GC size
begins to increase steadily.  For the extremely rare and very most luminous GCs at
$M_I \simeq -12.5$, the scale sizes are almost twice as large as the median for
the lower-luminosity range (Fig.~\ref{rh_allmag}).
\item{} At any luminosity, and at any galactocentric distance, the metal-poor (blue)
clusters are systematically larger than the metal-rich (red) ones by $(17 \pm 2)$
percent (Figs.~\ref{rh_allmag}, \ref{rh_rad}, and \ref{rh_rgc2}).
\item{} Lastly, there is a consistent correlation of GC scale size with projected
galactocentric distance, at any metallicity, equal to $r_{h,med} \sim R_{GC}^{0.1}$
(Figs.~\ref{rh_rad} and \ref{rh_rgc2}).
\end{itemize}

Given the limited material that was available from previous work,
the fact that all three correlations are comparably weak explains why it has
been difficult to separate them from one another, or indeed to decide whether
or not each one is even present.  

The systematic increase in scale radius with luminosity for the highest-mass clusters
already has clear evidence in its favor from the Local Group galaxies \citep{bar07}
and the brightest GCs and Ultra-Compact Dwarfs (UCDs) in the Virgo and Fornax
galaxies \citep{has05,evs07,evs08}.  \citet{spi06} also found a significant
increase in mean size with luminosity for the GCs in the Sombrero galaxy,
with a noticeable ``onset'' point near $M_V \simeq -9$ 
above which the effective radii begin
to increase most strongly.  These studies show repeatedly that
these compact, massive stellar systems have 
$\langle r_h \rangle \sim const$ for luminosities less than
$M_I \simeq -11 \equiv 10^6 L_{\odot}$, and then gradually swing 
upward to a scaling $\langle r_h \rangle \sim L_V^{0.7}$ 
which bridges them to other objects
such as dE nuclei and eventually dwarf galaxies (see, e.g., Fig.~2 of 
\citet{evs08} for a useful and up-to-date composite diagram).
Nevertheless, the current data show few luminous GCs that match the larger sizes
associated with UCDs.  The mean size/luminosity relation for UCDs,
taken from \cite{evs08} and converted to $M_I$ with $(V-I) = 1.0$, is
plotted as the dashed lines in Fig.~\ref{rh_allmag}.  
Most of the GC data from these giant ellipticals lie well below
the UCD curve even at the highest luminosities.  
In other words, most of the ``classice'' luminous GCs have compact structures
that are smaller than UCDs.
These GCs with $M_I \lesssim -11.5$ instead
more closely match the \emph{lower envelope} of the data discussed by
\citet{evs08}. This lower envelope is populated mostly by a handful of dE,N nuclei.
This comparison, perhaps, provides additional hints that the most luminous normal GCs
may have had their origins as the nuclei of dwarfs that were later stripped
as they were accreted by the giant elliptical.

It should be kept in mind, however, that in a normal GC luminosity
distribution with turnover at $M_I \simeq -8.3$ and dispersion $\simeq 1.4$ mag,
only $\sim$ 1\% of the entire GC population lies above $M_I < -11$.
For the lower-luminosity GCs that make up the vast majority, the
assumption $r_h \sim const$ continues to be a useful one.

The present study was not especially tuned to measuring the sizes of
larger UCD-like objects, so
a separate analysis of them in these fields will have to be conducted.
Work of this kind has begun and will appear in a later paper.

In the model of \citet{lar03}, the mean size difference between metal-poor and metal-rich
clusters is a geometric projection effect:  if the intrinsic
size distribution for all clusters is actually the same, and if they all
follow the same $r_h(r_{gc})$ trend, then the metal-rich ones will be
smaller on average because their spatial distribution is more centrally
concentrated to the galactic center.  The later discussions of
\citet{spi06} and \citet{gom07} supported this interpretation.  By contrast,
\citet{jor04} presents an alternative model whereby the metallicity-based 
size difference is the end result of many Gy of dynamical evolution and
mass segregation within
the GC combined with stellar-evolution timescales that depend on metallicity.
Yet another possibility is that metallicity may have a more direct
and much earlier effect on cluster size, in the sense that higher
heavy-element abundance in the protocluster gas may encourage more rapid
cooling and cloud contraction before the stars form, allowing a metal-rich
cluster to have a smaller scale size from the start.  Detailed models
along that line have yet to be pursued.  

\citet{lar03} suggest that if the projection-effect model is correct,
then the
sizes of high- and low-metallicity clusters should become the same at
large galactocentric distance well outside the galaxy core.  Under
these assumptions, their quantitative
models show that the red-vs.-blue size difference should be
greatest within one or two $R_{eff}$.  However, the data
here (see especially Fig.~\ref{rh_rgc2}) indicate that this is not the case:  metal-rich
GCs are systematically smaller than metal-poor ones by much the same
factor at all locations, favoring the view that the metallicity-based difference
is intrinsic.  
At the same time, both types of clusters follow a 
size/distance dependence $r_h \sim R_{gc}^{\alpha}$
with a similar power-law scaling exponent $\alpha$,
but just with different zeropoints.
This effect strongly suggests that metallicity is not the whole story.
The scale sizes of the clusters are at
least partly determined by the depth of the tidal field that they
are formed in, ending up with more extended structures 
in environments of shallower external potential wells, such as in
dwarf galaxies or the outer halos of large galaxies.  Notably,
in a recent survey of GCs in many nearby dwarf galaxies, 
Georgiev et al (2008, 2009) \nocite{geo08,geo09}
find a median $r_h = (3.2 \pm 0.5)$ pc, and $r_h \sim 5$ pc in the LMC.
Sizes this large are reached in our gE sample only for the outer
regions at several $ R_{eff}$.

A final overview of this part of the discussion is that the scale
sizes of globular clusters appear to be determined jointly and universally by
their metallicity, spatial locations, and (at very high luminosity)
their mass:  all three of these effects are physically
real and need to be included in their conditions of formation.  
Deeper modelling of the combined effects of formation, space distribution,
and dynamical evolution is far beyond the scope of
this paper, but the sum total of all available data now appears to provide
enough evidence to motivate such models.

\section{The Mass/Metallicity Relation \label{mmr}}

The final magnitudes and colors of the globular cluster populations in our six gE galaxies,
as shown in Figs.~\ref{cmd6} and \ref{cmd_oldnew}, are based on fully
size-corrected magnitudes and can now be used to
investigate the blue and red GC sequences in more detail.  The important questions,
beyond the first-order one of establishing the basic bimodality of the
GC color distributions, are to ask whether either sequence shows a
detectable mass/metallicity relation; and whether the slope or form of that
relation might itself depend on cluster luminosity.  Beyond that,
we would like to test its
behavior versus other external characteristics such as cluster size
or galactocentric distance.

\subsection{Bimodal Fitting to the GC Sequences
\label{bimodal}}

The approach to defining the MMR adopted here
is to measure the mean color of each sequence
independently at many points along the luminosity range
\citep{har06,mie06}.
In some other previous studies \citep{str06,spi06,can07,deg07,spi08},
mean linear relations in color versus magnitude were fit to each sequence
and the best-fit slopes were quoted. 
If the GC sequence has the form 
$\Delta({\rm color})/\Delta({\rm magnitude}) = const$,
then this translates readily to a power-law form in heavy-element
abundance versus GC luminosity, $Z \sim L^p$, as long as the
color index varies linearly with [Fe/H].
However, the linear-fit approach 
imposes an uncomfortably strong assumption on the data.
The raw data provide no \emph{a priori} guarantee that the mean GC colors
will vary linearly with magnitude, and a forced linear fit will miss
any nonlinear trend that might be present.  Whether or not the MMR 
disappears below a certain mass range, for example, is important 
for key features of the self-enrichment model \citep{bai08}.
Fewer assumptions are forced
on the material by finding a bimodal fit to the distribution of GC colors
in independent magnitude bins, and then inspecting the resulting mean
points for any trends, linear or otherwise.

The tradeoff is that \emph{if} the MMR actually has a linear
or near-linear form \emph{over the entire luminosity range}, 
then its slope can be determined by the linear-fit method with
a smaller number of objects than is required for the
bimodal fitting in independent bins.
The approach I adopt here is that this linear-full-range 
assumption first needs to be  tested independently on
large, statistically unambiguous datasets before being
carried over to smaller galaxies.

Another factor capable of biasing the deduced MMR is that of sample field
contamination.  In the present study, contamination by 
(mostly) faint background galaxies and a few
foreground stars over a wide range of colors is nearly negligible
over the brighter part of the GC population but grows to significant
levels for $I \gtrsim 24.5$ (see Fig.~\ref{cmd6}).
The contamination on the blue side is typically larger and thus
may leave the artificial impression of a too-steep MMR slope
if a linear fit is imposed.  A notable exception 
from previous literature is the database
of \citet{mie06} where the GCs in the Virgo galaxies were individually selected and
the sample has low contamination.  In general, however, bimodal
fitting to the color distribution will be less affected by
contamination than the linear-fit method
since it is designed to pick out the peak color of each
component regardless of the scatter of outlying points.
The level of field contamination for the present data is
very small for our main luminosity range of interest
($M_I \lesssim -8.5$, brighter than the GCLF turnover) but
the issue will be discussed in greater detail
in Section \ref{contam} below.

The RMIX multicomponent fitting code \citep{macd07} is used here to
determine the bimodal fits. The code and its application for
this problem are described
in more detail in \citet{weh08}.  In short, starting from either the individual
datapoints or a previously defined histogram, RMIX finds the
best-fit multimodal description of the color distribution.  
The user-specified input consists of the number
of modes to be solved for; initial guesses at their means and
standard deviations; and the basic fitting function to be used.
The usual Gaussian model is most common, but half a dozen other well known functional 
forms are available within RMIX.  The mean, standard deviation, and proportion of
each component can either be fixed by the user or solved for 
in any combination.

The solutions for the six individual galaxies are summarized in 
Table \ref{rmixfits} and shown in Figure \ref{cmd_rmix}.
Here, magnitude bins $\Delta I$ of at least 0.5 mag were used, but
with the proviso that a minimum of roughly 100 
objects fell into each bin.  To keep the fewest possible external constraints
on the solutions, RMIX was required only to assume a bimodal Gaussian distribution
in $(B-I)_0$ and then was left
free to solve for the mean $(B-I)$ colors $\mu_1(blue), \mu_2(red)$, the
standard deviations of each mode $\sigma_1, \sigma_2$, and their proportions
of the total population $p_1+ p_2 = 1$.  The input data included all objects
within the color range $1.5 < (B-I) < 2.5$, an
interval chosen to reject the largest part of the
outlying field contaminants while safely including both GC components.
In a very few cases (individual bins here and there) it was necessary to constrain 
the fit in order to get convergence, usually by fixing the 
proportions of one or the other of the $\sigma$'s;
these few cases are marked in Table \ref{rmixfits} with
internal uncertainties of zero.  Overall, however, this
solution technique
differs very noticeably from most previous ones in the literature
because the large GC populations allow 
these minimal constraints on the fits.

The maximum statistical weight can be obtained by combining all six systems
into a single dataset, where everything is translated to the
common scale of absolute magnitude and intrinsic color $M_I, (B-I)_0$.
Bimodal RMIX fits for the combined data are listed in Table \ref{rmixall}.
These are also plotted in Figure \ref{cmd_rmixall}, as are the mean
points for the six individual systems.  

Assuming that the photometry itself is systematically correct,
the clear conclusions from these bimodal fitting results are that

\noindent (a) 
A nonzero relation between luminosity and color exists along the
blue-GC sequence particularly;  

\noindent (b) The trend is nonlinear in form; and

\noindent (c) Along the red sequence, any such relation
is gentler or absent.  

I show in Figure \ref{cmdsize} the
CMDs for those objects with positive, nonzero size measurements
from the ISHAPE fits (that is, objects with $r_h > 0$), and 
then separately for all
the rest of the objects that were either found to be starlike ($r_h = 0$)
or too faint for successful ISHAPE fitting.  The vast majority of
objects in the second category populate the area of the CMD fainter than
$M_I = -8.5$ and thus do not affect anything but the faintest bin of
the RMIX bimodal sequence fits.  Removing the starlike objects entirely
(left panel of Fig.~\ref{cmdsize}) does not change the best-fit 
red and blue GC sequences.

\subsection{Background Field Contamination \label{contam}}

Although the GCS populations clearly dominate the color-magnitude
diagrams in the six galaxies studied here, it is useful to gauge
the actual level of field contamination.  The raw data in this HST/ACS
program contain no true ``control field'' exposures pointed at 
adjacent blank fields.  However, three of the galaxies imaged in the
original program turned out to have small GC populations, and
the outskirts of their fields can be used for an estimate of
the field populations.  These are NGC 5322, 5557, and 7049
\citep{har06}.  The advantages of using these fields are that
they were taken with the same filter combinations and to the
same absolute-magnitude limits, and the galaxies themselves have
similar distances, foreground reddenings, and Galactic latitudes 
to the other six.

The fields of these other three target galaxies were all remeasured
with the same procedures through daophot and ISHAPE as described
above.  Inspection of the color-magnitude diagrams showed that
GC populations around them are definitely present in the inner regions,
but die off with increasing radius until for $R_{gc} \gtrsim 125''$
the GC numbers begin to fall below the general density of field objects 
that are widely distributed in color.
Thus to generate a composite ``control field'' I have combined the regions beyond
$R_{gc} = 125''$, which add up to a total area 11.2 arcmin$^2$, 
conveniently equal to the area of a single ACS/WFC field.  

The CMD for this synthetic control field is shown in Figure \ref{cmd_field}.
Here the raw $(I, B-I)$ data have been converted to $(M_I, (B-I)_0)$
assuming $(m-M)_I \simeq 33.1$ and $E_{B-I} \simeq 0.05$, which are the averages
for NGC 5322, 5557, and 7049.  The data for ``starlike'' objects ($r_h = 0$)
scatter uniformly across the CMD and should represent genuine field objects.
By contrast, the data for the objects with $r_h > 0$ should include 
some genuine GCs that belong to these three
galaxies (at a distance of 40 Mpc, $R_{gc} = 125''$ corresponds to a 
linear projected distance of only 25 kpc,
which is still well within the bounds of typical gE halos), along with
a variety of small background galaxies.
Traces of the red and blue GC sequences can indeed be seen in the middle color range.
In other words, the composite field data can represent
only an approximate \emph{upper limit} to the true level of field
contamination in the six main target galaxies.

The distributions in color for the background data are shown
in Figure \ref{bkgd_color}.  Here, the data are divided into three
$M_I$ luminosity intervals and compared with the data for the same
ranges in the six main program galaxies that have the largest GC
populations.  In each case the ``background'' histograms have been
multiplied up by a factor of 6 to match the total area coverage. 

The presence of GCs in the field sample can be seen most noticeably
in the upper two panels $(M_I < -9.4)$, where low peaks
appear in the shaded histograms at $(B-I)_0 \sim 1.6$ and 2.0,
near the normal positions of the blue and red sequences.
Otherwise, the field populations are consistent with a
low and roughly flat, uniform distribution in color across our main color
range of interest $(B-I)_0 = 1.3 - 2.3$ where the GC population lies.
Subtraction of a uniform pedestal 
of only a few percent amplitude does not change the identification of
the GC sequence peak colors in the bimodal fits, 
and has the disadvantage of simply adding more noise to the data.
The histograms have therefore not been explicitly corrected
for this small and near-uniform background.

\subsection{Characterizing the MMR}

The next step in evaluating the color/metallicity distributions 
is to characterize their detailed structure.
In Fig.~\ref{cmd_rmixall},
the \emph{solid lines} come from the fits to the entire database
and thus are weighted towards the galaxies with the biggest
GC populations (in decreasing order of importance, they are
NGC 4696, 7626, 3258, 3268, 1407, and 3348; the first two
make up half the total and the last two only 17\%).
By comparison, the \emph{individual points} in the figure 
give equal weight to all galaxies.  Nevertheless, both
approaches form a consistent pattern.  
The blue sequences in the galaxies show a trend to become redder at
brighter magnitude \emph{and} more strongly so along the upper
half of their run.  No equally strong trend appears along the red
sequences.  

Before attempting to parametrize either sequence,
we first note that the most obvious transition region appears
to be in the range $-10 > M_I > -11$, below which no significant MMR
appears along either sequence, and above which the blue 
sequence is much closer to the red sequence:
\begin{itemize}
\item{} Along the blue sequence, the mean color for the
three brightest bins in Table \ref{rmixall} for $M_I < -10.4$
is $\langle B-I \rangle_0 = 1.689 \pm 0.035$.  For the five fainter
bins $M_I > -10.4$, we have $\langle B-I \rangle_0 = 1.573 \pm 0.009$.
The color difference between the brighter blue clusters and
the fainter ones is thus $\Delta(B-I) = 0.116 \pm 0.036$, which is significant
at the 3.2-sigma level.
A standard Kolmogorov-Smirnov test between these two histograms
shows that they are significantly different at much higher than the
99\% confidence level.  The largest deviation between the two 
cumulative distributions comes at $(B-I)_0 = 1.6$, precisely in
the middle of the blue sequence.
\item{} Along the red sequence, $\langle B-I \rangle_0 = 1.946 \pm 0.026$
for the bright interval $M_I < -10.4$, while 
$\langle B-I \rangle_0 = 1.985 \pm 0.009$ for $M_I > -10.4$.
The difference $\Delta(B-I) = -0.039 \pm 0.028$ is significant at 
only the 1.4-sigma level.
\end{itemize}

Another way to illustrate this luminosity-based trend is shown
in Figure \ref{cmdsplit} and returns to the very first papers
on the subject, by \citet{ostrov98} and \citet{dirsch03}.  These authors
used GC photometry in the Fornax giant NGC 1399 to suggest that
the brightest clusters made up a broad and perhaps unimodal color
distribution, seemingly counter to the bimodality paradigm that
had already become well established.  
In Figure \ref{cfiti}, the color distributions
over these two biggest magnitude intervals are shown for our
six BCGs, along with the double-Gaussian deconvolutions for each one.
The quantitative RMIX fits show that (a) for the brighter interval
$M_I < -10.4$ where the two modes are more strongly overlapped,
a \emph{unimodal} fit is nevertheless strongly rejected in favor 
of two components; (b) \emph{only} two modes are necessary for an
excellent fit to any part of the magnitude range; that is, the GC
population in these galaxies is bimodal rather than multimodal.  

As a demonstration that these fits are not dependent on the particular
magnitude being used, Figure \ref{cfitb} shows the same
comparison where now the sample is sorted by absolute $B$ magnitude
instead of $I$.  The redder component is less prominent here and
less certainly fit because
the red-sequence GCs are much fainter in $B$ than in $I$, but the same two
modes are present with very much the same color difference as
in the $M_I-$based fits.  For the bright interval $M_B < -9.0$,
we have
$\mu_1 = 1.650 \pm 0.022, \mu_2 = 1.886 \pm 0.030$, while for the
faint interval $M_B > -9.0$,  we have
$\mu_1 = 1.556 \pm 0.004, \mu_2 = 1.971 \pm 0.006$. 
The fits in $M_I$ are to be preferred, first because the $I-$band is
much more representative of the true bolometric luminosities
of the clusters \citep{mie06}, and second because (in the current
dataset) the $I-$band photometry is slightly deeper and of higher precision.

Perhaps the cleanest version of the current GC data
is presented in Figure \ref{cmdbest}.  Here, the 7831 datapoints from
all six galaxies, for the objects that have $r_h > 1.5$ pc,
are shown for the purpose of selecting a ``best'' dataset that
minimizes field contamination, as discussed above.
The MMR along each of the two components is gradual enough that
it can be conveniently represented by simple polynomial fits.
Useful interpolation equations in $(M_I, B-I)$ are
\begin{equation}
(B-I)_0 = (2.543 \pm 0.513) \, + \, (0.237 \pm 0.103) M_I \, + \, (0.0143 \pm 0.0051) M_I^2 ,
\end{equation}
for the blue sequence, and
\begin{equation}
(B-I)_0 = (2.142 \pm 0.089) \, + \, (0.017 \pm 0.009) M_I .
\end{equation}
for the red sequence.  These are derived from the mean points in Table \ref{rmixall}
and are shown in Fig.~\ref{cmdbest} as the solid blue lines.
The colors can be converted to metallicity through the
standard transformation from the Milky Way clusters \citep{har06},
\begin{equation}
(B-I)_0 \, = \, 2.158 \, + 0.375 \, {\rm [Fe/H]} \, .
\end{equation}
However, these curves should not be used to extrapolate to
fainter levels $M_I > -8$.

The red sequence has a barely significant slope, and
is scarcely distinguishable from a vertical
line at constant color at $\langle B-I \rangle_0 = 1.97$.
In fact, the brightest
two points along the red sequence are the least well 
determined in the whole dataset because of their relatively low
population. Very minor differences in the adopted RMIX
constraints (or lack of them) can shift the mean colors of
these two bins by a few hundredths of a 
magnitude.\footnote{See also \citet{mie06},
who demonstrate that the bimodal-Gaussian fitting procedure can
end up producing a slight artificially positive slope to the red sequence
if the two sequences are partially overlapped and not equally populated;
see their section 3.2.  The fitted slope deduced here is consistent
with their results.}

An alternate form of the interpolation equations can be estimated
more crudely by direct polynomial fits to the individual datapoints,
bypassing the RMIX bimodal solutions entirely.  To do
this, I define the blue GCs to be the 4110 objects that
fall within the color range $1.3 < (B-I)_0 \leq 1.8$,
and the red GCs the 3625 objects that fall within $1.8 < (B-I)_0 < 2.4$.
Then the polynomial fits become
\begin{equation}
(B-I)_0 = 2.186 \, + \, 0.152 M_I \, + \, 0.0094 M_I^2 
\end{equation} 
for the blue sequence, and 
\begin{equation}
(B-I)_0 = 2.183 \, + \, 0.018 M_I 
\end{equation}
for the red sequence. These are shown as the dotted lines
in Fig.~\ref{cmdbest}.  The uncertainties on the fitted
coefficients are the same as for the previous pair of equations.
This second pair of interpolation equations
generally agrees well with the first pair, but 
differs most noticeably at the highest-luminosity part of the diagram
where the data are sparsest.  The key potential problem with
the second approach is that it ignores the effects of mutual
contamination of the two sequences on each other:  that is, some
points belonging to the red sequence scatter blueward of the dividing
line at $(B-I)_0 = 1.8$, and conversely some blue-sequence points fall redward
of the same line.  If the relative numbers of objects and the intrinsic
widths of each sequence are not equal (which is certainly the
case here), then the amounts of mutual contamination will not be
the same and may bias the solutions for the equations.  For this
reason, I recommend that the first pair
of equations (5,6) (the ones starting from the mean RMIX points) should
be preferred because they more accurately pick out the centers
of each sequence.

\subsection{Discussion and Comparisons}

To summarize this part of the analysis, the very most basic
conclusion is that the mean color 
of the blue-sequence clusters begins to change, with high statistical
significance, at around the luminosity level $M_I \sim -10.5$.
For a typical mass-to-light ratio $(M/L)_I = 2$ \citep{mcl00},
this level translates to a GC mass $M \simeq 10^6 M_{\odot}$.
\emph{Below} this level, the present data give no compelling
evidence that there is a significant correlation between
luminosity and color.  \emph{Above} this level, the blue GCs
become significantly redder in the mean, causing the blue
and red modes to overlap more strongly.

\begin{figure}
\plotone{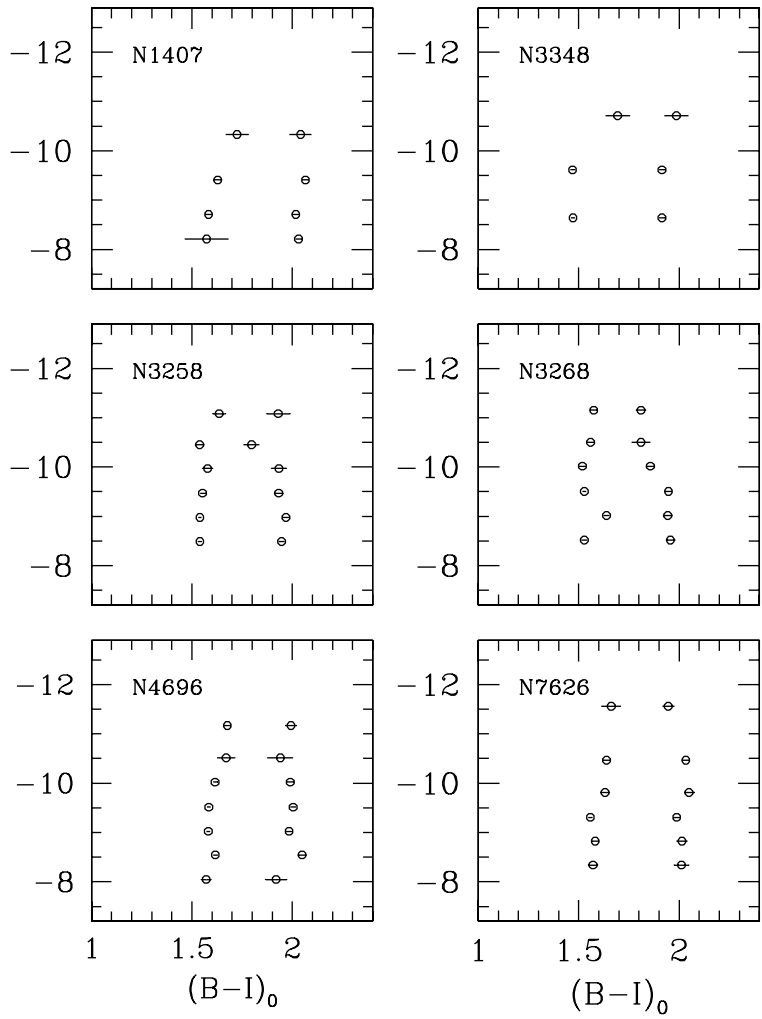}
\caption{Deconvolution of the $(B-I)$ color distributions for the GC populations
in the six individual galaxies.  The fitting code RMIX (see text) has been used
to isolate the best-fit mean colors of the blue and red sequences, as listed
in Table \ref{rmixfits}.
}
\label{cmd_rmix}
\end{figure}

\begin{figure}
\plotone{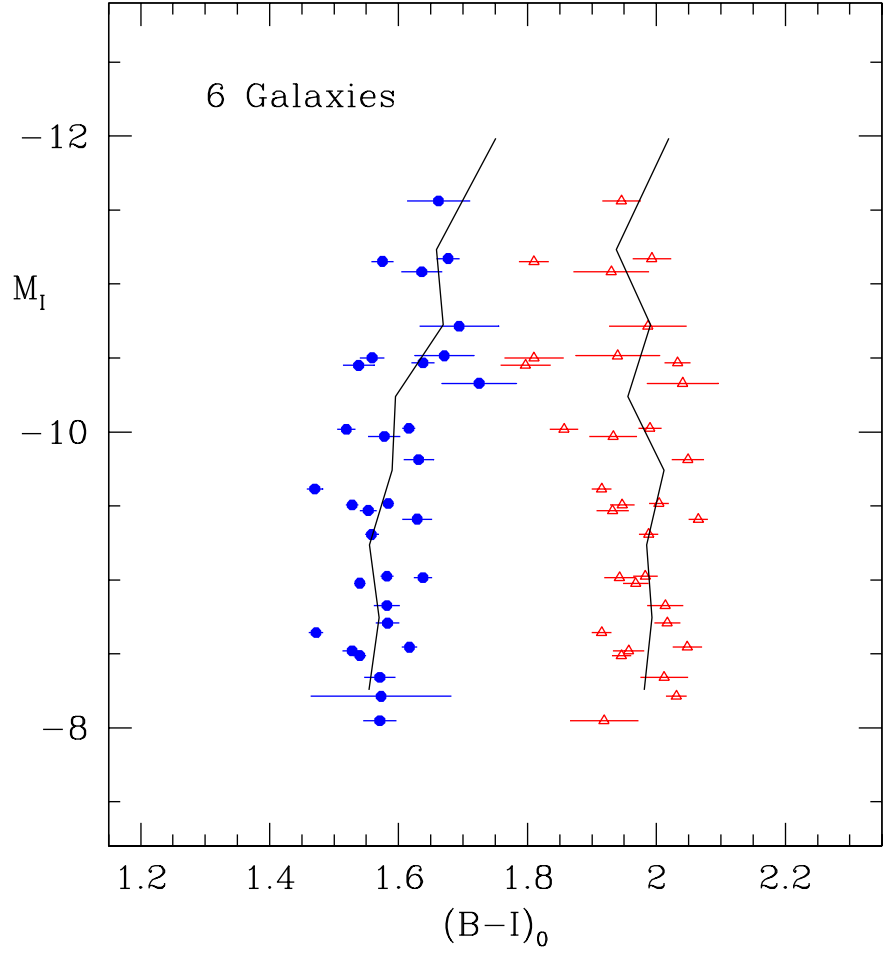}
\caption{Mean points in $(M_I, (B-I)_0)$ for all six galaxies,
as determined from the bimodal Gaussian fits
listed in Table \ref{rmixfits}.  \emph{Solid dots} are the fiducial
points for the blue sequence, \emph{open triangles} for the red 
sequence.  The \emph{solid lines} through each sequence connect the
mean points as determined from the combined dataset in Table \ref{rmixall}.
}
\label{cmd_rmixall}
\end{figure}

\begin{figure}
\plotone{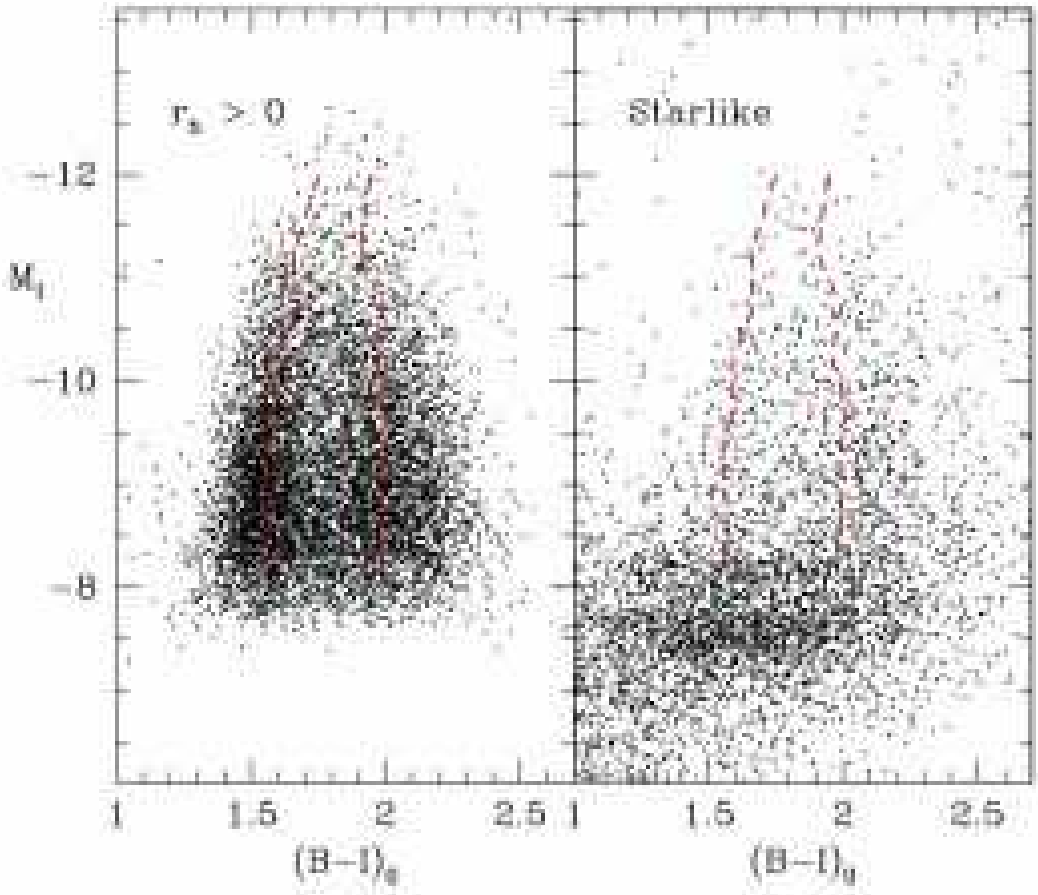}
\caption{\emph{Left panel:} CMD for all objects with positive measurements of
effective radius $r_h$.  \emph{Right panel:}  CMD for all objects with 
either ``starlike'' size measurements ($r_h = 0$) or those too faint for
successful ISHAPE fits.  The heavy dashed lines showing the red and blue
GC sequences are the same as in previous figures.
}
\label{cmdsize}
\end{figure}

\begin{figure}
\plotone{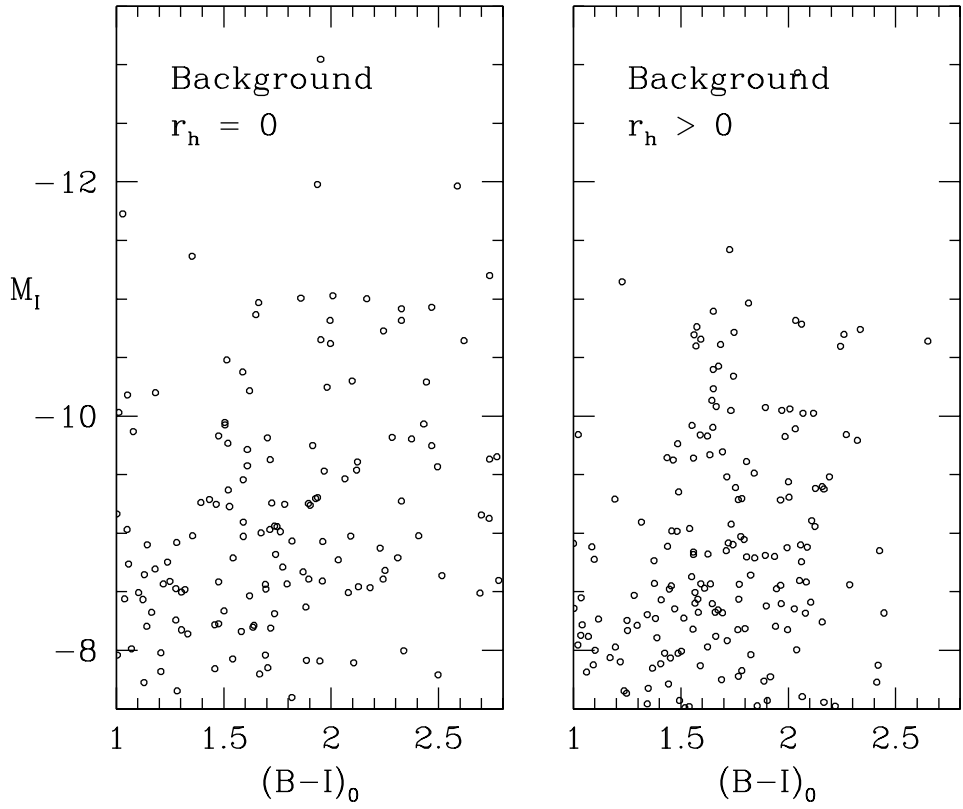}
\caption{\emph{Left panel:} CMD for all starlike objects in
the outer regions of the three ``control field'' galaxies,
NGC 5322, 5557, and 7049.  The data have been measured and
converted to absolute magnitude and color as described in
the text.  \emph{Right panel:}  CMD for all control-field
objects with positive size measurements.  Note the presence
of globular clusters along the normal blue and red sequences
in the outskirts of these three galaxies.
}
\label{cmd_field}
\end{figure}

\begin{figure}
\plotone{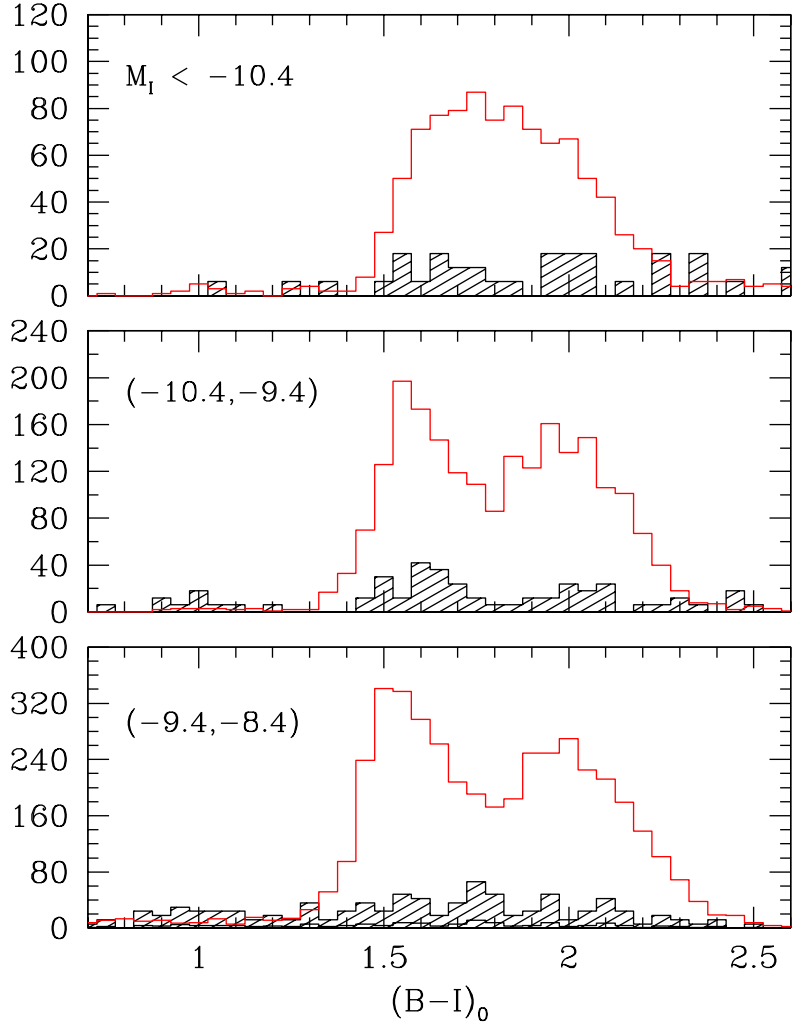}
\caption{Histograms in $(B-I)_0$ for the measured objects in all
the galaxy fields combined, subdivided by GC luminosity $M_I$.
The \emph{unshaded} histograms show the numbers of objects
per 0.05-mag color interval in the six program galaxies.
The \emph{shaded} histograms show the numbers in the composite
``control field'' (see text), multiplied by a factor of 6 to
match the total areas.  The control field data represent
only an upper limit to the true background contamination level.
}
\label{bkgd_color}
\end{figure}

\begin{figure}
\plotone{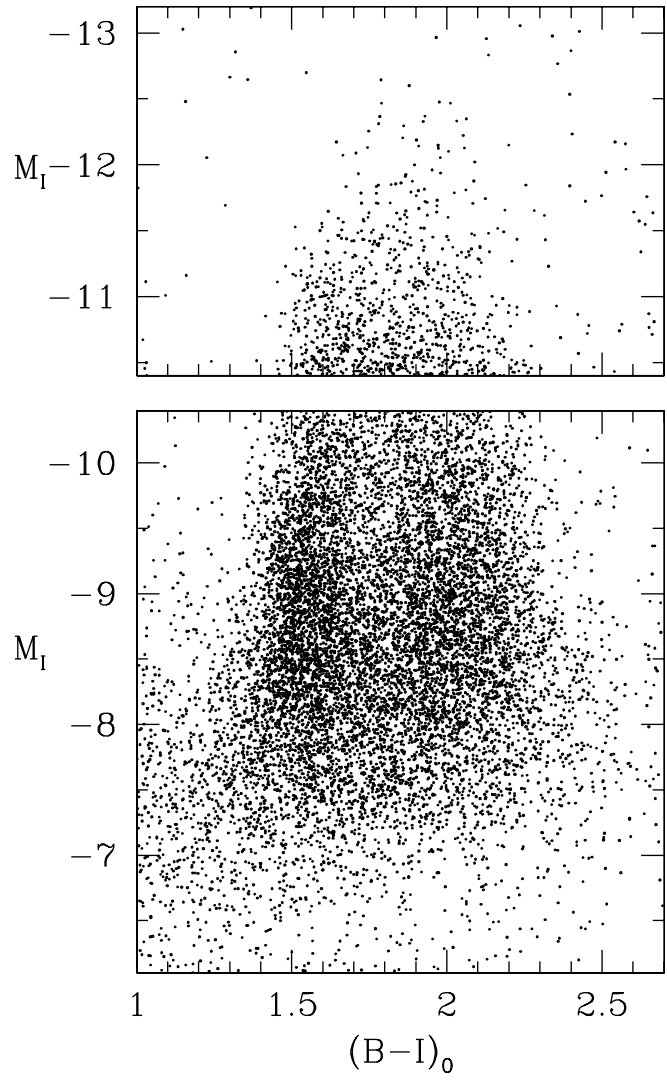}
\caption{\emph{Upper panel:} color-magnitude distribution for the GCs in
all six galaxies, in the luminosity interval $M_I < -10.4$.
\emph{Lower panel:} color-magnitude distribution for the fainter objects,
$M_I > -10.4$.
The reader is invited to block off each panel in turn to view the other
one independently.
}
\label{cmdsplit}
\end{figure}

\begin{figure}
\plotone{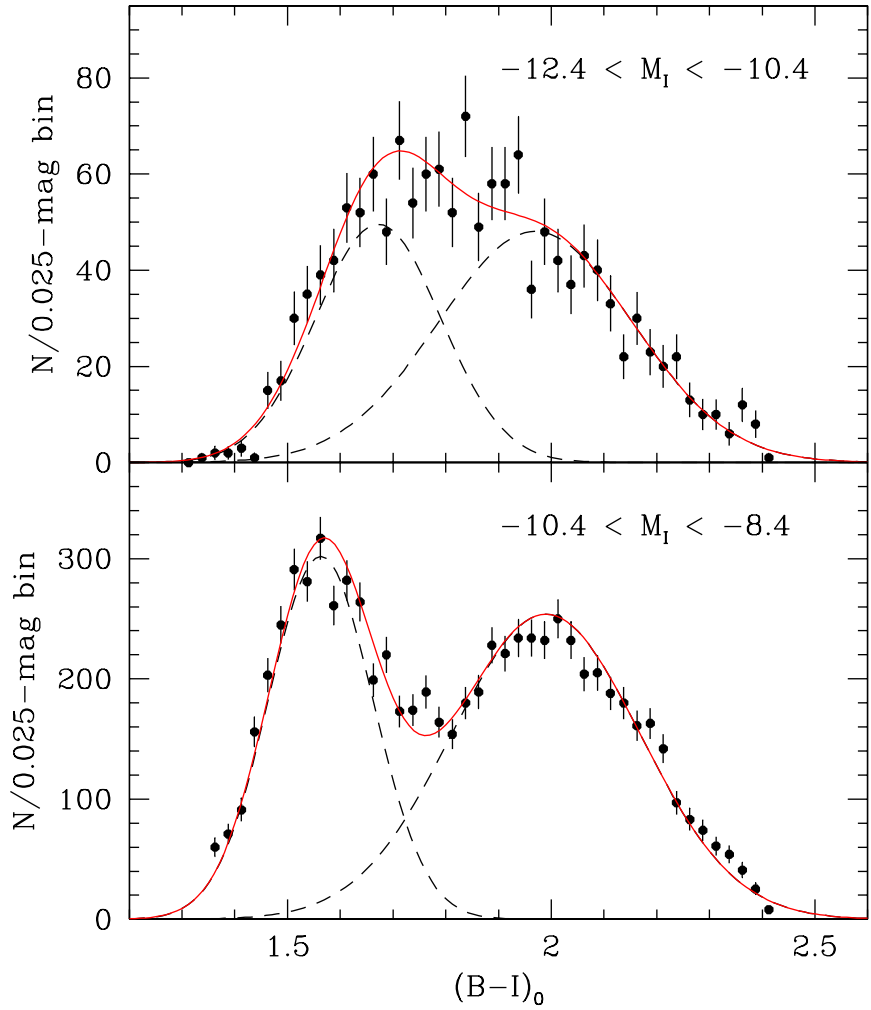}
\caption{Deconvolution of the $(B-I)$ color distributions for the GC populations
in the combined data sample.  The upper panel shows the histogram and the RMIX
two-gaussian fit for the brightest clusters, in the luminosity range
$M_I = (-12.4, -10.4)$.  The lower panel shows the same data for the less luminous
clusters in the range $M_I = (-10.4, -8.4)$.  Note that the faint limit $M_I=-8.4$ 
is approximately the level of the GCLF turnover point (see text).
}
\label{cfiti}
\end{figure}

\begin{figure}
\plotone{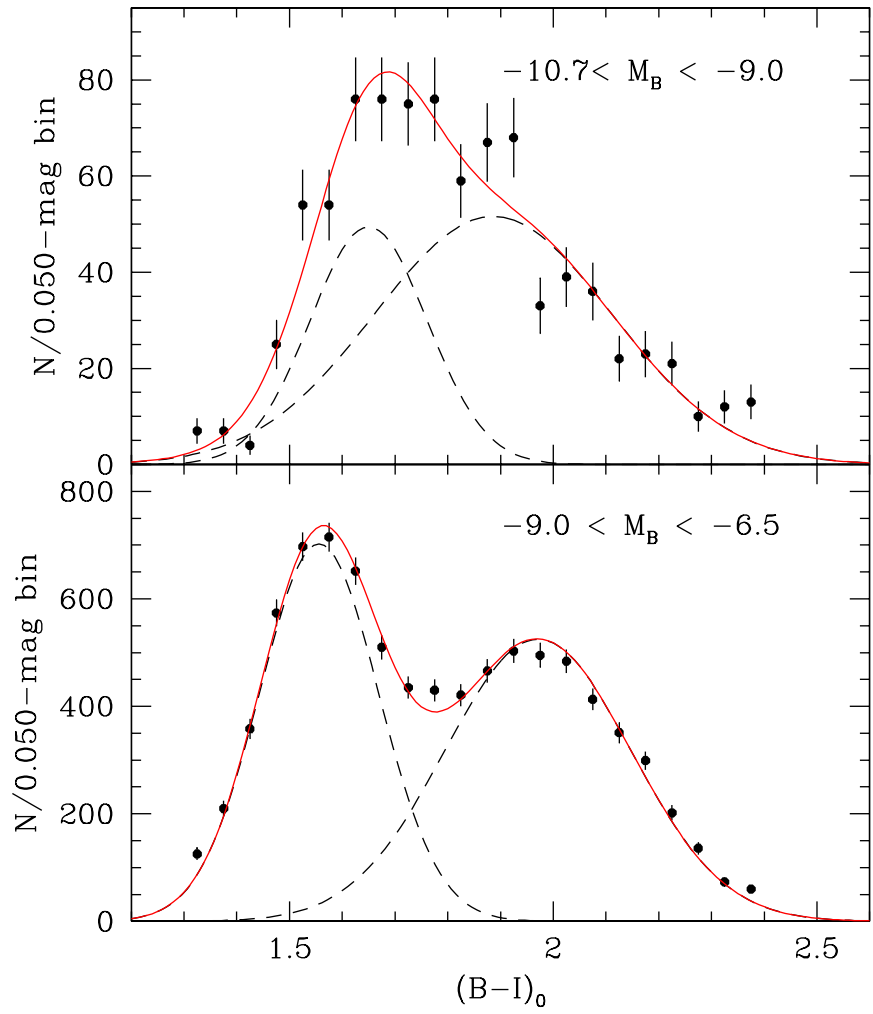}
\caption{Deconvolution of the $(B-I)$ color distributions for the GC populations
in the combined data sample, as in the previous figure but now plotted in
intervals of absolute $M_B$ magnitude.
}
\label{cfitb}
\end{figure}

\begin{figure}
\plotone{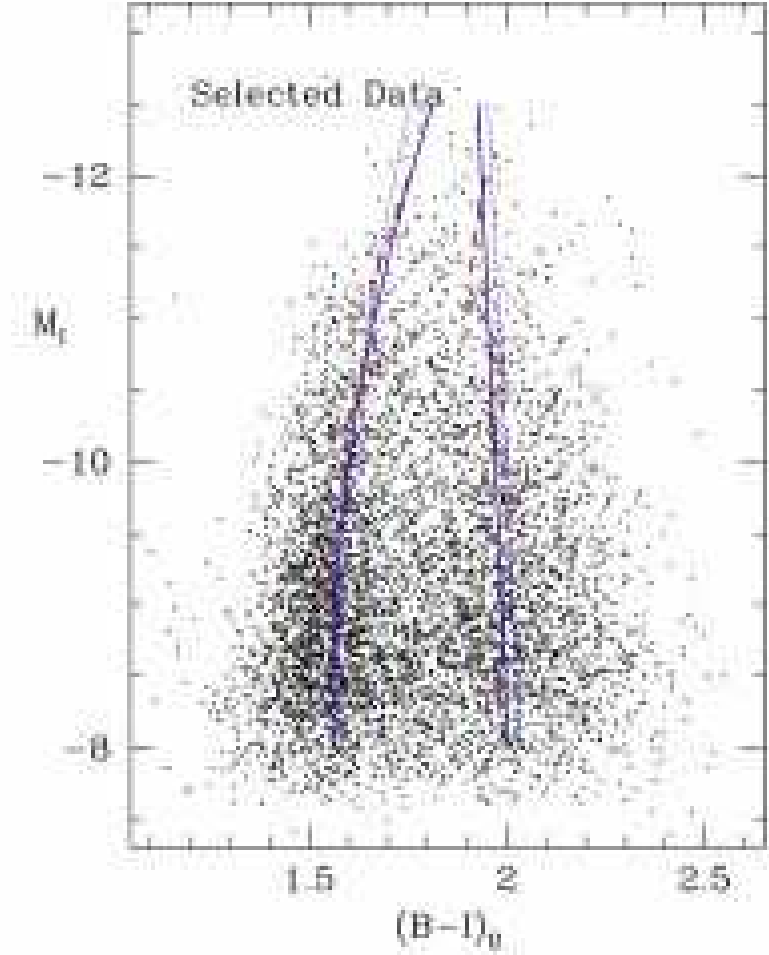}
\caption{CMD for the 7831 measured objects with 
size measurements $r_h > 1.5$ pc, drawn from the previous Fig.~\ref{cmdsize}.
This sample minimizes field contamination and provides the best
visual representation of the blue and red GC sequences.  
The heavy \emph{dashed lines} show the mean points along the
two sequences, as listed in Table \ref{rmixall}.
The \emph{solid lines} shows the polynomial fits to these mean
points as given in equations (5,6), and the \emph{dotted lines}
show the same polynomial fits determined directly from the individual
cluster data divided at $(B-I)_0 = 1.8$ (see text).
}
\label{cmdbest}
\end{figure}

At the same time, this transition point does not seem to be
abrupt.  Figs.~\ref{cmd_rmix} and \ref{cmd_rmixall} suggest
that the blue sequence makes a gradual changeover from a near-vertical
sequence at fainter magnitudes, to a sloped one at brighter
levels.  The polynomial fits stated above reflect this changeover.
For the range $M_I < -9.5$ crossing
the transition region, if we simply approximate the
trend by a linear fit, the mean slope is 
$\Delta(B-I)/\Delta(M_I) = -0.072 \pm 0.018$.  Since
$\Delta(B-I)/\Delta({\rm [Fe/H]}) = 0.375$ \citep{har06},
this translates to a heavy-element abundance scaling
$Z \sim L_{GC}^{0.48 \pm 0.12}$.  Extrapolating upward
to the very most massive GCs known, this correlation suggests
that the blue-sequence GCs at $10^7 M_{\odot}$ should have
an average $\langle$Fe/H$\rangle \simeq -1.0$ instead of 
the ``normal'' mean [Fe/H] $\simeq -1.5$ which the vast
majority of the lower-mass ones have.

An ``anomalous'' feature in the bright-end distribution appears
in the $M_I$ range around $-10.5$ to $-11$, where in
Fig.~\ref{cmd_rmixall}, three of the red-sequence points stand
noticeably off the main distribution.  These three points are
at $(B-I)_0 \simeq 1.8$ and are all from the two galaxies
NGC 3258 and 3268 (see Fig.~\ref{cmd_rmix}).  Referring back
to the raw CMDs (Figs.~\ref{cmd6} and \ref{cmd_oldnew})
we see that this is the magnitude range where the overlap
and blurring between the blue and red sequences becomes
quite noticeable.  The heavier overlap decreases the stability
of the bimodal fitting and increases the internal uncertainties
in all of the fitted parameters $(\mu_{1,2}, \sigma_{1,2}, p_1/p_2)$.
My provisional interpretation is therefore that these anomalous points
are primarily the result of small-sample statistics; the fact that
these intermediate-color points do not show up in the bimodal fits
to the combined sample of all six galaxies with its higher
statistical weight (Table \ref{rmixall}) provides
some support for that conclusion.  Nevertheless, 
the possibility should be kept in mind that our basic assumption of a
two-sequence model -- the assumption that is built in to almost
every previous discussion of GC metallicity distributions --
may not capture the entire nature of the
color/metallicity distribution for these extremely high-mass
clusters.

Because the photometry of the program objects is specifically
corrected for their individual effective radii, any
concerns that the basic existence of the MMR is merely an
artifact of the measurement process are removed.  This study,
and the one by \citet{mie06} for the Virgo galaxies, now provide the
major HST-based datasets in which the GC photometry is accurately
size-corrected.  

Extensive comparisons of the different forms of MMR published
in earlier papers are beyond the aims of the present work,
though more such intercomparisons are done by
\citet{coc09}.  They combine the results from most of these previous
studies by converting the several different photometric color systems
that have been used for different galaxies
into a homogeneous metallicity scale, and correlate the results
for the MMR slopes with galaxy luminosity and GCS specific
frequency.  

A still newer study by \citet{wat09} of the single galaxy M87 is 
of particular interest for the unusual photometric characteristics of
the data.  They use extremely deep HST/ACS imaging in 
$V$ and $I$ covering the inner 10 kpc ($1.5 R_{eff}$) of M87
to construct aperture-corrected photometry of about
2000 GCs, deep enough to cover almost the entire GCLF.
Of these, $\sim 1000$ are brighter than the 
GCLF turnover point (compared with more than 8000 from
the six galaxies in this paper)
and about 130 are brighter than $M_I = -10.5$ (compared
with over 750 from the present study).
They find the normal, first-order bimodality of the MDF 
although with a large internal spread in the $(V-I)$ colors, but
also claim that that the blue GC sequence has no
significant MMR slope.  This conclusion relies on a forced
linear-fit model over the whole magnitude range 
in $(I, V-I)$ and is thus
dominated by the 94\% of the sample that is fainter than
the actual MMR ``turn-on'' point at $M_I \simeq -10.5$.  
In fact, their data do not rule out the type of MMR that has been found
here:  that is, a near-vertical blue sequence on the fainter
part of the magnitude range, moving upward to a stronger
slope towards higher metallicity with a transition region
above a million Solar masses.  The data in the present study
draw from a statistical sample more than 6 times larger,
and in addition, use a photometric index $(B-I)$ that is twice as
sensitive to GC metallicity as is $(V-I)$. 

Of the previous studies, the most closely comparable one is probably
\citet{mie06}, who analyze GC samples from 79 Virgo ellipticals
and perform bimodal-Gaussian fits to the GC color distributions
with techniques very similar to the ones used here.
Their database has similar size (a total of a few thousand GCs brighter than
the GCLF turnover luminosity), uses a color index $(g'-z')$ with
similar metallicity sensitivity, and has a low level of field
contamination.  They divide the galaxies into four different
luminosity groups (divided roughly speaking into
supergiants, giants, normal ellipticals, dwarfs)
and then discuss the trends of mean GC color
versus magnitude in each group.  The most important differences
between their analysis and the present one are that (a) the Virgo
data extend about 1.5 mag fainter in GC luminosity, below the
GCLF turnover; and (b) they apply the forced linear-fit assumption
to the mean points from the bimodal fits, which I have argued
against above.  The blue GC sequence
is found to have a significant slope (MMR) for the three bins 
containing the large ellipticals, but no significant slope is
found for the GCs in the dwarf ellipticals.  In all cases the red GC sequence
is indistinguishable from vertical.  Referring particularly
to their Figure 2 where the mean points are plotted, I note that 
the significance of the blue MMR slope will be very much reduced
if the brightest two points ($M_z < -10$, close to the $10^6 M_{\odot}$
level) are excluded and only the fainter bins are used.
The strongest case in their data for a nonzero MMR slope is
from the combined GC sample for the three brightest giants,
M87, M49, and M60 (their Fig.~2a), though even there the blue
sequence becomes essentially vertical for $M_z > -9$.

\begin{figure}
\plotone{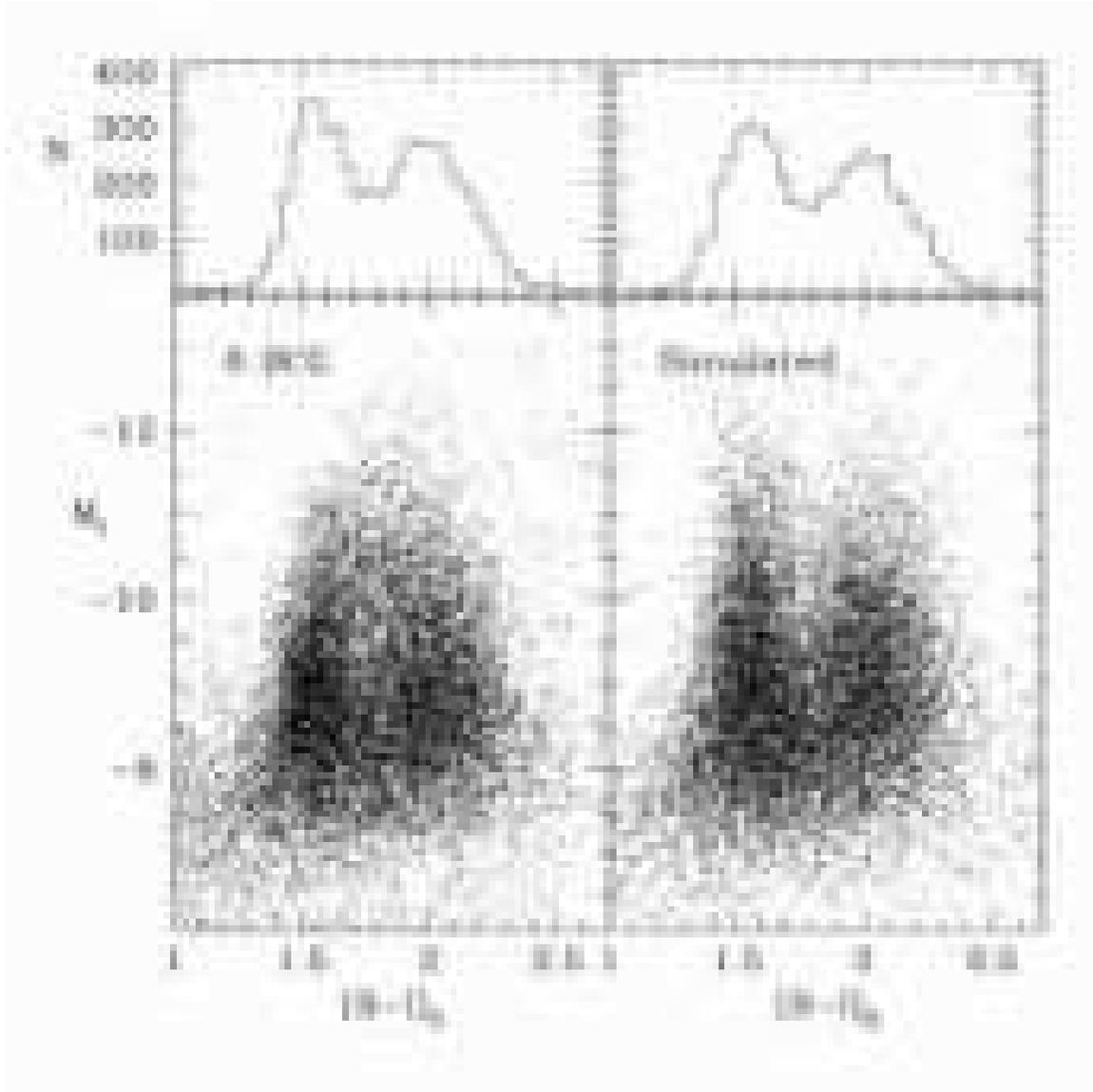}
\caption{\emph{Left panel:} Combined color-magnitude data for the
six BCGs, from Fig.~13a.  The histogram at top shows the distribution
in $(B-I)_0$ for the luminosity range $-10.0 < M_I < -8.5$, a range
fainter than the main region of the MMR but brighter than the
region where the colors are broadened significantly by observational
measurement scatter.
\emph{Right panel:} Simulated data for a bimodal GC population with
parameters as described in the text.  Both the red and blue sequences
are purely vertical (no MMR), with mean colors $(B-I)_0 = 1.56$
and 2.00.  The histogram above shows the color distribution for
the same luminosity region $(-10.0, -8.5)$.
}
\label{simsample}
\end{figure}

Additional tests for any biasses and internal 
uncertainties in the fitting procedures 
can be carried out through simulated CMDs 
\citep[see, e.g.][for recent examples]{mie06,wat09}.
Figure \ref{simsample} shows one illustration of a model CMD tuned to the 
present $(B-I)$ data, in which the total
numbers of GCs and contaminating field objects are closely similar
to the real data for the six BCGs combined.  Here, the 
simulated blue sequence
is purely vertical with a mean color $\langle B-I \rangle = 1.56$
and Gaussian standard deviation $\sigma_{B-I} = 0.09$, and follows
a Gaussian luminosity function with $\langle M_I\rangle = -8.3$, $\sigma_I = 1.45$.
The red sequence is purely vertical with $\langle B-I \rangle = 2.00$,
$\sigma_{B-I} = 0.15$, and its LF follows $\langle M_I \rangle = -8.3$,
$\sigma_I = 1.25$.  Field stars are added following a uniform
distribution in color and a power-law increase
in magnitude. All the artificial data are broadened in color by 
measurement scatter that increases with magnitude, and lastly have
photometric completeness
functions in $B$ and $I$ applied to them that mimic the real data.
The simulated CMD in Fig.\ref{simsample}
includes 17500 GCs, 55\% of which are
on the red sequence and 45\% on the blue sequence.
Lastly, the histograms above each CMD show that the intrinsic
distributions in $(B-I)$ \emph{for the luminosity range below
the onset of the MMR} are similar.
The simulated CMD clearly fails to reproduce the
structure at the high-luminosity end ($M_I \lesssim -10$)
that we see in the real data.  The most noticeable effect is that
the real data do not have the very metal-poor \emph{and} very luminous
GCs that would lie on the direct upward extension of a vertical
blue sequence.  If these were actually present in the real galaxies, 
they would be more easily found than GCs anywhere else in the 
color-magnitude diagram.
RMIX bimodal fits to the simulated data show no significant trend for a
spurious MMR to appear, or for significant biasses in
the deduced sequence colors.  

\subsection{Implications for Theoretical Models}

The currently published analyses that draw from
large, deep, and metallicity-sensitive photometry point to
an emerging consensus around the view that the MMR has a
nonlinear form:  the blue sequence is 
nearly vertical at low GC luminosities,
but continuously changes to a stronger increase of
metallicity with cluster mass as we go upward along the
sequence.  On the red sequence, however, the current data are consistent with the
interpretation that it stays nearly
vertical (metallicity is uncorrelated with luminosity).

These observations, if correct, already 
have implications for the different theoretical
models based on the central concept of self-enrichment.
\citet{str08} construct a model built on
the assumption of a continuous linear form of the MMR, which
as noted above corresponds to a scaling $Z \sim M^p$ over
the entire GC mass range.  In their model
this form turns out to require that the \emph{star formation efficiency}
$f_{\star}$ within the protocluster cloud
also scales continuously with cluster mass in
the same way.  This particular form of a self-enrichment model
is therefore ruled out by the observations, which show that
the blue sequence is nearly vertical (metallicity is uncorrelated
with mass) for $M \lesssim 10^6 M_{\odot}$, and that more generally
the MMR is nonlinear.

In contrast, \citet{bai08} assume 
$f_{\star} = const$ with GC mass (specifically $f_{\star} = 0.3$, though the
precise value is not important).  The important quantity that
changes with cluster mass is instead the fraction of SNe-ejected gas 
held back in the protocluster, i.e. the
heavy-element \emph{retention efficiency} $f_Z$ which
drives the enrichment of the lower-mass stars in the protocluster.
The retention $f_Z$ is a strong nonlinear function of $M$ 
(that is, the depth of the protocluster potential well). 
This approach leads to
the prediction that the blue GC sequence should show
a fairly distinct visible ``onset'' of the MMR at around one
million Solar masses.  The same sort
of self-enrichment onset should also appear on the red GC sequence,
though it would only become obvious at still higher mass
because the mean pre-enrichment of the red GCs (that is, their
baseline mean metallicity) is an order of magnitude higher.
This theoretical form of the MMR is more in agreement with the
observations.

Both these model approaches are admittedly idealized, and a
synthesis of their ideas might lead to an
improved match to the data.  The Bailin/Harris model predicts
a MMR onset that is probably too abrupt (a rapid changeover to
a sloped MMR appears just at the point where the amount of internal
self-enrichment becomes larger than the baseline pre-enrichment
metallicity).  The data here (see particularly Fig.~\ref{cmd_rmixall})
favor a smoother and more gradual change in slope over a wide
mass range.  Clearly the enriched-gas retention efficiency $f_Z$ should 
increase with protocluster mass, but the star formation efficiency
may also change systematically with increasing mass.  These models
also treat the protocluster in isolation.  If it is placed within
a surrounding lower-density gaseous region (giant molecular cloud
or dwarf galaxy), then the addition of some external pressure
confinement could increase the effective $f_Z$ and extend the
visible onset of the MMR more smoothly to lower masses (see the model papers
for additional discussion).  

\subsection{The MMR and Cluster Size}

A final stage of the MMR discussion in this paper is to ask
whether the details of the mass/metallicity correlation might
depend on other external factors.  Two factors that can be
investigated here are the \emph{characteristic size} of the
GC and its \emph{location in the galaxy}, both of which may
affect the efficiency of self-enrichment.

If the MMR itself is a \emph{second-order} feature of the GC metallicity
distribution, requiring $\sim 10^3$ clusters to clearly reveal it,
then any correlations of the MMR with other parameters are
still more subtle \emph{third-order} features that we might expect to require samples of
$\sim 10^4$ clusters or more.  The database in the present paper is
therefore just barely large enough to begin this kind of search.
Indeed, the results to be described below are suggestive of
certain trends but in the end must be considered as no more than
tantalizing hints.

The slope of the MMR should in principle depend on the depth of the cluster
potential well, and its effective radius helps determine that depth. 
That is, for a given cluster mass
we might expect the self-enrichment process to be less 
effective for more extended GCs,
which would have lower escape energies and thus be less able
to hold back their early SNe ejecta.  If this idea is correct,
then along the blue GC sequence the clusters with the biggest
radii should be systematically bluer (more metal-poor) than those with small radii.

\begin{figure}
\plotone{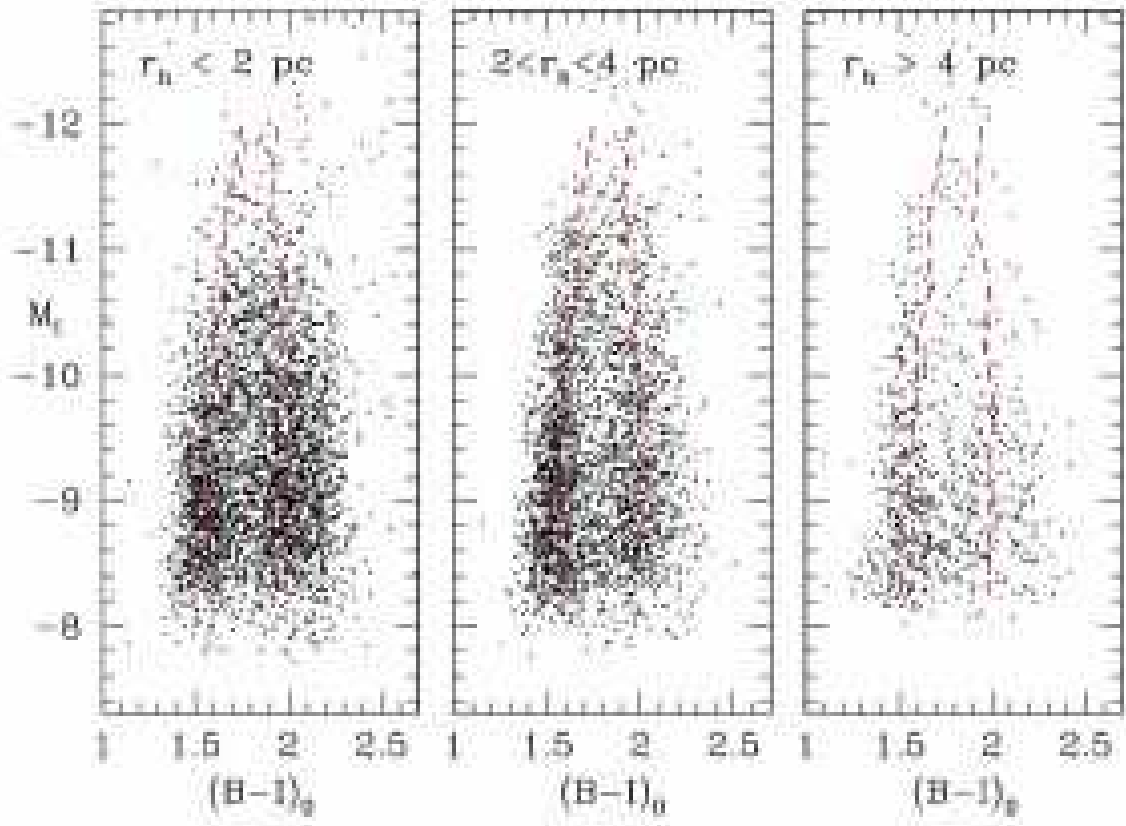}
\caption{\emph{Left panel:} Color-magnitude distribution for the
GCs with $S/N > 20$ and measured half-light radii smaller than 2 pc.
\emph{Middle panel:} The same data for objects between 2 and 4 pc.
\emph{Right panel:}  The same data for the biggest objects, 
$r_h > 4$ pc.  In all three panels, the dashed lines show the
mean GC sequences derived from the entire data sample (see
Fig.~\ref{cmd_oldnew} and accompanying text).
}
\label{cmd3_rh}
\end{figure}

A first look for any such effect is shown in Figure \ref{cmd3_rh},
where the total sample is subdivided by $r_h$.  The only pre-selection
applied here is to restrict the list to those objects with 
ISHAPE $S/N > 20$ to ensure some minimal quality in the 
cluster size estimate.  
\begin{itemize}
\item{} As discussed above, the subsample with $r_h < 2$ pc  
is likely to have the highest proportion of
field contaminants of various kinds.  Nevertheless, many 
of the objects in this range are likely to be GCs:  the 
bimodal sequences remain well defined and their locations
track the mean loci defined from the whole sample (dashed
lines in the Figure).
\item{} The objects in the range $2 < r_h < 4$ pc
make up the classic, ``normal'' globular clusters whose sizes
fall comfortably into the baseline pattern set by the Milky Way.  
This part of the sample is likely to be relatively free from contamination,
and defines clear bimodal sequences.  They accurately
follow the mean lines for the entire sample as well.
Interestingly, there are relatively more \emph{high-}luminosity
clusters ($M_I < -11$) in this group than in the previous
one, but this is largely due to the radius/luminosity correlation
discussed earlier (Fig.~\ref{rh_allmag}).
\item{}  The biggest objects are in the third panel of the figure.
These fall in the size range $4 < r_h < 10$ pc and would include
the most extended known GCs with luminosities higher than the
GCLF turnover point, except for a few
still bigger UCD-like objects.  
One feature to note is that, at all luminosities, 
there are relatively fewer red-sequence clusters
in this size range than at the lower sizes;
this is the result of the size/metallicity correlation discussed
earlier (Fig.~\ref{rh_rad}).  The details of the blue GC sequence
are potentially more interesting, because if
the predictions of the self-enrichment model are correct, we 
might expect to begin seeing its effects here.  Indeed, 
these large clusters fall steadily about 0.1 mag to
the blue of the mean line for the whole sample, corresponding
to $\Delta$[Fe/H] $\simeq -0.25$ dex in metallicity.
The self-enrichment model is likely not to be the only possible
explanation of this offset, but it is a tantalizing piece of
evidence that at very least is consistent with the model.
\end{itemize}

A Kolmogorov-Smirnov test was used to compare the color distributions
for the three groups of clusters shown in Fig.~\ref{cmd3_rh}, 
restricted to the upper luminosity range $M_I < -10.0$ where the
differences appear to be most noticeable.  The third group ($r_h > 4$ pc)
is systematically bluer than the first group ($r_h < 2$ pc) with
very high significance, $> 99$\%.  The middle group ($r_h = 2-4$ pc)
is different from either of the other two at the $\gtrsim$ 90\% level,
suggestive but not conclusive.  

In the third panel of Fig.~\ref{cmd3_rh}(c), the very high 
luminosity range $-11.2 > M_I > -11.8$ also shows a dozen objects that fall
between the two main sequences.  Two are from the NGC 3268 field,
seven from NGC 7626, and three from NGC 4696, all have
$S/N > 200$, and all have $r_h$ values from 4 to 7 pc.
Careful visual inspection on the original images shows that one of these has
a comparably bright neighbor $0.45''$ away, not close enough to
invalidate the profile fits.  All the others are isolated objects
with small, GC-like ellipticities and no morphological peculiarities.
There is no reason to reject them from the GC sample.

\subsection{Trends versus Galactocentric Distance}

\begin{figure}
\plotone{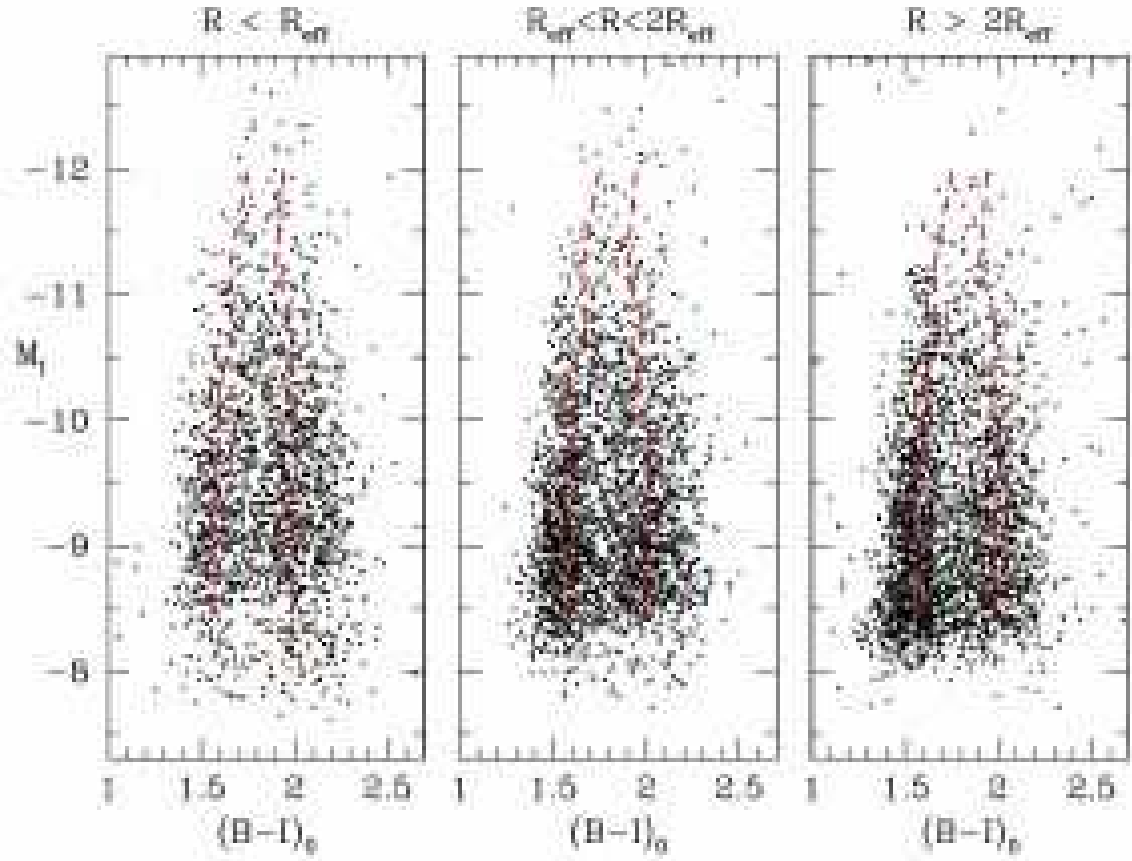}
\caption{Color-magnitude distribution for the globular clusters 
with $S/N > 20$ in all
six galaxies, subdivided by radial zone.  The projected galactocentric
distance of each object is expressed in units of the effective radius
$R_{eff}$ of the galaxy's spheroid light.  In each panel, the 
dashed lines show the mean GC sequences derived from the entire
data sample.  The most remote measured objects are at radii $\simeq 5 R_{eff}$
(see text).
}
\label{cmd_zone}
\end{figure}

\citet{mie06}, in their study of the Virgo galaxy GCs, suggested
that the MMR may become less noticeable at larger galactocentric distance.
Such a trend might be the mark of
different enrichment or formation histories at different locations
in the parent galaxy's halo.  However, their data were restricted
to the inner radial region
$R_{gc} \lesssim 1.5 R_{eff}$ (for the bigger galaxies).
The present sample covers a much bigger range in $R_{gc}$ and thus
should be able to provide a more sensitive test of any such trend.
Figure \ref{cmd_zone} shows the color-magnitude diagrams in
three successive radial zones divided by galactocentric distance:
the inner region $R_{gc} < R_{eff}$, a middle zone $R_{eff} < R_{gc} < 2 R_{eff}$,
and the broad outer region $2 R_{eff} < R_{gc} \lesssim 5 R_{eff}$.  
These radial subdivisions leave
about the same numbers of objects in each
zone, so the mean red and blue sequence lines that were
defined previously (the heavy
dashed lines in the figure) represent essentially a straight average over all 
radial zones. 

The correlation of MMR with $R_{gc}$ also provides an implicit
secondary test that the GC photometry is not systematically affected
by the background light gradient of the galaxy.  For $R_{gc} \gtrsim R_{eff}$,
the gradient becomes low and any residual effects on the measurements
will be small.
 
\begin{figure}
\plotone{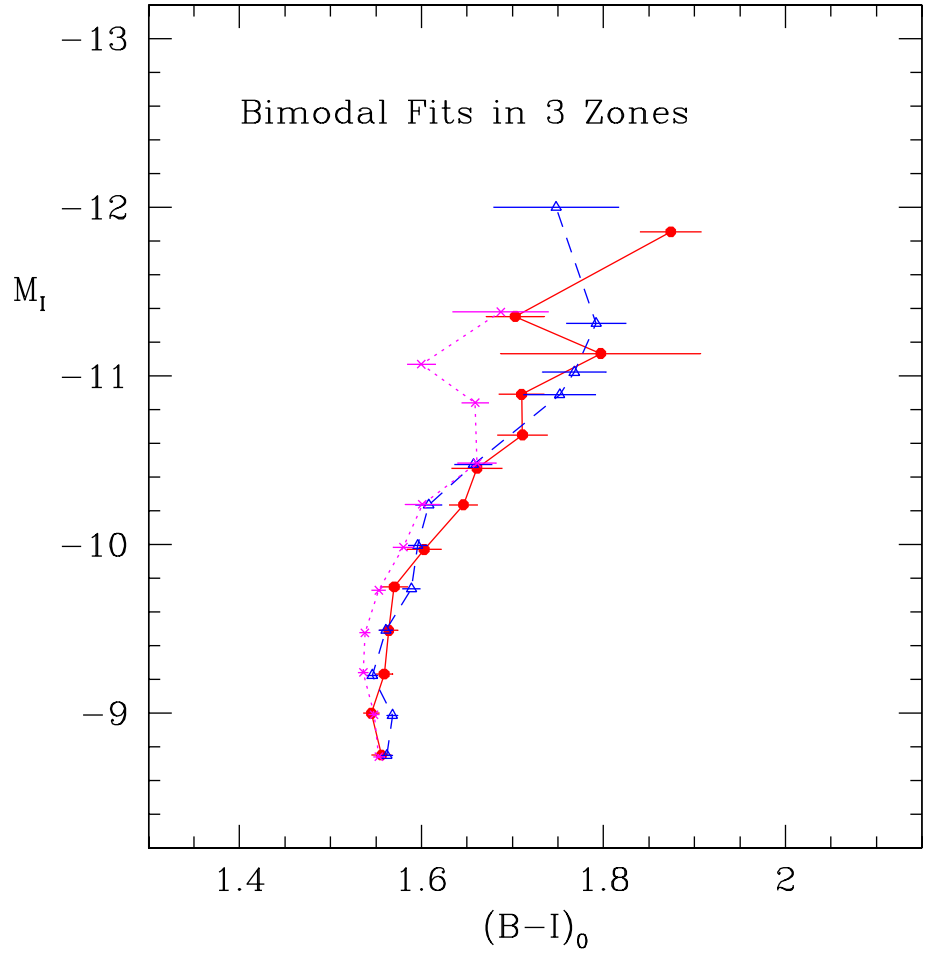}
\caption{Mean points determined from the RMIX bimodal-Gaussian fits 
for the blue cluster sequence in 
three radial zones as defined in the previous figure.  Solid
dots and line are for the inner zone, the triangles and
dashed line for the middle zone,
and the crosses and dotted line for the outer zone.  
}
\label{cmd_zonelines}
\end{figure}

To search for clues that the MMR slope might
depend on location, 
the RMIX bimodal fits described above were repeated separately for
each zone.  The results are shown
in Figure \ref{cmd_zonelines}.  Here, because the sample sizes
in each luminosity bin are smaller, it was not possible to
obtain fully unconstrained fits to the same level of confidence
for all five parameters ($\mu_1, \mu_2, \sigma_1, \sigma_2, p_1/p_2$) 
in the high-luminosity, sparsely populated bins that are the key
to the MMR slope.  However, because we are 
looking especially for differences in the mean colors of 
the \emph{blue} sequence
from one zone to the next, the RMIX fits were deliberately
constrained so that the mean color and standard deviation
of the \emph{red} sequence were fixed, since it is nearly vertical
to begin with.  The standard
deviation is fixed at $\sigma_2(red) = 0.18$ (see Table \ref{rmixall}), 
and the mean color fixed at $\mu_2(red) = 1.994$ for the inner
zone, 1.979 for the middle zone, and 1.949 for the outer zone.
These values were adopted to account for the shallow radial metallicity
gradient in the system (discussed in the next section).
In addition, to smooth the mean sequences numerically, the RMIX
fits were done in $\simeq 0.5-$mag wide luminosity bins
but stepped by $\Delta M_I = 0.25$ mag between bins.

In Fig.~\ref{cmd_zonelines}, from bottom to top the same gradual
changeover as noted previously is apparent, from a nearly vertical
sequence at $M_I \lesssim -10$ to a progressively steeper MMR
at higher luminosities.  The internal errorbars on each mean point
also grow dramatically at higher luminosity as the sample size
in each bin decreases; and for $M_I \lesssim -11$, the bimodal fits
become rather unstable as the blue sequence overlaps more and more
heavily with the red sequence and even the proportions $(p_1,p_2)$
in each component become quite uncertain.  There is no convincing
difference between the mean lines for the inner (solid line) and
middle (dashed line) zones.  A more noticeable difference appears
for the outer zone, which stays offset to the blue by $\sim 0.1$ mag
relative to the inner zones.  Taken at face value, this means that
the MMR decreases in amplitude further out in the halo.

\begin{figure}
\plotone{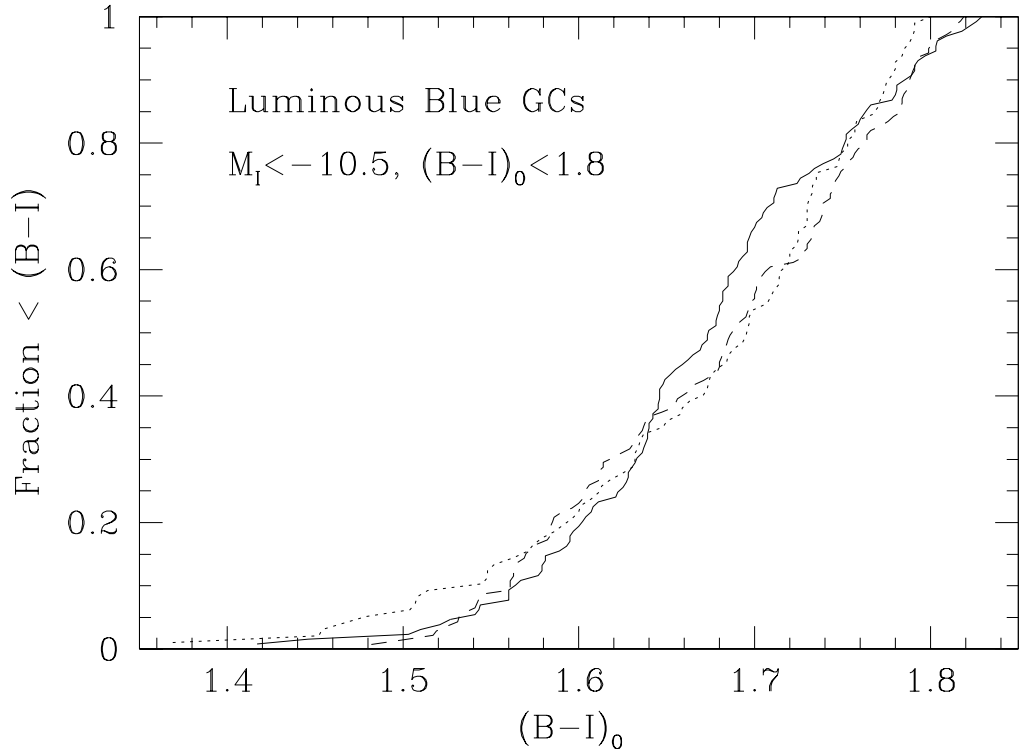}
\caption{Fraction of luminous, blue-sequence globular clusters bluer than
$(B-I)_0 = 1.8$, plotted against color.  \emph{Dotted line} is for the
inner-zone clusters ($R< R_{eff}$), \emph{dashed line} is for the
mid-zone ($R_{eff} < R < 2 R_{eff}$), and \emph{solid line} is for
the outer zone ($R > 2 R_{eff}$).
}
\label{ks_zone}
\end{figure}

Another test that 
avoids some of the problems of binning the data, is to look directly at
the color distribution in $(B-I)_0$ for the blue-sequence GCs
brighter than $M_I = -10.5$.
The fiducial dividing line between the 
red and blue sequences is $(B-I)_0 = 1.8$:  the inner zone has 97 objects 
bluer than that boundary, the middle zone 139, and the outer zone 129.
Figure \ref{ks_zone} shows the cumulative color distributions for
these bright, blue GCs.  The existence of a global, weak radial
gradient in mean metallicity (see next section) first requires
adding $\Delta(B-I) = 0.02$ mag to the middle zone and 0.03 mag
to the outer zone before comparison with the inner zone.  
The color distributions becomes steadily
more weighted to the bluer side at larger $R/R_{eff}$,
particularly in the color range $(B-I)_0 \simeq 1.6 - 1.7$ that
is crucial for determining the MMR.
However, a normal Kolmogorov-Smirnov two-sample test shows that none
of the zones are different from any of the others at higher than
70\% significance.  This is not sufficient to demonstrate any
radial change in the MMR convincingly and is left only as
another tantalizing hint.  

A tentative conclusion from these tests is that (a)
the MMR exists at all the galactocentric distance zones
that the present data are capable of exploring; but that 
(b) there is \emph{very weak} evidence that the MMR
becomes gradually less noticeable at larger galactocentric distance.
The direction of this effect is the same as found by \citet{mie06}
(see their Table 2 and section 4.2 discussion).
Bigger samples of high-luminosity clusters
from other galaxies will be needed to probe this effect
conclusively.

\citet{bai08} suggest two effects that might end up producing
a radial gradient in the MMR.
One of these would be relevant in large galaxies that
hosted active galactic nuclei (AGN's) at an early stage
in their evolution:  the AGN would dramatically increase the
external ultraviolet radiation field and, possibly, inhibit
the self-enrichment process within a protoglobular cluster
of a given mass.  Taken at face value, this model would predict
that the MMR should be stronger at larger
galactocentric distance farther from the AGN, which is opposite to what we observe.  
However, there is no guarantee that
the timing of GC formation would automatically overlap with the
appearance of the central AGN. The UV radiation field might even produce
the opposite gradient if, for example, the inner GCs had already
finished forming by the time the AGN reached its peak luminosity
but the outer GCs had not.  

The second possible effect is
related to the difference between the
\emph{initial} masses of the GCs (which determine their degree
of self-enrichment) and their \emph{present-day} masses, which are
smaller because of mass loss due to early SNe ejecta, stellar
winds, and the slower processes of
tidal stripping and evaporation of stars over $10^{10}$ y.
The inner GCs
should have been subjected to a higher degree of dynamical mass loss
due to tidal stripping than the outer GCs over their lifetimes, 
so an inner-halo GC
that has the same mass \emph{now} as an outer-halo one would have started
out at higher mass and thus have been able to self-enrich more.
Testing this idea more quantitatively and finding out specifically
how it affects the final MMR will require knowing, at a minimum,
the average amount of GC mass loss from the protocluster stage
onward, as a function of both $M_{GC}(init)$ and $R_{gc}$.  
This direction may prove to be a rich area for detailed modelling.

Another GC phenomenon that may be linked to these arguments is
the long-standing puzzle that the GC luminosity (or mass) function is
nearly independent of galactocentric distance $r$.
If the initial GCMF were nearly the same
throughout the early galactic halo, then the strong dependence of
dynamical evolution time on $r$ should build in a present-day
GCMF whose low-mass end and ``turnover point'' depend on $r$.
A recent discussion by \citet{mf08} shows that the observed
shape of the GCMF can be explained if its initial shape depends
appropriately on the characteristic cluster density $\rho_h$, because the mass
loss rates increase with $\rho$.  The fact that the outer clusters
have lower average densities would also go along with less self-enrichment
and a weaker MMR at larger distances from the galactic center.

\section{Metallicity Gradients}

Careful inspection of the mean positions of the 
near-vertical sections of both blue and
red sequence lines shows that at larger radius, the mean color
of each sequence shifts slightly to the blue compared with the loci
at smaller radii.  That is, both types of GCs appear to show a
metallicity gradient. 
It has long been realized that the GCS in a normal
large galaxy taken \emph{as a whole} 
(that is, combining all types of clusters) shows a negative metallicity gradient,
but the majority of the effect is created simply by the decreasing ratio of
red versus blue GCs with increasing galactocentric distance 
\citep{gei96,harris98,dirsch03,rhode04,har06}.  That is, 
the steeper spatial distribution of the red, metal-rich clusters
produces a \emph{population} gradient, which has the byproduct of
an overall metallicity gradient.  In \citet{har06} we discussed
the population gradients in the BCGs studied here.
By contrast, much less published material
exists that is directed at finding any true metallicity 
gradient \emph{within} each of
the two types of clusters, and indeed any such gradients are subtle.

A large body of literature exists for abundance gradients in the
integrated light of large galaxies, measured from 
line-strength indices.  These studies are usually based on data from
the high-surface-brightness region within $1 R_{eff}$
of the galaxy center 
\citep[see][for a sampling of such studies, along with extensive guides 
to the literature]{fis95,kob99,meh03,san06}.  The analyses typically
show that heavy-element abundance within this bulge light
scales with projected galactocentric
distance as anywhere from $Z \sim R^{-0.15}$ to $R^{-0.4}$, favoring interpretations
in which the central regions of these galaxies formed by a combination
of dissipative collapse (favoring steep gradients) and subsequent mergers
(favoring flatter gradients).  The results are
rather strongly model-dependent in the sense that each line index
must be transformed to some tracer of heavy-element abundance.

Globular clusters can provide valuable extensions to these integrated-light
measurements.
GCs are easy to measure at radii well beyond $R_{eff}$, and transformation
of their colors to metallicities is based directly on calibration from
the Milky Way clusters.  These transformations need 
only the reasonable assumption that
the GCs  are ``old'' ($\gtrsim 10$ Gyr) and thus in the regime where
the colors are quite insensitive to minor age differences. 
The only practical issue is that a large statistical sample of GCs,
over a large radial range,
is needed to measure the metallicity gradient accurately.  A large radial
range, and large sample size, are 
precisely the advantages of the database in the present study.

\begin{figure}
\plotone{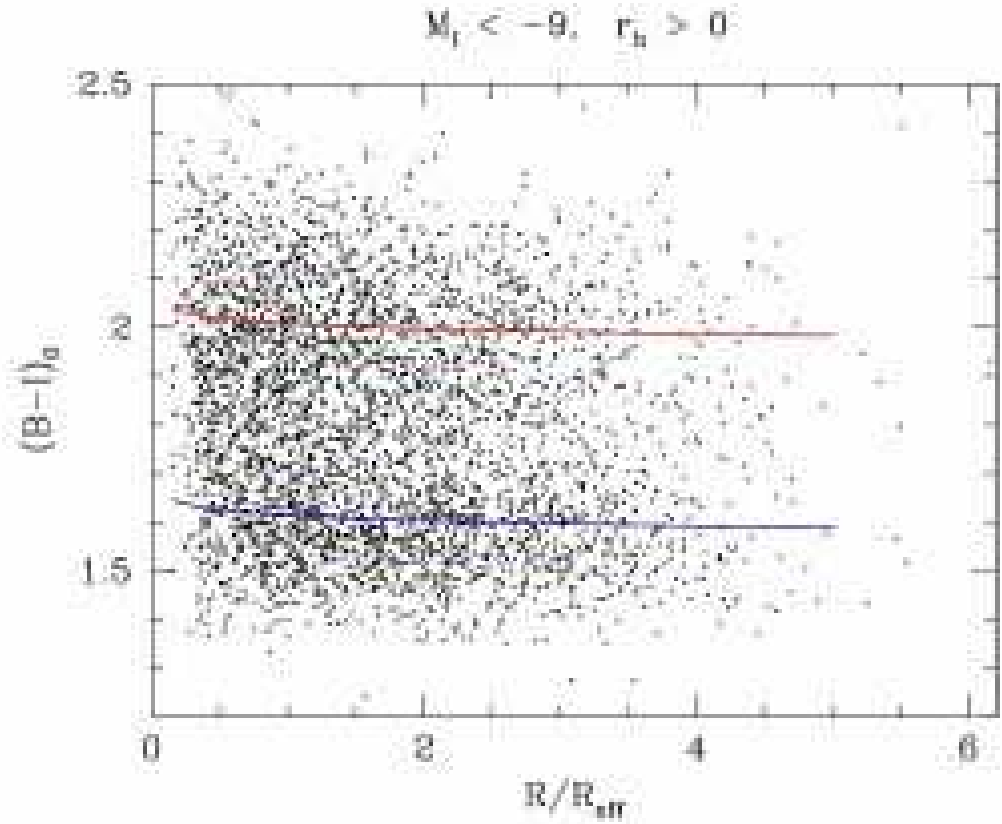}
\caption{Colors of the bright ($M_I < -9$) objects with positive
size measurements ($r_h > 0$), plotted against galactocentric distance.
Data from all six galaxies are combined.  The individual distances $R$
are normalized to the effective radius $R_{eff}$ of the spheroid light.
The solid lines show the best-fit power-law solutions for the
metallicity gradients, as listed in the last line of Table \ref{radialz}.
}
\label{gradient}
\end{figure}

Figure \ref{gradient} shows the intrinsic colors $(B-I)_0$ of the
GCs in all six galaxies plotted directly against galactocentric
distance, again normalized to $R_{eff}$ for each one.  
For this purpose the sample
is restricted to the brighter objects ($M_I < -9$) with 
positive size measurements ($r_h > 0$), to minimize the effects
of field contamination.  As above, the dividing line
between blue and red GCs is defined as $(B-I)_0 = 1.80$ and we
then solve directly for the best-fitting relation for mean color
versus radius within each group, in the conventional form
$\langle B-I \rangle_0 = a\, +\, b\, {\rm log} (R/R_{eff})$.

Table \ref{radialz} lists the solutions for the individual galaxies,
along with the solutions for the combined sample.
The slopes $b$ in all cases are shallow, and essentially equal for
both the red and blue components.  In addition, they are not significantly
different between individual galaxies, so to gain statistical weight,
we can combine all of them into a single solution. 
The combined sample yields a nonzero signal for the slopes   
at the $> 4.5-\sigma$ level for both types of clusters,
and therefore the existence of a metallicity gradient appears to be convincingly real.
Given the conversion ratio $\Delta(B-I) = 0.375 \Delta$[Fe/H] (see above),
both of the gradients convert to a heavy-element abundance
scaling $Z_{GC} \sim R^{-0.10 \pm 0.022}$.

The slope $\Delta {\rm log} Z /\Delta {\rm log} R = -0.10$ is shallower
than is conventionally found in the integrated-light studies (see the
references cited above), but it is also dominated by more distant regions of
the halo, $R > R_{eff}$.  Few other GCS studies exist for which direct
comparisons can be made.  \citet{gei96}, in their wide-field study of
the GCS around the Virgo giant NGC 4472, found slopes of 
$Z \sim R^{-0.15 \pm 0.03}$ for the blue GCs and
$Z \sim R^{-0.12 \pm 0.06}$ for the blue GCs, 
out to a limiting radius
of $\sim 5 R_{eff}$.  These are clearly similar to the results here.
Geisler et al. also point out that the gradient for
the red GCs closely matches the color gradient for the integrated halo
\emph{light} of the host galaxy; and direct evidence from halo-star
studies in giant ellipticals \citep[e.g.][]{hh02,rej05} shows that
the field-star population is dominated by the metal-richer stars whose
MDF matches that of the red GCs \citep[see also][]{forte05,forte07}.
\citet{for01} determined the gradient for the blue GCs
in two Fornax ellipticals, NGC 1399 and 1427, finding $Z(blue) \sim R^{-0.2}$
for both systems.  Their data extended to $12 R_{eff}$ for NGC 1399 and
$10 R_{eff}$ for NGC 1427.   Recently, \citet{lee08} have found shallow
but significant $(C-T_1)$ color gradients in both the blue and red GCs in the Virgo giant
M60 that are quite similar to those for M49.  
Lastly, in M31, \citet{per02} and \citet{fan08} find that the
metal-poor GCs exhibit a metallicity gradient out to 20 kpc (or about
$4 R_{eff}$) but suggest that the metal-rich GCs show no gradient.

Perhaps the earliest modern investigation of this question is
by \citet{zin85} for the Milky Way itself. He found that
the inner, metal-poor GCs were significantly more enriched than
the outer-halo ones.  Using contemporary metallicity data for 105
metal-poor clusters with [Fe/H] $ < -0.95$ 
\citep[with data taken from the 2003 edition of][]{har96},
I find a best-fit solution
$\langle$ Fe/H $\rangle = (-1.446 \pm 0.063) - (0.165 \pm 0.062) {\rm log} r_{gc}$
where $r_{gc}$ is the true three-dimensional distance
from the Galactic center.  
This GC sample extends
out to more than 100 kpc from the Galactic center.
A solution in which the same data are
artificially projected onto two dimensions (making them directly comparable
with the other galaxies above) gives
$\langle$ Fe/H $\rangle = (-1.490 \pm 0.047) - (0.146 \pm 0.051) {\rm log} R_{gc}$.  
In this case the 3D slope is already so shallow that projection
into two dimensions does not decrease it much.

The available data sketch out a consistent pattern whereby both the
metal-poor and metal-rich globular cluster subsystems have 
shallow but real radial metallicity gradients, in the range $Z \sim R^{-(0.1-0.2)}$.
Particularly in the large spiral galaxies, the gradients for the metal-rich
clusters are harder to determine because they are less numerous and spatially more
centrally concentrated, but for the metal-poor clusters
the Milky Way scaling of $Z \sim R^{-0.15\pm0.05}$ 
is typical of the mean for all the galaxies.  The most internally precise
value for any of the systems is the one determined here ($b = -0.10 \pm 0.022$)
for the six BCGs.

Interpretation of these metallicity gradients in terms of the past history
of the parent galaxy is an intriguing and incompletely explored area.
Major mergers can produce extensive spatial mixing and reduce any
pre-existing steep metallicity gradient that was the result of
dissipative early star formation in
the progenitors.  Gas-rich mergers will produce a new population of
stars and clusters especially in the inner regions and may thus
help rebuild the gradient, at least for the \emph{metal-rich} component.
However, two intriguing features of the observational evidence so far are
that (a) the gradient slope $b \simeq -(0.1-0.2)$ appears to be very much the
same for both giant elliptical and spiral galaxies, and (b) it is also
the same for the metal-rich
and metal-poor clusters, even though they may have experienced very
different histories as well as formation times.  More extensive
N-body simulations of mergers specially tuned to trace metallicity gradients
well outward into the halo will be needed to constrain the range of possibilities
and help decide how much of the gradients we observe now are relics of
the primordial conditions of formation.

\section{Summary}

In this study the globular cluster populations in six giant elliptical
galaxies have been investigated through their distributions in
size, metallicity, and luminosity.  A total of about 15000 objects, 
of which about $\sim 8000$ are high-probability 
globular clusters belonging to these six galaxies, have
been measured down to a luminosity $M_I \simeq -8$, slightly fainter
than the turnover point of the standard GC luminosity function.
Particular emphasis is given to
characterizing the relation between GC mass and metallicity (the MMR)
that becomes most noticeable for the blue, metal-poor cluster population
at high luminosity.  A summary of the findings in this study can
be made as follows:

\begin{enumerate}
\item The target galaxies in this study are at typical distances
$d \sim 40$ Mpc and at these distances, their globular clusters 
can be barely resolved as nonstellar with our HST/ACS imaging data.
Analytic King-model fits convolved with the stellar point-spread
function have been used successfully to determine total magnitudes,
colors, and effective (half-light) radii for each detectable object.
\item The new photometry of the clusters based on individually
size-corrected total magnitudes is more accurate than the previous
data based either on PSF-fitting or fixed-aperture photometry
\citep{har06}, but it does not lead to substantial changes in
the color-magnitude distribution.  The GCs accurately define
the classic bimodal red and blue components at all luminosities,
with the blue component at $\langle B-I \rangle_0 = 1.6$
and the red component at $\langle B-I \rangle_0 = 2.0$.
\item The distribution in intrinsic GC scale size in these
giant ellipticals accurately matches that of the Milky Way clusters,
with a peak frequency near $r_h \simeq 2$ pc and an extended tail of
larger objects.  Field contamination of the total sample consists
mostly of very faint or starlike objects that can be objectively
culled out of the GC data.
\item The range of intrinsic $r_h$ sizes is large for all types
of clusters, metal-poor and metal-rich alike.  However,
the blue, lower-metallicity GCs are systematically larger on average
than the red, higher-metallicity ones by about 17\%.  This difference is
statistically highly significant, and holds up at all GC luminosities
and all galactocentric distances.  These results suggest that the
correlation between cluster scale size and metallicity is intrinsic
to the clusters, a result of their formation or later internal 
dynamical evolution, rather than reaction to the external tidal field.
\item The median cluster sizes, of both types, also scale with
projected galactocentric distance as $r_h \sim R_{gc}^{0.11}$.
The data extend outward to approximately $5 R_{eff}$ into the halos
of the galaxies, and thus probe more extended regions of the halos
than do most previous studies.
These results fit in well with data from other galaxies indicating
that clusters forming in shallower, less extreme tidal fields
end up with larger characteristic sizes.
\item The median scale size of the clusters also depends weakly
and nonlinearly on cluster luminosity, in the sense that GCs more
luminous than $M_I \lesssim -11$ become progressively bigger.
The effective radii of the very most luminous clusters known
are almost twice as large as the bulk of the clusters at $M_I > -11$.
Most of these luminous GCs, however, are smaller than UCDs of
the same luminosity.
\item The blue GC sequence exhibits a clear mass/metallicity relation
(MMR), but it is not linear in color versus magnitude.  Its shape
is rather that of a gradual changeover from nearly vertical
(for the lower-luminosity range $M_I \gtrsim -9.5$) to a steeper
slope at higher luminosity.  The uppermost luminosity range corresponds
to a heavy-element scaling with cluster mass of $Z \sim M^{0.5}$.
\item The red GC sequence is nearly vertical at all luminosities.
\item The shape of the MMR for both the blue and red sequences
is consistent with the basic interpretation that it is due to
cluster self-enrichment during their formation epoch.
Though theoretical models of this type are still in early stages,
the data rule out models in which the star formation efficiency
$f_{\star}$ within a protocluster scales strongly with cluster mass.
Instead, models in which $f_{\star} \sim const$ and the heavy-element
retention efficiency $f_Z$ increases with cluster mass produce a
more realistic match to the observations.
\item The slope of the MMR along the blue sequence does not depend
strongly on cluster size $r_h$.  The data contain a hint that the
biggest GCs ($r_h > 4$ pc) follow a slightly bluer sequence at the
high-luminosity end.  If so, it would be consistent with the idea
that self-enrichment was less effective (for a given GC mass) in
more extended clusters with shallower potential wells.
\item The data from these giant elliptical galaxies also show that
the MMR depends weakly, if at all, on projected galactocentric distance.
If a trend exists, it is in the sense that the MMR has lower amplitude
at larger $R_{gc}$; this may indicate that the inner GCs have lost more
mass than the outer ones due to tidal stripping and dynamical evolution,
so that they have a larger contrast between their initial masses 
(which determine the amount of self-enrichment) and their present-day
masses (which we directly observe).
\item Both the red and blue GC subsystems exhibit shallow metallicity
gradients:  they become systematically more metal-poor at larger
galactocentric distance.  The gradients in both cases scale as
$Z \sim R_{gc}^{-0.10 \pm 0.022}$ and, though small, are clearly
measurable in this large dataset.  These results agree well with
the limited data available for other elliptical and spiral galaxies,
and are likely to point to a combination of the amount of initial
enrichment plus later merging histories.
\end{enumerate}

\section{Final Caveats}

Phenomena such as the MMR, the GC size distributions, 
and the metallicity gradients within 
the components of the globular cluster systems, have the potential
to provide rich grounds for modelling and interpretation, but
are still imperfectly understood.
All the attempts to date to describe what is happening at the high-mass end
of the globular cluster distribution continue to be data-limited.
Even in the present
study, only a few hundred GCs brighter than $M_I = -10.5$ 
($10^6 M_{\odot}$) are available to work with, and these are spread
over several individual galaxies within which the conditions
of formation could have differed.  Breaking down that
part of the sample further by metallicity, luminosity, or galactocentric
distance then leads us quickly into a shadowy zone where conclusions
are tentative.  

On strictly observational grounds, the correlation between GC color
and luminosity called the MMR 
continues to be a difficult effect to measure, and
should still be considered as incompletely established.
Even the data discussed here are
not yet enough to address important questions such as whether or not
the MMR may differ in amplitude from galaxy to galaxy, or if it is
genuinely absent in some galaxies.
Careful attempts have been made here to
reduce or eliminate several worries in previous studies, including
small sample size; systematic errors in 
PSF-fitting or fixed-aperture photometry due to differing
cluster sizes; low signal-to-noise; and background light
gradients.  If the present data still hide systematic errors, they must be
due to some subtle and ill-understood features of the image
preprocessing or measurement algorithms that would preferentially
affect the photometry of the brightest objects instead of the faintest ones, and
the blue GCs rather than the only-slightly-redder ones.  
In the longer term, probably the best security 
against such systematics will be to
carry out more GCS studies with different telescopes and cameras,
still larger samples, and especially metallicity-sensitive color indices.
Some comparative work of this type is already in the literature,
but there are still only a few large galaxies in which the MMR can
be described as plainly visible.

On theoretical grounds, the central concept of self-enrichment 
is an attractive one, but its most important assumption is that it requires the 
main star formation period in a
proto-GC to last long enough for the most massive stars to form,
evolve, and eject their enriched material
before most of the low-mass stars have finished forming themselves.  
The often-used arbitrary assumption of an ``instantaneous starburst'' cannot
be correct in real star-forming regions, but for this 
model to work, the star formation period
needs to stretch over at least 10 Myr and perhaps as long as
$20 - 30$ Myr \citep[see][]{bai08}.
Contemporary observations of dense, massive star-forming regions point
encouragingly in that direction, but are not yet definitive.  
If the average formation time in the most massive GCs 
is as short as $\sim 5$ Myr, for example, then the self-enrichment model would
be in serious trouble.

Many other giant and supergiant ellipticals whose globular cluster
systems have not yet been studied are well within reach
of the HST and large ground-based telescopes. 
These hold out obvious prospects
for increasing the GCS sample sizes further by large factors.
Major strides can be made towards understanding this intriguing 
correlation further.

\acknowledgments
This work was supported by the Natural Sciences and Engineering
Research Council of Canada through research grants to WEH, and
by the Killam Foundation of the Canada Council through a research fellowship.
Many thanks to S{\o}ren Larsen, who provided advice about the ISHAPE
code and implemented several detailed improvements to the code that helped
this work.  I am grateful to Gretchen Harris for a close reading of
the manuscript.


\clearpage


\begin{deluxetable}{ccccccccc}
\tabletypesize{\footnotesize}
\tablecaption{Brightest Cluster Galaxies Imaged with the ACS \label{basicdata}}
\tablewidth{0pt}
\tablehead{
\colhead{Galaxy} & \colhead{Cluster} & \colhead{Redshift} & 
\colhead{$M_V^T$} & \colhead{$E(B-I)$} & \colhead{$(m-M)_I$} & 
\colhead{$R_{eff}$} & \colhead{$B,I$ Exposures} \\
& \colhead{or Group} & \colhead{(km s$^{-1}$)} & & & \colhead{($M_I<-8.4$)} & 
\colhead{(arcmin)} & \colhead{(sec)}  \\
}

\startdata
NGC 1407 & Eridanus & 1627 & $-22.35$ & 0.16 & 31.96 & $1.17'$ & $2 \times 750, 2\times 340$ \\
NGC 3348 & CfA 69 & 2837 & $-22.13$ & 0.17 & 33.18 & $0.44'$ & $6 \times 1200, 4 \times 530$ \\
NGC 3258 & Antlia & 3129 & $-21.87$ & 0.20 & 33.23 & $0.50'$ & $4 \times 1340, 4 \times 570$ \\
NGC 3268 & Antlia & 3084 & $-21.96$ & 0.24 & 33.27 & $0.60'$ & $4 \times 1340, 4 \times 570$ \\
NGC 4696 & Cen30 & 2926 & $-23.31$ & 0.23 & 33.29 & $0.74'$ & $4 \times 1360, 4 \times 580$ \\
NGC 7626 & Pegasus I & 3405 & $-22.35$ & 0.17 & 33.57 & $0.64'$ & $6 \times 1310, 2 \times 1300$ & \\

\enddata

\end{deluxetable}

\begin{deluxetable}{ccccc}
\tabletypesize{\footnotesize}
\tablecaption{Characteristics of the Point-Spread Functions \label{psftab}}
\tablewidth{0pt}
\tablehead{
\colhead{Galaxy} & \colhead{FWHM($B$)} & \colhead{$N_{\star}(B)$} & 
\colhead{FWHM($I$)} & \colhead{$N_{\star}(I)$} \\
}

\startdata
NGC 1407 & $0.091''$ & 10 & $0.090''$ & 5 \\
NGC 3348 & $0.081''$ & 8 & $0.089''$ & 10 \\
NGC 3258 & $0.098''$ & 6 & $0.091''$ & 18 \\
NGC 3268 & $0.096''$ & 8 & $0.084''$ & 15 \\
NGC 4696 & $0.102''$ & 25 & $0.095''$ & 52 \\
NGC 7626 & $0.102''$ & 6 & $0.096''$ & 11 \\
\\
Mean & $0.095''$ & 10 & $0.091''$ & 19  \\
     &$\pm 0.0033''$ && $\pm0.0018''$ & \\

\enddata

\end{deluxetable}

\begin{deluxetable}{ccccc}
\tabletypesize{\footnotesize}
\tablecaption{Median Sizes of Globular Clusters versus Luminosity\label{rhmed}}
\tablewidth{0pt}
\tablehead{
\colhead{$M_I$ Range} & \colhead{$n(blue)$} & \colhead{$\langle r_h \rangle (blue)$} & 
\colhead{$n(red)$} & \colhead{$\langle r_h \rangle (red)$} \\
& & \colhead{(pc)} & & \colhead{(pc)} \\
}

\startdata
$(-12.0, -12.7)$ & 10  & 4.63 & 15 & 3.86 \\
$(-11.5, -12.0)$ & 17  & 2.21 & 32 & 2.85 \\
$(-11.0, -11.5)$ & 105 & 3.10 & 81 & 2.63 \\
$(-10.5, -11.0)$ & 186 & 2.80 & 150 & 2.36 \\
$(-10.0, -10.5)$ & 301 & 2.65 & 263 & 2.21 \\
$(-9.5, -10.0)$  & 419 & 2.51 & 313 & 2.21 \\
$(-9.0, -9.5)$   & 289 & 2.80 & 139 & 2.65 \\
\\
Milky Way & 48 & 3.06 & 18 & 2.40 \\
\enddata

\end{deluxetable}

\begin{deluxetable}{ccccc}
\tabletypesize{\footnotesize}
\tablecaption{Median Sizes of Globular Clusters versus $R_{GC}$ \label{rhmed_rgc}}
\tablewidth{0pt}
\tablehead{
\colhead{$R_{GC}/R_{eff}$ } & \colhead{$n(blue)$} & \colhead{$\langle r_h \rangle (blue)$} & 
\colhead{$n(red)$} & \colhead{$\langle r_h \rangle (red)$} \\
& & \colhead{(pc)} & & \colhead{(pc)} \\
}

\startdata
$(0.0 - 0.5)$ & 50  & 2.28 & 56 & 1.95 \\
$(0.5 - 1.0)$ & 193 & 2.51 & 185 & 2.06 \\
$(1.0 - 1.5)$ & 232 & 2.50 & 175 & 2.06 \\
$(1.5 - 2.0)$ & 227 & 2.36 & 161 & 2.18 \\
$(2.0 - 2.5)$ & 221 & 2.63 & 141 & 2.36 \\
$(2.5 - 3.0)$ & 159 & 2.95 & 127 & 2.33 \\
$(3.0 - 3.5)$ & 119 & 2.95 & 55 & 2.36 \\
$(3.5 - 4.0)$ & 56 & 2.95 & 32 & 2.58 \\
$(4.0 - 5.0)$ & 37 & 3.15 & 31 & 2.77 \\

\enddata

\end{deluxetable}

\clearpage

\begin{deluxetable}{lcccccl}
\tabletypesize{\footnotesize}
\tablecaption{RMIX Bimodal Fits to the Individual Galaxies \label{rmixfits}}
\tablewidth{0pt}
\tablehead{
\colhead{$I$ range} & \colhead{$N$(bin)} &
\colhead{$\mu_1$} & \colhead{$\sigma_1$} & \colhead{$\mu_2$} & 
\colhead{$\sigma_2$} & \colhead{$p_1$} \\
}

\startdata
NGC 1407 \\
$21.0-22.0$ & 113 & $1.885 \pm 0.058$ & $0.102 \pm0.033$ & $2.201\pm 0.055$ & $0.120\pm 0.033$ & 0.413 \\
$22.0-23.0$ & 287 & $1.789 \pm 0.023$ & $0.123 \pm0.017$ & $2.225\pm 0.015$ & $0.121\pm 0.011$ & 0.367 \\
$23.0-23.5$ & 219 & $1.743 \pm 0.018$ & $0.078 \pm0.016$ & $2.177\pm 0.020$ & $0.159\pm 0.016$ & 0.298 \\
$23.5-24.0$ & 216 & $1.733 \pm 0.109$ & $0.109 \pm0.018$ & $2.191\pm 0.016$ & $0.158\pm 0.015$ & 0.331 \\
\\
NGC 3348 \\
$21.0-23.0$ &  96 & $1.864\pm 0.061$ & $0.142\pm 0.035$ & $2.157\pm 0.060$ & $0.104\pm 0.031$ & 0.663 \\
$23.0-24.0$ & 214 & $1.640\pm 0.012$ & $0.074\pm 0.011$ & $2.085\pm 0.015$ & $0.148\pm 0.012$ & 0.318 \\
$24.0-25.0$ & 348 & $1.642\pm 0.011$ & $0.078\pm 0.010$ & $2.085\pm 0.015$ & $0.165\pm 0.012$ & 0.313 \\
\\
NGC 3258 \\
$21.0-22.5$ &  91 & $ 1.836\pm 0.032$ & $0.097\pm 0.021$ & $2.130\pm 0.059$ & $0.112\pm 0.035$ & 0.614 \\
$22.5-23.0$ & 113 & $ 1.738\pm 0.024$ & $0.063\pm 0.022$ & $1.997\pm 0.039$ & $0.176\pm 0.020$ & 0.267 \\
$23.0-23.5$ & 197 & $ 1.778\pm 0.025$ & $0.090\pm 0.017$ & $1.133\pm 0.037$ & $0.134\pm 0.025$ & 0.483 \\
$23.5-24.0$ & 279 & $ 1.753\pm 0.013$ & $0.085\pm 0.009$ & $2.132\pm 0.025$ & $0.137\pm 0.019$ & 0.534 \\
$24.0-24.5$ & 370 & $ 1.740\pm 0.009$ & $0.090\pm 0.007$ & $2.168\pm 0.020$ & $0.137\pm 0.013$ & 0.550 \\
$24.5-25.0$ & 402 & $ 1.740\pm 0.010$ & $0.097\pm 0.008$ & $2.146\pm 0.015$ & $0.131\pm 0.012$ & 0.513 \\
\\
NGC 3268\\
$21.0-22.5$ &  84 &  $1.815\pm 0.017$ & $0.026\pm 0.018$ & $2.050\pm 0.023$ & $0.154\pm 0.015$ & 0.128 \\
$22.5-23.0$ & 101 &  $1.799\pm 0.019$ & $0.030\pm 0.066$ & $2.050\pm 0.046$ & $0.170\pm 0.028$ & 0.107 \\
$23.0-23.5$ & 162 &  $1.759\pm 0.014$ & $0.045\pm 0.013$ & $2.097\pm 0.022$ & $0.198\pm 0.015$ & 0.156 \\
$23.5-24.0$ & 249 &  $1.768\pm 0.010$ & $0.081\pm 0.008$ & $2.187\pm 0.019$ & $0.152\pm 0.015$ & 0.471 \\
$24.0-24.5$ & 334 &  $1.878\pm 0.014$ & $0.096\pm 0.010$ & $2.183\pm 0.024$ & $0.165\pm 0.018$ & 0.449 \\
$24.5-25.0$ & 377 &  $1.768\pm 0.015$ & $0.108\pm 0.011$ & $2.197\pm 0.024$ & $0.163\pm 0.017$ & 0.477 \\
\\ 
NGC 4696 \\
$21.0-22.5$ & 127 & $1.907\pm 0.018$ & $0.100\pm 0.000$ & $2.223\pm 0.030$ & $0.150\pm 0.000$ &0.517 \\
$22.5-23.0$ & 192 & $1.901\pm 0.046$ & $0.084\pm 0.027$ & $2.170\pm 0.065$ & $0.141\pm 0.031$ &0.378 \\
$23.0-23.5$ & 297 & $1.846\pm 0.010$ & $0.074\pm 0.008$ & $2.220\pm 0.018$ & $0.152\pm 0.014$ &0.389 \\
$23.5-24.0$ & 374 & $1.814\pm 0.009$ & $0.081\pm 0.007$ & $2.234\pm 0.015$ & $0.153\pm 0.012$ &0.441 \\
$24.0-24.5$ & 636 & $1.812\pm 0.010$ & $0.091\pm 0.007$ & $2.213\pm 0.019$ & $0.174\pm 0.013$ &0.445 \\
$24.5-25.0$ & 658 & $1.847\pm 0.012$ & $0.123\pm 0.008$ & $2.278\pm 0.023$ & $0.156\pm 0.016$ &0.589 \\
$25.0-25.5$ & 516 & $1.801\pm 0.026$ & $0.122\pm 0.022$ & $2.149\pm 0.053$ & $0.234\pm 0.026$ &0.319 \\
\\
NGC 7626 \\
$21.0-22.5$ &  76 & $1.832\pm 0.049$ & $0.100\pm 0.000$ & $2.116\pm 0.030$ & $0.150\pm 0.000$ &0.194 \\
$22.5-23.5$ & 218 & $1.808\pm 0.018$ & $0.106\pm 0.012$ & $2.203\pm 0.020$ & $0.137\pm 0.015$ &0.437 \\
$23.5-24.0$ & 228 & $1.801\pm 0.023$ & $0.120\pm 0.017$ & $2.219\pm 0.025$ & $0.132\pm 0.017$ &0.494 \\
$24.0-24.5$ & 323 & $1.728\pm 0.010$ & $0.074\pm 0.007$ & $2.158\pm 0.015$ & $0.160\pm 0.012$ &0.344 \\
$24.5-25.0$ & 385 & $1.752\pm 0.020$ & $0.112\pm 0.014$ & $2.184\pm 0.028$ & $0.173\pm 0.019$ &0.426 \\
$25.0-25.5$ & 375 & $1.741\pm 0.024$ & $0.115\pm 0.017$ & $2.182\pm 0.037$ & $0.180\pm 0.024$ &0.464 \\

\enddata

\end{deluxetable}

\clearpage

\begin{deluxetable}{lcccccl}
\tabletypesize{\footnotesize}
\tablecaption{RMIX Bimodal Fits to the Combined Sample\label{rmixall}}
\tablewidth{0pt}
\tablehead{
\colhead{$M_I$ range} & \colhead{$N$(bin)} &
\colhead{$\mu_1$} & \colhead{$\sigma_1$} & \colhead{$\mu_2$} & 
\colhead{$\sigma_2$} & \colhead{$p_1$} \\
}

\startdata
$(-12.7, -11.5)$ & 285 & $1.757 \pm 0.038$ & $0.102 \pm0.032$ & $1.944\pm 0.029$ & $0.209\pm 0.013$ & $0.181 \pm 0.109$ \\
$(-11.5, -11.0)$ & 359 & $1.642 \pm 0.017$ & $0.100 \pm0.000$ & $1.902\pm 0.019$ & $0.180\pm 0.000$ & $0.286 \pm 0.057$ \\
$(-11.0, -10.5)$ & 627 & $1.669 \pm 0.031$ & $0.121 \pm0.016$ & $1.991\pm 0.043$ & $0.140\pm 0.020$ & $0.531 \pm 0.114$ \\
$(-10.5, -10.0)$ & 1039 & $1.595 \pm 0.014$ & $0.098 \pm0.009$ & $1.956\pm 0.020$ & $0.176\pm 0.012$ & $0.345 \pm 0.048$ \\
$(-10.0, -9.5)$ & 1448 & $1.590 \pm 0.008$ & $0.107 \pm0.005$ & $2.012\pm 0.009$ & $0.146\pm 0.006$ & $0.439 \pm 0.021$ \\
$(-9.5, -9.0)$ & 2022 & $1.555 \pm 0.005$ & $0.091 \pm0.004$ & $1.985\pm 0.008$ & $0.164\pm 0.006$ & $0.414 \pm 0.017$ \\
$(-9.0, -8.5)$ & 2526 & $1.570 \pm 0.006$ & $0.108 \pm0.005$ & $1.993\pm 0.010$ & $0.165\pm 0.006$ & $0.438 \pm 0.020$ \\
$(-8.5, -8.0)$ & 2333 & $1.554 \pm 0.009$ & $0.113 \pm0.006$ & $1.981\pm 0.013$ & $0.170\pm 0.008$ & $0.445 \pm 0.027$ \\
\\
$(-12.4, -10.4)$ & 848 & $1.672 \pm 0.008$ & $0.119 \pm0.006$ & $1.970\pm 0.008$ & $0.184\pm 0.007$ & $0.400 \pm 0.011$ \\
$(-10.4, -8.4)$ & 4111 & $1.563 \pm 0.003$ & $0.099 \pm0.002$ & $1.990\pm 0.005$ & $0.179\pm 0.004$ & $0.397 \pm 0.011$ \\
\\
\enddata

\end{deluxetable}

\begin{deluxetable}{rrccrcc}
\tabletypesize{\footnotesize}
\tablecaption{Radial Metallicity Gradients \label{radialz}}
\tablewidth{0pt}
\tablehead{
\colhead{Galaxy} & \colhead{$N$(blue)} &
\colhead{$a$} & \colhead{$b$} & \colhead{$N$(red)} & 
\colhead{$a$} & \colhead{$b$} \\
}

\startdata
NGC1407 & 115 & $1.630 \pm 0.010$ & $-0.033 \pm 0.035$ & 225 & $2.036 \pm 0.009$ & $-0.027 \pm 0.028$ \\
NGC3268 & 450 & $1.597 \pm 0.007$ & $-0.042 \pm 0.019$ & 292 & $1.980 \pm 0.009$ & $+0.002 \pm 0.025$ \\
NGC3258 & 386 & $1.604 \pm 0.008$ & $-0.033 \pm 0.025$ & 306 & $1.970 \pm 0.007$ & $-0.030 \pm 0.024$ \\
NGC3348 & 141 & $1.581 \pm 0.012$ & $-0.017 \pm 0.035$ & 148 & $1.989 \pm 0.011$ & $-0.036 \pm 0.038$ \\
NGC4696 & 590 & $1.628 \pm 0.004$ & $-0.038 \pm 0.014$ & 576 & $2.010 \pm 0.006$ & $-0.011 \pm 0.018$ \\
NGC7626 & 310 & $1.615 \pm 0.007$ & $-0.060 \pm 0.023$ & 326 & $2.035 \pm 0.007$ & $-0.068 \pm 0.026$ \\
\\
All 6& 1992 & $1.615 \pm 0.003$ & $-0.0379 \pm 0.0068$ & 1873 & $2.008 \pm 0.003$ & $-0.0371 \pm 0.0079$ \\
\\
\enddata

\end{deluxetable}

\clearpage

\clearpage

\clearpage

\clearpage

\clearpage

\clearpage

\clearpage

\clearpage

\clearpage

\clearpage

\clearpage

\clearpage

\clearpage

\clearpage

\clearpage

\clearpage

\clearpage

\clearpage

\clearpage

\clearpage

\clearpage

\clearpage

\clearpage

\clearpage

\clearpage

\clearpage

\clearpage

\clearpage

\clearpage

\clearpage

\clearpage

\clearpage

\clearpage

\clearpage

\end{document}